\documentclass{aa}  
\usepackage{hyperref}
\hyperlink{}
\hypersetup{
    colorlinks=true,
    citecolor=blue,   
    linkcolor=blue,   
    urlcolor=blue     
}
\usepackage{pdflscape}
\usepackage{multirow}
\usepackage{booktabs}
\usepackage{array}
\usepackage{tabularx}
\usepackage{ragged2e}
\usepackage{graphicx}
\usepackage{pgffor} 
\usepackage{orcidlink} 
\usepackage{academicons}
\usepackage{afterpage}

\usepackage{txfonts}
\usepackage{xcolor}
\usepackage{ragged2e}
\usepackage[utf8]{inputenc}
\definecolor{gr}{rgb}{0.2, 0.20, 0.80}

\newcommand{\gsim}{\ \raise -2.truept\hbox{\rlap{\hbox{$\sim$}}\raise 5.truept\hbox{$>$}\ }}

\definecolor{emerald}{rgb}{0.4,0.66,0.31}

\DeclareUnicodeCharacter{2212}{-}

\begin{document}

   \title{VEGAS-SSS: Tracing Globular Cluster Populations in the Interacting NGC\,3640 Galaxy Group }

   \subtitle{}

\author{Marco Mirabile\inst{1,2,3}\orcidlink{https://orcid.org/0009-0007-6055-3933}
          \and
          Michele Cantiello\inst{1}\orcidlink{https://orcid.org/0000-0003-2072-384X}
          \and
        Pratik Lonare\inst{1}\orcidlink{ https://orcid.org/0009-0000-0028-0493}
        \and
        Rossella Ragusa\inst{4}
        \and
        Maurizio Paolillo\inst{1,4}
        \and 
        Nandini Hazra\inst{1,2}
        \and
        Antonio La Marca\inst{5,6}\orcidlink{https://orcid.org/0000-0002-7217-5120}
        \and
        Enrichetta Iodice\inst{4}
        \and 
        Marilena Spavone\inst{4}
        \and
        Steffen Mieske\inst{7}\orcidlink{https://orcid.org/0000-0003-4197-4621}
        \and
        Marina Rejkuba\inst{8}\orcidlink{https://orcid.org/0000-0002-6577-2787}
        \and 
        Michael Hilker\inst{8}
        \and 
        Gabriele Riccio\inst{1}\orcidlink{https://orcid.org/0000-0002-6399-2129}
        \and 
        Rebecca A. Habas\inst{1}\orcidlink{https://orcid.org/0000-0002-4033-3841}
        \and 
        Enzo Brocato\inst{1}
        \and 
        Pietro Schipani\inst{4}
        \and
        Aniello Grado\inst{4}
        \and
        Luca Limatola\inst{4}\orcidlink{https://orcid.org/0000-0002-1896-8605}
          }

\institute{INAF Osservatorio Astr. d’Abruzzo, Via Maggini, 64100 Teramo, Italy\\
              \email{marco.mirabile@inaf.it}
         \and
             Gran Sasso Science Institute,  Viale Francesco Crispi 7, 67100 L’Aquila, Italy
         \and
             University of Naples “Federico II”, C.U. Monte Sant’Angelo, Via Cinthia, 80126 Naples, Italy
        \and
        INAF − Astronomical Observatory of Capodimonte, Salita Moiariello 16, I-80131, Naples, Italy
        \and 
        SRON Netherlands Institute for Space Research, Landleven 12, 9747 AD Groningen, The Netherlands
        \and
        Kapteyn Astronomical Institute, University of Groningen, Postbus 800, 9700 AV Groningen, The Netherlands
        \and
        European Southern Observatory, Alonso de Cordova 3107, Vitacura, Santiago, Chile
        \and
        European Southern Observatory, Karl-Schwarzschild-Strasse 2, 85748 Garching bei München, Germany
}

   \date{}

  \abstract
   {Globular clusters (GCs) are among the oldest stellar systems in the universe. As such, GCs population are valuable fossil tracers of galaxy formation and interaction history. This paper is part of the VEGAS-SSS series, which focuses on studying the properties of small stellar systems (SSSs) in and around bright galaxies.}
   {We use the multi-band wide-field images obtained with the VST to study the properties of the globular cluster (GC) population in an interacting pair of galaxies.}
   {We derived $ugri$ photometry over $1.5\times 1.5$ sq. degrees centered on the galaxy group composed by two elliptical galaxies: NGC\,3640 and its fainter companion NGC\,3641. We studied the GC system properties from both the $ugri$ and $gri$ matched catalogs. GC candidates were identified based on a combination of photometric properties (colors, magnitudes) and morphometric criteria (concentration index, elongation, FWHM, etc.), using sources with well-defined classifications from spectroscopic or imaging data available in the literature and numerical simulations as references. The selection criteria were also applied to empty fields to determine a statistical background correction for the number of identified GC candidates.
   }
   { The 2D density maps of GCs appear to align with the diffuse light patches resulting from merging events of the galaxies. The highest density peak of GCs is observed to be on NGC\,3641 rather than NGC\,3640, despite the latter being the more massive galaxy. The azimuthal averaged radial density profiles in both galaxies reveal that the GC population extends beyond the galaxy light profile and indicate the likely presence of an intra-group GC component. A color bimodality in $(u{-}r)$ and $(g{-}i)$ is observed for NGC\,3641, whereas NGC\,3640 shows a broad unimodal distribution. Analysis of the GC Luminosity Function indicates that both galaxies are roughly located at the same distance ($\sim$ 27 Mpc). We provide an estimate of the total number of GCs, and determine the specific frequency for NGC\,3640, $S_{\rm N}$ =2.0$\pm$0.6, which aligns with expectations, while for NGC\,3641 we find a large $S_{\rm N}$ = 4.5$\pm$1.6. }
  {}

\maketitle

\section{Introduction}
A globular cluster (GC) is among the simplest astrophysical stellar systems in the universe. It is a very dense system of old stars with a typical age t $\geq$ 10 Gyr. Its half-light radius is typically  $\sim2.5pc$, and it exhibits a mean absolute magnitude in the visual passband of $M_V\sim$ -7.5mag, and with stellar masses between $\sim$ $M_\odot^4-M_\odot^6$ \citep{brodie06}. 

Initially, GCs were thought to host single-age and single-metallicity stellar populations. However, studies of the Milky Way (MW) GCs and old star clusters in the Magellanic Clouds and other nearby galaxies have revealed the ubiquitous presence of chemical inhomogeneities indicative of quite complex chemical evolution characterised by multiple stellar populations as recently reviewed by \citet{gratton19}. Although they are more than originally believed, GCs generally exhibit simpler stellar populations compared to their host galaxies in terms of metallicity and age distribution, allowing for more precise analysis of their properties due to their less intricate star-formation histories. Additionally, the old age and high luminosity of GCs make them valuable tools for extragalactic studies. The old age makes GC systems relics that trace the formation and evolution of a galaxy and its surrounding environment. Indeed, they are tracers of the main star formation events in the early universe and later trace the dynamical evolution of their host galaxy \citep{brodie06}. Thanks to their brightness, GCs can be observed at large distances, especially in spheroidal galaxies where GCs appear as bright point sources against the smooth light profile of the galaxy. Systematic investigations of GC systems in external galaxies have provided valuable insights into their properties such as: the luminosity function, spatial distribution, color profiles and kinematics. These properties serve as effective indicators of the past formation and evolution of galaxies \citep{harris01,brodie06}.
Despite their attractiveness as tracers of galaxy evolution, much remains to be understood \citep{forbes2021}.

The GC systems in external galaxies typically exhibit a universal luminosity function that is well approximated by a Gaussian, characterized by a peak -- known as the turnover magnitude (TOM) --at a nearly constant absolute magnitude of $M_V^{TOM}\sim$ -7.5 mag \citep{harris01,rejkuba12}. The width of the Gaussian distribution depends on the mass (luminosity) of the galaxy \citep{villegas10} with more massive galaxies showing wider luminosity/mass distribution function. The near-universality of the GC luminosity function (GCLF) has led to its use as a standard candle, serving as a secondary distance indicator \citep{ferrarese20}. 

With the advent of deep wide-field surveys, studies of the spatial distributions of GCs have revealed the existence of rich GC substructures within galaxy groups and clusters \citep{Geisler96,cantiello20,lambert2020,ragusa2022FrASS...952810R,Marleau24a,Saifollahi24}.
Thanks to their intrinsic luminosity, GCs can be studied out to very large galactocentric radii (far beyond 10 $R_{e}$\footnote{$R_e$ is the effective radius of the galaxy, defined as the distance within which half of the galaxy's light is concentrated.}), where the galaxy's surface brightness fades away  \citep[e.g][]{durrell2014,dabrusco16, iodice17a, bassino2017,dabrusco2022}. The association of GCs with the host cluster/group allows for a direct exploration of the substructures therein, such as their distribution and density. These findings can then be used to validate predictions from galaxy evolution models.

In the last decade, the VEGAS (VST Survey Telescope Early-type GAlaxy Survey) survey has been collecting deep multi-band optical imaging data for galaxy groups and clusters \citep[e.g.][]{,cantiello15,spavone18,ragusa2021A&A...651A..39R,ragusa2023A&A...670L..20R,lamarca22a,lamarca22b}. Its aim is to study the faintest low-surface brightness features in the group/cluster and to characterize the GC population around the observed systems \citep{capaccioli15,iodice20,ragusa2022FrASS...952810R}.

In this study, we present an analysis of the GC system in the interacting pair dominated by NGC\,3640. NGC\,3640 and its companion NGC\,3641 were classified as an E3 and an E galaxy, respectively, by \citet{devaucouleurs91}. They are $\sim27$ Mpc away based on surface brightness fluctuation (SBF) distance estimates \citep{tonry01,tully13}. NGC\,3640 is part of a loose group composed of at least eight galaxies \citep{Madore2004}. This group is considered to be dynamically young, as evidenced by the absence of X-ray emissions \citep{Osmond2004}.

\citet[P88 hereafter]{Prugniel1998}  used both photometric and spectroscopic data to investigate the morphology and kinematics of NGC\,3640. Through the analysis of the surface brightness profile, they identified signatures indicative of past merging events, primarily manifested as shells. Additionally, P88 observed a likely dust lane along the minor axis of the galaxy (from north to south), pointing toward NGC\,3641. With spectroscopic data, they determined that NGC\,3640 is a fast rotator along its major axis (from east to west) with a $V/\sigma$ ratio of 1.5. P88 proposed that NGC\,3640 is in an advanced stage of merging, although not yet completed due to the absence of strong nuclear radio emission, which is a typical post-merger sign. They suggest that NGC\,3640 likely merged with a gas-poor disk galaxy.

\citet{Schweizer1992} conducted an analysis on a sample of E and S0 galaxies, estimating heuristic merger ages based on $UBV$ colors for each galaxy. In this work, NGC\,3640 is identified as a young elliptical candidate, indicating formation through mergers less than $\sim$ 7 Gyr ago.
Near-IR color studies of NGC\,3640 carried out by \citet{Silva1998a} argue against a major starburst occurring within the last 3 Gyr. Additionally, \citet{denicolo05} and \citet{Brough2007} investigated the stellar populations of NGC 3640 using long slit spectra. \citet{denicolo05} reported an age of the center of 2.5 Gyr, whereas \citet{Brough2007} found an older (4 Gyr) and less metal-rich center when examining the same region. Moreover, \citet{Brough2007} concluded that NGC\,3640 is consistent with having undergone a dissipational merger up to $\sim$7.5 Gyr ago.
Consistent with that, \citet{Hibbard2003} observed no HI emission from NGC\,3640, but detected it from an inclined low surface brightness (LSB) dwarf spiral galaxy located at coordinates 11h21m51.95s, +03°24'17'' (J2000), with a radial velocity consistent with that of NGC\,3640\footnote{\citet{Hibbard2003} used a radial velocity of $V_{opt}=1314~{\rm km~s^{-1}}$ for NGC\,3640, resulting in $\Delta V=62~{\rm km~s^{-1}}$.} ($V_{\rm H}=1180~{\rm km~s^{-1}}$).

\citet{gebhardt99} conducted the first and most detailed study, to the best of our knowledge, about the GC population of NGC\,3640. They observed NGC\,3640 with the \textit{Hubble} Space Telescope (HST) and estimated a lower limit for the total GC population of $\sim50$ GCs. The GC candidates showed a color distribution peaked at $V-I=1.08^{+0.080}_{-0.035}$ mag.

Despite the extensive literature on NGC\,3640, none of the analyses have definitively confirmed or ruled out any interactions with the less massive companion galaxy, NGC\,3641, which is at a projected angular distance of only $\sim 2.5$ $arcmin$ ($\sim20$ kpc) from NGC\,3640. The two galaxies show a quite high relative radial velocity of $\sim500\ km/s$ which may suggest that if they interacted, it should have been a high speed encounter.

\begin{table}[h!]
\small
\centering
\caption{Main properties for NGC 3640 and NGC 3641.}

\begin{tabular}{cccc}
\toprule
\multirow{2}{*}{Quantity} & \multicolumn{2}{c}{Galaxy} & \multirow{2}{*}{Note} \\
\cmidrule(lr){2-3}
 (units) & NGC 3640 & NGC 3641 & \\
\midrule
\vspace{0.1cm}
R.A. (deg) & 170.278542 & 170.286725 & (1) \\
\vspace{0.1cm}
Dec. (deg) & 3.234833 & 3.194593 & (2) \\
\vspace{0.1cm}
Morphology & E3 & E & (3) \\
\vspace{0.1cm}
$v_{hel}$ (${\rm km~s^{-1}}$) &    1296 $\pm$ 5    & 1789$\pm$ 5   & (4)  \\
\vspace{0.1cm}
D (Mpc) & $27 \pm 2$ & $27 \pm 3$ & (5) \\
\vspace{0.1cm}
$R_e$ (arcsec) & 30.9 & 9.33 & (6) \\
\vspace{0.1cm}
$m_V$ (mag) & $10.4 \pm 0.1$ & $13.2 \pm 0.2$ & (7) \\
\vspace{0.1cm}
$m_z$ (mag) & $9.787 \pm 0.002$ & $12.499 \pm 0.002$ & (8) \\
\vspace{0.1cm}
$M_V$ (mag) & $-21.7 \pm 0.2$ & $-19 \pm 0.3$ & (9) \\
\vspace{0.1cm}
$M_z$ (mag) & $-22.4 \pm 0.2$ & $-19.7 \pm 0.3$ & (10) \\
\vspace{0.1cm} 
$\sigma_{g}^{GCLF}$ (mag) & $1.14 \pm 0.2$ & $0.9 \pm 0.3$ & (11) \\
$\sigma_{z}^{GCLF}$ (mag) & $1.12 \pm 0.2$ & $0.9 \pm 0.3$ & (12) \\
\bottomrule
\end{tabular}

\begin{justify}
\textbf{Notes}: (1)-(2) coordinates taken from NASA/IPAC Extragalactic Database (NED); (3) morphological type taken from \citet{Vaucouleurs1991}; (4) heliocentric velocity taken from \citet{Cappellari2011}; (5) distance of the two galaxies obtained from the SBF measurements reported in \citet{tully09,tully23}; (6) effective radii taken from \citet{Cappellari2011}; (7) apparent magnitude in the $V$-band taken from \citet{galex2007}; (8)  apparent magnitude in the $z$-band taken from the Sloan Digital Sky Survey \citep{sdss2009};
(9)-(10) absolute magnitudes in the $V-$ and $z-$bands assuming the apparent magnitudes in (7)-(8) and distances in (5); 
(11)-(12) expected widths of the luminosity function in the $g$- and $z$-bands, estimated using Eqs. 4 and 5 in \citet{villegas10}. 
\end{justify}

\label{tab:properties}
\end{table}

In this study, we investigate the GC population to gain further insights into the history of this system of galaxies.
The paper is structured as follows. In Sect. \ref{sec:2}, we outline the data used, detailing the procedures for source detection, photometric calibration, and completeness estimation. Section \ref{sec:3} focuses on the selection criteria for GC candidates based on their morphological and photometric properties. We also provide the final GC candidates catalogs as well as the single candidates in each observed passbands. In Sect. \ref{sec:anlaysis_gc}, we present the analysis conducted using the identified GC candidates. Section \ref{sec:lsb} delves into the investigation of GC content in dwarf galaxies across the observed field. A brief summary of our conclusions is
presented in Sect. \ref{sec:summarize}.


\section{Observations and data analysis}
\label{sec:2}

In this section, we provide a brief overview of the imaging data used in this work and the image processing procedure.

\subsection{Observations and image processing }
\label{sec:data_descr}

The observations used in this work are part of the \textit{VST Elliptical GAlaxy Survey} (VEGAS, P.I. E.Iodice, \citealt{capaccioli15,iodice21}). The survey was performed with the INAF VLT Survey Telescope (VST), which
is a 2.6 m diameter optical telescope located at Cerro
Paranal, Chile \citep{Schipani2010}. The imaging is in the $u$, $g$, $r$
and $i$-band using the 1$\times$1 square degree field of view camera
OmegaCAM \citep{kuijken11}. 
A detailed description of the survey and procedures adopted for data acquisition and reduction can be found in \citet{grado12}, \citet{capaccioli15}, and \citet{iodice16fds}. Here, we only provide a brief overview. 
The data were processed with VST-tube \citep{grado12}, a pipeline specialized for the data reduction of VST-OmegaCAM, executing pre-reduction (bias subtraction, flat field normalization), illumination and fringe corrections (for the $i$-band), photometric and astrometric calibrations. The final co-added images were normalized to a Zero Point (ZP) of 30.00 mag, calibrated using the PSF magnitude within 4$\farcs$0 (19.04 pixel) circular aperture of reference standard stars observed with the Sloan Digital Sky Survey \citep[SDSS]{Blanton2017}. Table \ref{tab:obs} summarizes the main characteristics of the observations. Figure \ref{fig:color_comp} shows a color composite image obtained with the python package $make\_lupton\_rgb$ \citep{lupton04} of the observed field. The figure shows the two studied galaxies at its center and the two other galaxies in the field, NGC\,3643 and NGC\,3630, along with the dwarfs identified in this work (white boxes, see Sect. \ref{sec:lsb}).

\afterpage{
\begin{figure*}
    \centering
    \includegraphics[width=\textwidth-2.5cm]{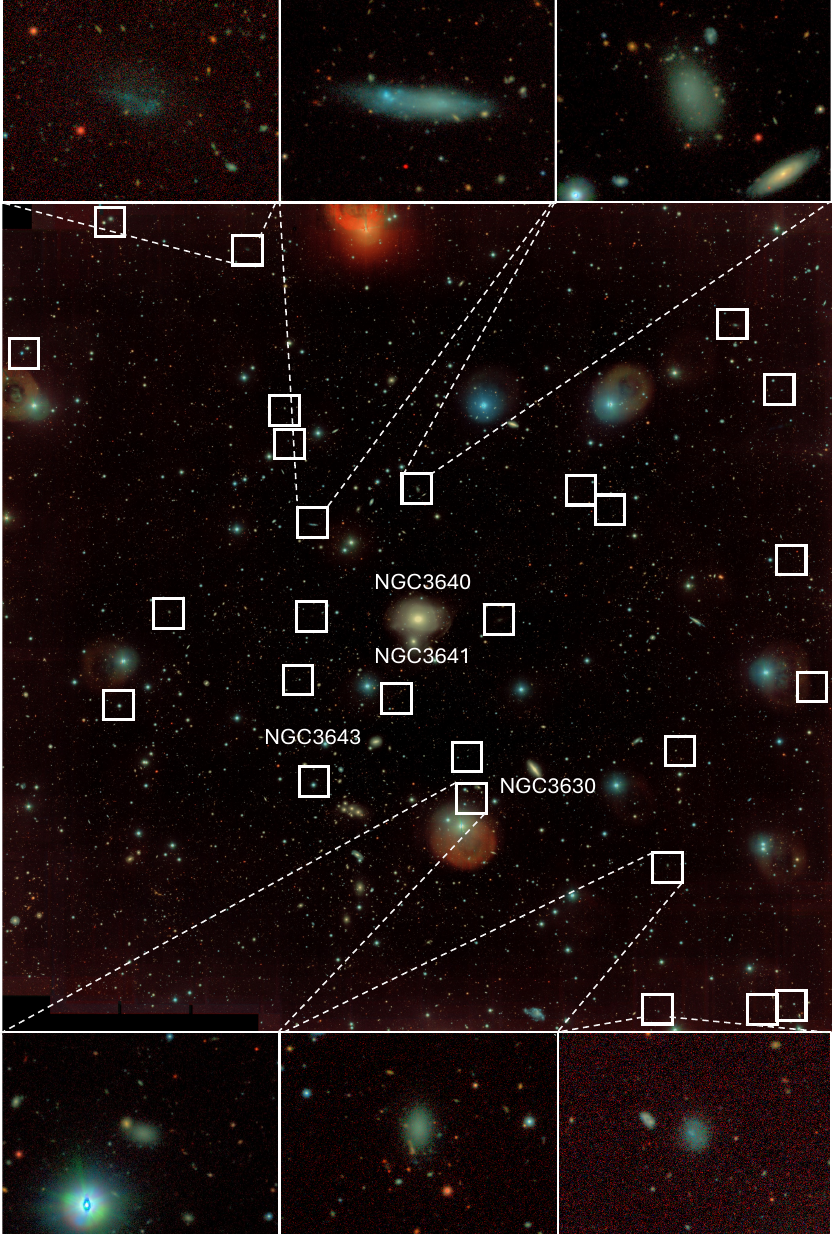}
    \caption{Color composite image of the observed 1.5$\times$1.5 sq. degrees around NGC\,3640. The large galaxies in the group are labelled and with white boxes we mark the location of the dwarf galaxies identified in this work (Sect. \ref{sec:lsb}). Several of the dwarfs are displayed in the cutouts that span $\sim 2\times1.4\ arcmin^2$ ($\sim$ 15$\times$11~$kpc^2$).}
    \label{fig:color_comp}
\end{figure*}
\newpage
\begin{figure*}
    \centering
    \includegraphics[width=\columnwidth,height=7.8cm]{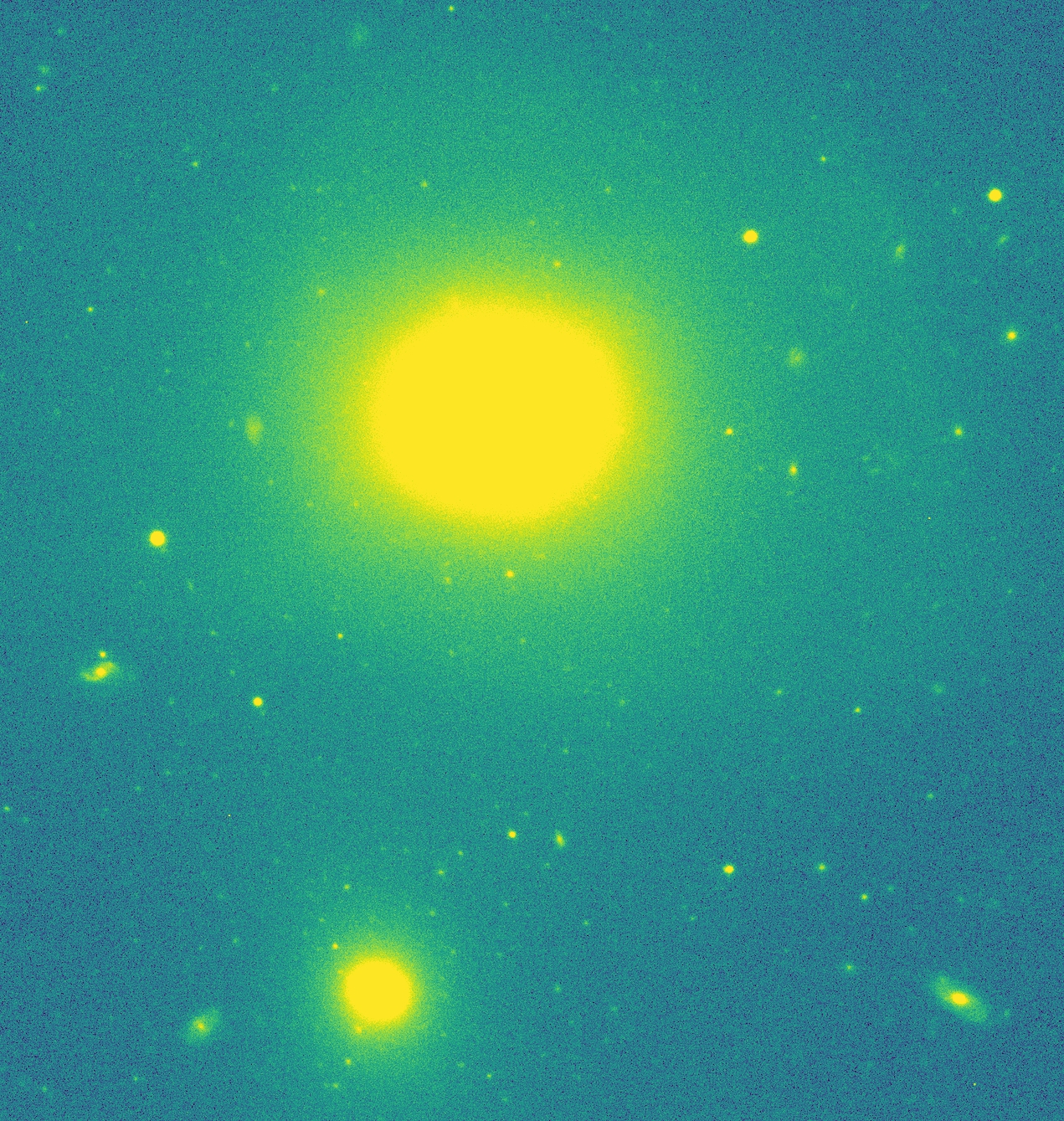}
    \includegraphics[width=\columnwidth,height=7.8cm]{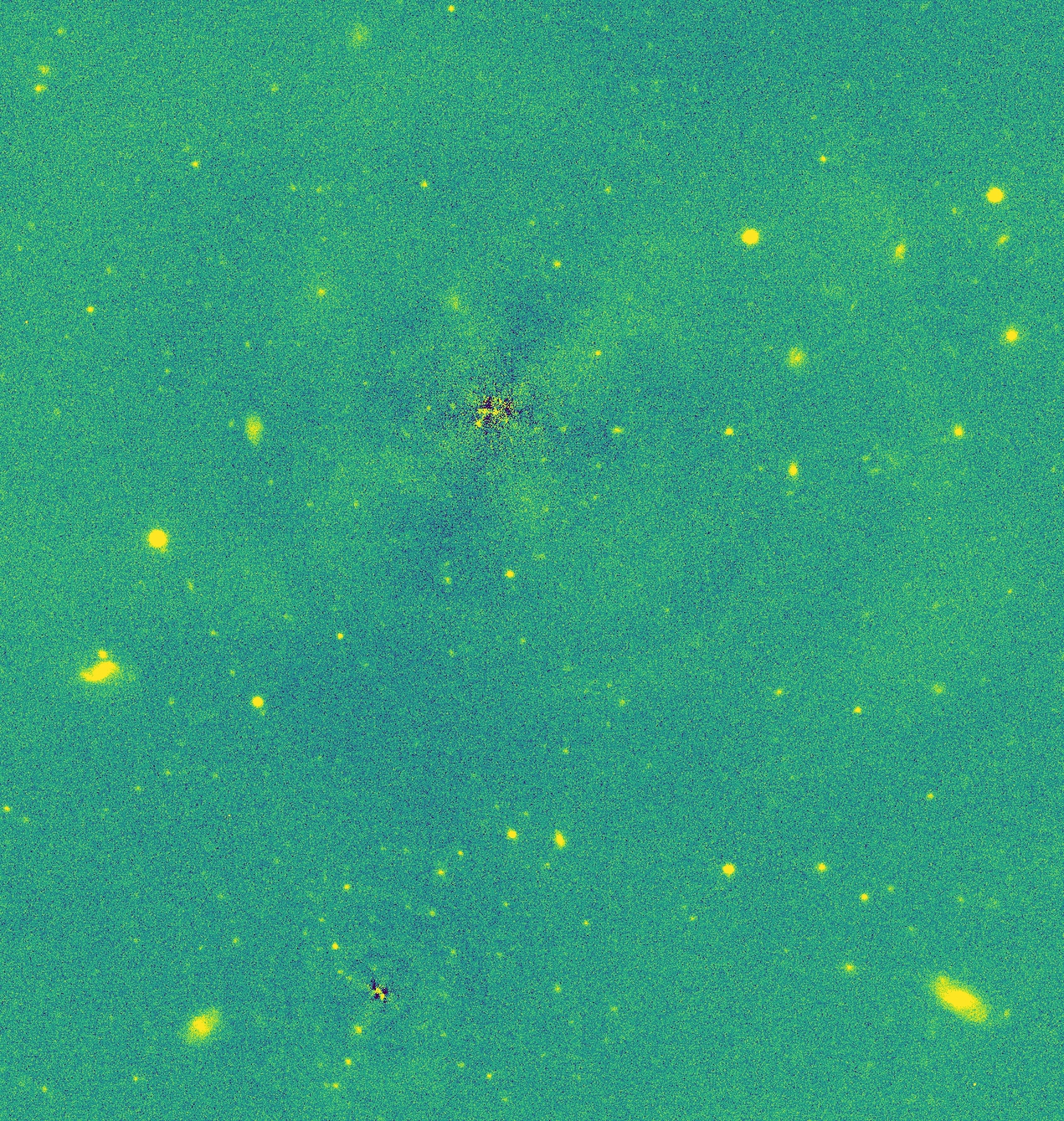}
        \includegraphics[width=\columnwidth,height=7.8cm]{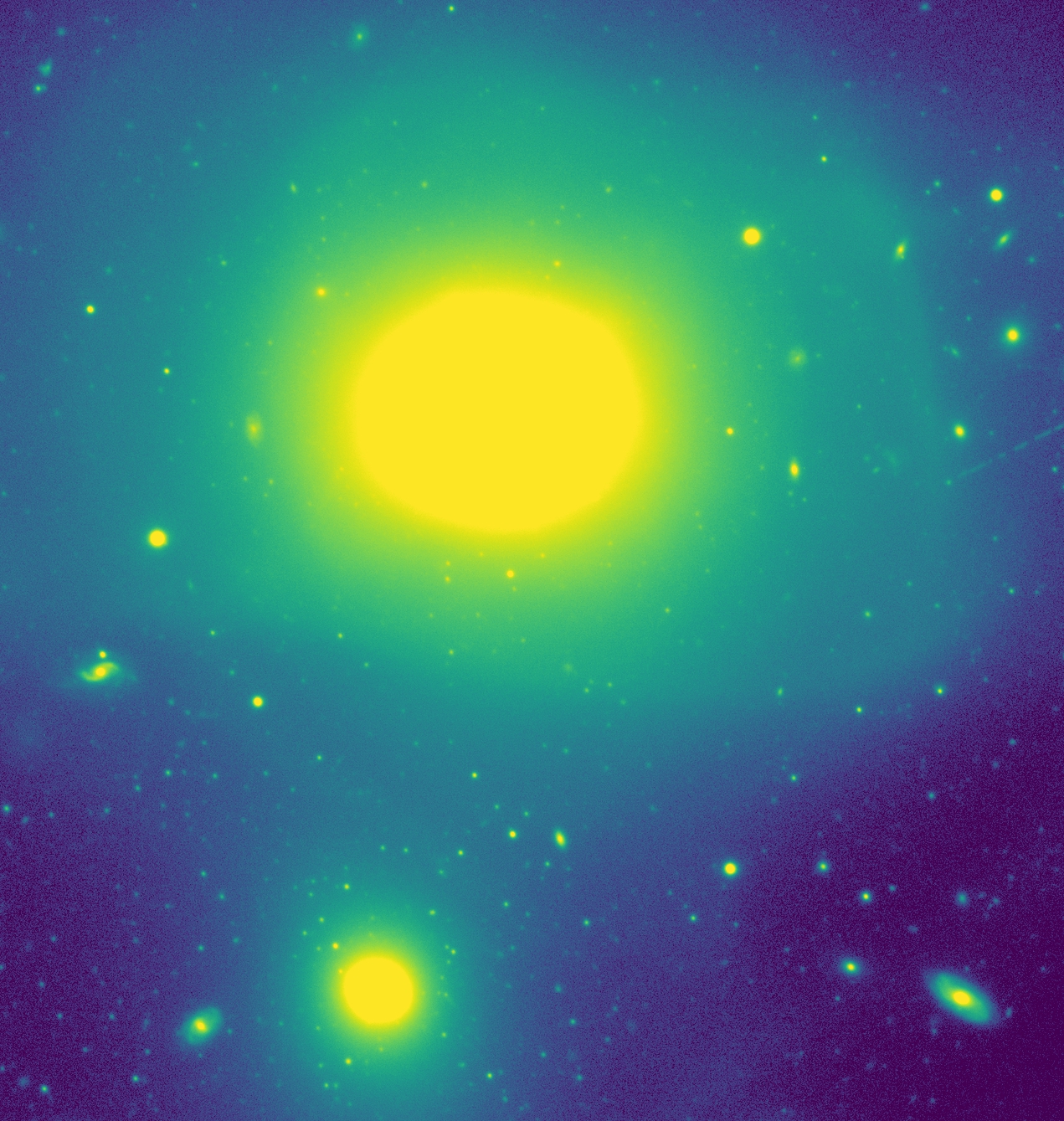}
    \includegraphics[width=\columnwidth,height=7.8cm]{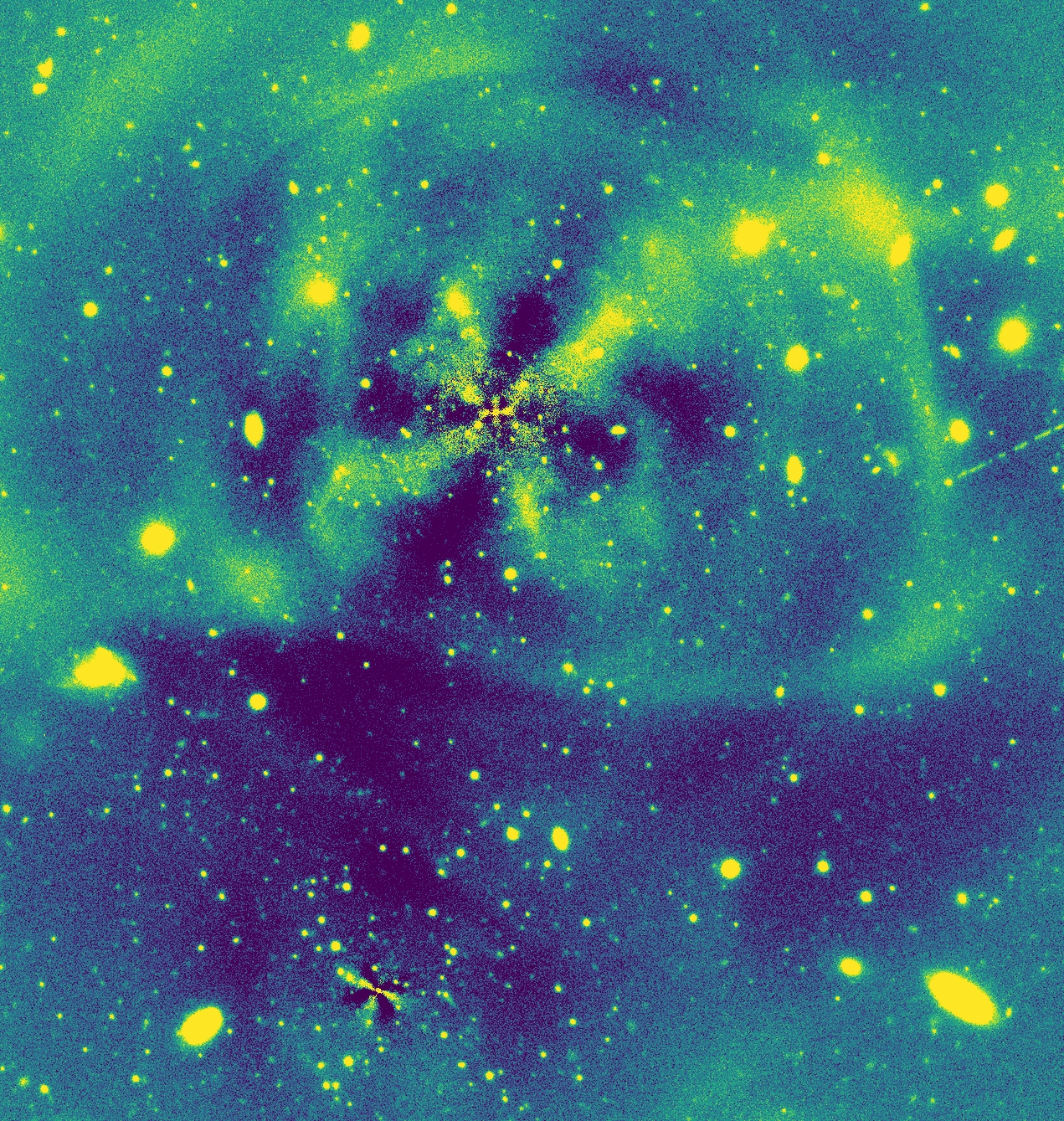}
        \includegraphics[width=\columnwidth,height=7.8cm]{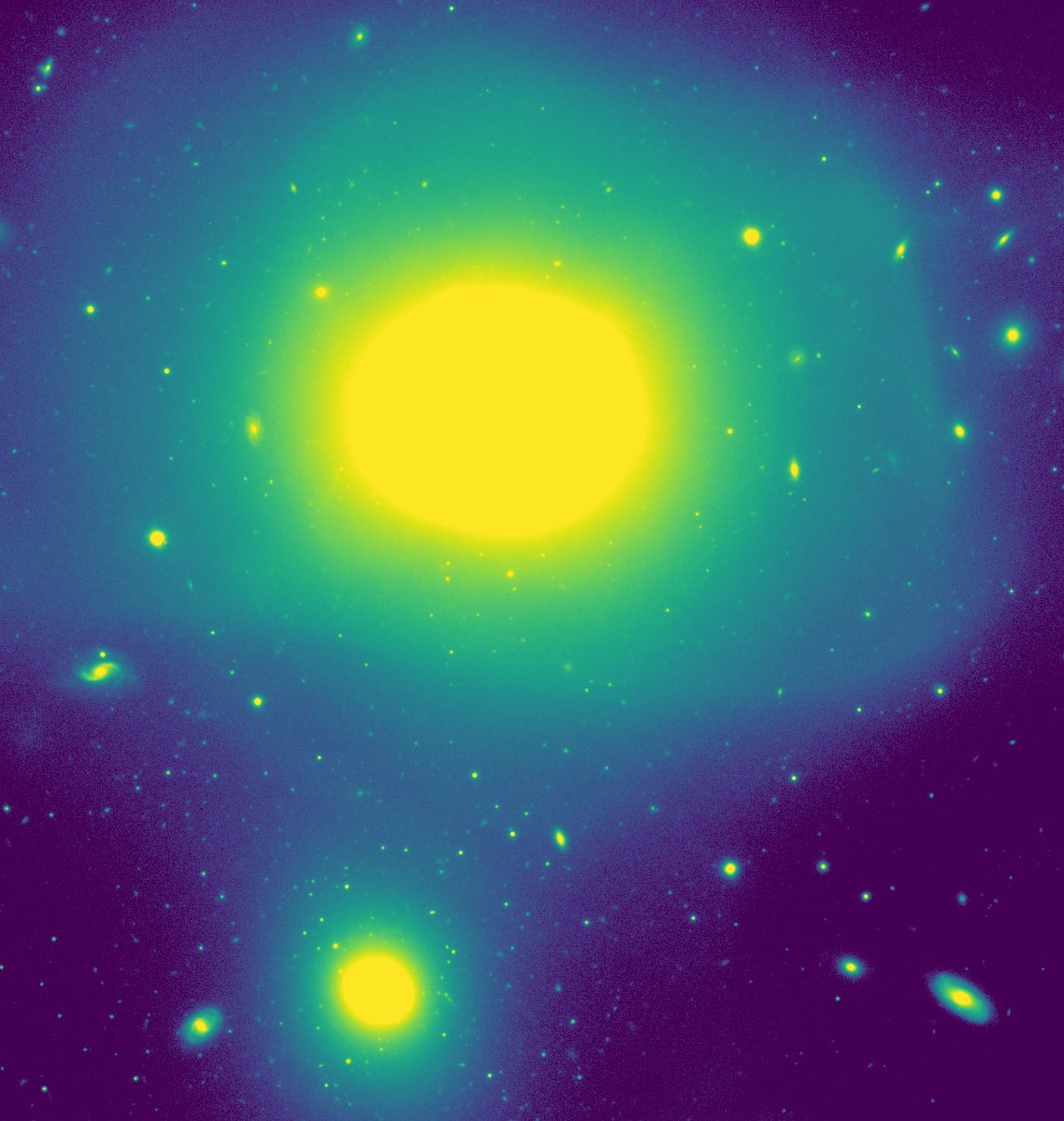}
    \includegraphics[width=\columnwidth,height=7.8cm]{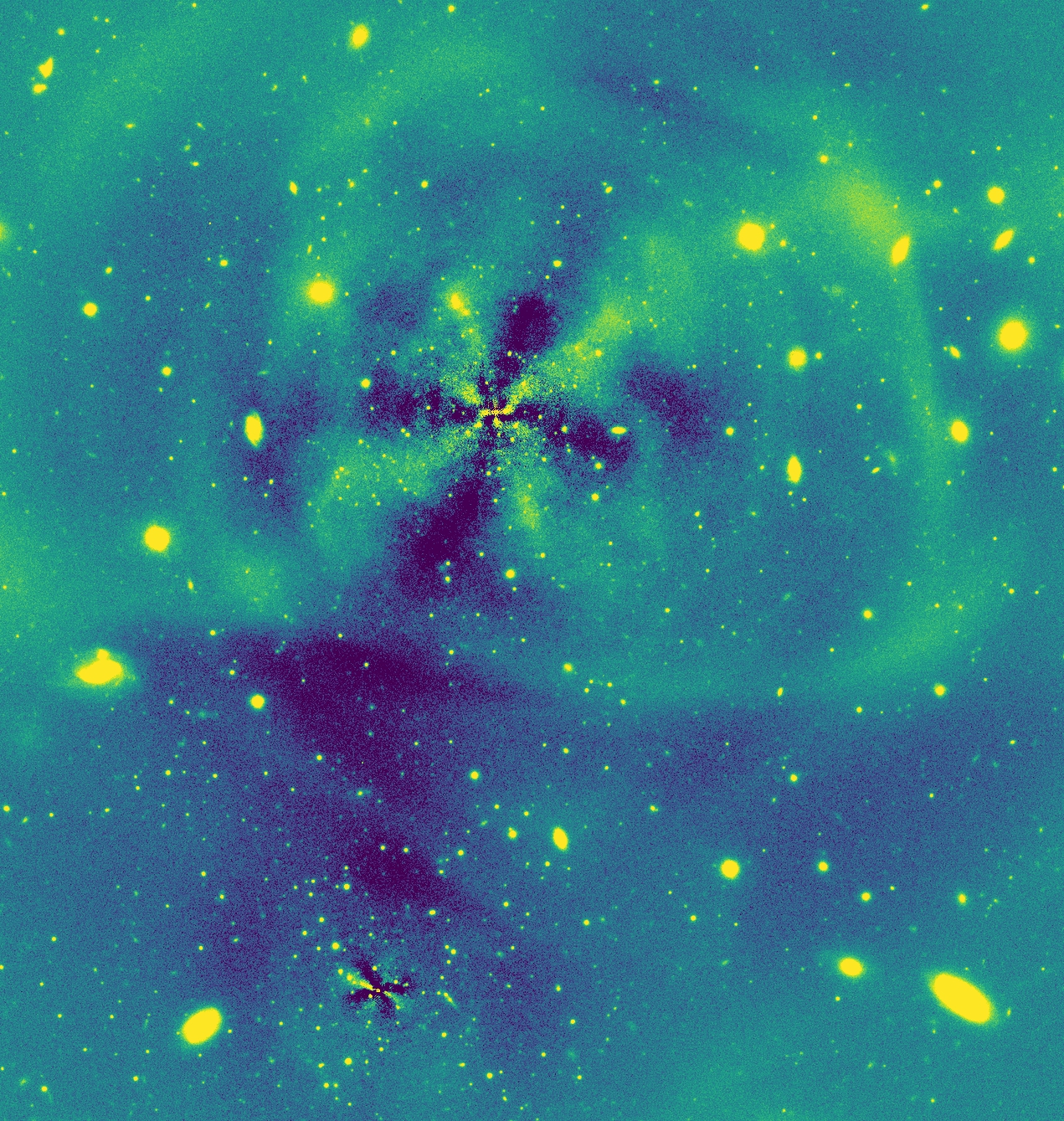}

    \caption{Galaxy images and residual cutouts  around the two galaxies, $\sim$4$\times$5 $arcmin^2$ ($\sim$ 30 $\times$ 40 $kpc^2$) on a side. North is up and east is left. $Left$: NGC\,3640 (upper galaxy) and NGC\,3641 (lower galaxy) in the $u$, $g$ and $r$-band, from the top to the bottom, respectively. $Right$: galaxies-subtracted images. The observation of the $u$ (first row) and $g$ (second row) residuals do not show obvious dust features, contrary to previous findings \citep{Prugniel1998}.}
    \label{fig:residual}
\end{figure*}
}

\begin{table}
    \small
    \centering
    \caption{Images properties.}
    \begin{tabular}{cccccc}
    \hline \hline
    
     Filter & Exp. Time& FWHM & A.C.& $\Delta mag_{VST-SDSS}$ \\
      & (h) & ($arcsec$) & (mag) & (mag) \\
      & (1) & (2) & (3) & (4) \\
\hline
    $u$& 3.1&0.95 $\pm$  0.03  &  0.421 $\pm$0.005 &   0.02 $\pm$ 0.05    \\ 
    $g$& 1.8&0.90 $\pm$  0.03  &  0.382 $\pm$0.003 & -0.01 $\pm$ 0.03      \\ 
    $r$& 1.8&0.56 $\pm$  0.04  &  0.218 $\pm$0.002 & -0.02 $\pm$ 0.02     \\ 
    $i$& 1.8&0.70 $\pm$  0.06  &  0.235 $\pm$0.004 & -0.01 $\pm$ 0.02        \\ 
    \hline
    \end{tabular}

    \begin{justify}
    \textbf{Notes}: Col. 1 reports the exposure time; Col. 2 shows the median point-like FWHM along with the $rms$; Col. 3 reports the aperture corrections for point source photometry and the uncertainties; Col. 4 lists the median offset between our photometry and the SDSS one for bright and compact sources.
    \end{justify}
    
    \label{tab:obs}
\end{table}


\subsection{Photometry and photometric calibration}
\label{sec:photometry}
\sloppy
In order to detect point-like sources close to the centre of the galaxies, where the background surface brightness is high, we first modeled and subtracted the light profile of all the brightest galaxies in the observed field. This is because the steepness of the galaxy surface brightness profile is reflected as drastic changes in the background, reducing the efficiency of detection for faint sources in these regions. The luminosity profiles of NGC\,3640, NGC\,3641, NGC\,3630, and NGC\,3643 were independently derived in all available passbands. To model the light distribution of the galaxies we used the python Elliptical Isophote Analysis package\footnote{\url{https://photutils.readthedocs.io/en/stable/isophote.html.}}. A description of how the model is generated can be found in \citet{Hazra2022}. Figure \ref{fig:residual} shows an example of the original image and the residual (original minus model), for NGC\,3640 and NGC\,3641 in the $u$, $g$ and $r$-band. When examining the residuals in the $u$ and $g$ wavelength range, we did not observe any dust lane, contrary to the observations made by P88. After modelling and subtracting the light of
the bright galaxy in the field, NGC\,3640, and its companion, NGC\,3641, the field clearly shows the underlying GCs as point sources, as well as a rich web of low-surface brightness features. The system of shells previously reported by P88 is
strongly enhanced in the galaxies-subtracted images (Fig. \ref{fig:residual}, right panels), but there are no obvious signs of localized dust patches/lanes.

To obtain photometry of the sources in the field, we used SExtractor \citep{bertin96} on the galaxy-subtracted images for each filter independently. For compact sources, we chose as reference the aperture magnitude within an eight-pixel diameter ($\sim$ 1$\farcs$68 at OmegaCAM pixel scale). This choice relies on maximizing the signal vs. noise within the aperture. The aperture correction (AC), required to take into account the missing flux beyond such aperture, was determined by selecting $\sim100$ bright, isolated point-like sources, then measuring the magnitude difference between our reference aperture (8 pixels) and the pipeline calibration diameter (i.e. 19.04 pixels). 
We investigated potential spatial variations of the AC value across the four bands by dividing the observed fields into four regions. Estimates of the AC value for each region were obtained and found to be consistent with each other within the error.
Table \ref{tab:obs} presents the aperture corrections derived for each passband along with their uncertainties (estimated using the median absolute deviation MAD). The same  bright and compact sources were also used to estimate the median FWHM (Full Width at Half Maximum), reported in Table \ref{tab:obs}.

To give a rough idea of the difference in depth between the available passbands, we note that the number of detections ranged between $\sim50000$, in the $u$-band, and $\sim320000$ in the $r$-band catalog above $1\sigma$ the background level. 

The catalogs were then matched using a matching radius of 1$\farcs$0. The matching radius was chosen after several tests with various radii, balancing between completeness (most affected in the shallower $u$-band) and contamination (most affected in our deepest and richest image, the $r$-band). We produced both $ugri$- and $gri$-matched catalogs, because of the much shallower $u$-band data. On these catalogs, we applied the following color corrections provided by the VST-Tube pipeline \citep{grado12}:

\begin{equation}
\begin{aligned}
m_{u_{new}}& =m_u +(m_u{-}m_g)\times0.025(\pm 0.002) \\
m_{g_{new}}&=m_g +(m_g{-}m_i)\times0.023(\pm 0.001)  \\
m_{r_{new}}&=m_r +(m_r{-}m_i)\times0.054 (\pm 0.001)  \\
m_{i_{new}}&=m_i -(m_g{-}m_i)\times0.003(\pm 0.001)\\
\end{aligned}
\end{equation}

As sanity check, we matched and compared our photometry with the SDSS PSF-magnitude measurements \citep{Blanton2017}. The two photometric systems are equivalent, and we identified $\sim18000$ matching sources in the $ugri$ catalog (within a 1$\farcs$0 matching radius). Table \ref{tab:obs} reports the median magnitude offset, $\Delta mag=m_{VST}-m_{SDSS}$,  of the matched source together with the $rms_{MAD}$, showing good photometric consistency within errors in all bands.

For the foreground extinction correction, we obtained the extinction values from the IRSA (NASA/IPAC Infra-Red Science Archive) dust query module of Astropy, which provides $E_{B-V}$ at the position of the source. The adopted values to multiply $E_{B-V}$ to obtain the extinction are  4.239, 3.303, 2.285, 1.583, for $ugri$, taken from \citet{sf11}.
The complete single band catalogs, as well as the $ugri$ and $gri$ matched catalogs, are available on the VEGAS portal webpage\footnote{\url{http://old.na.astro.it/vegas/VEGAS/Welcome.html}}, and on the CDS.

\begin{figure}
    \centering
    \includegraphics[width=\columnwidth]{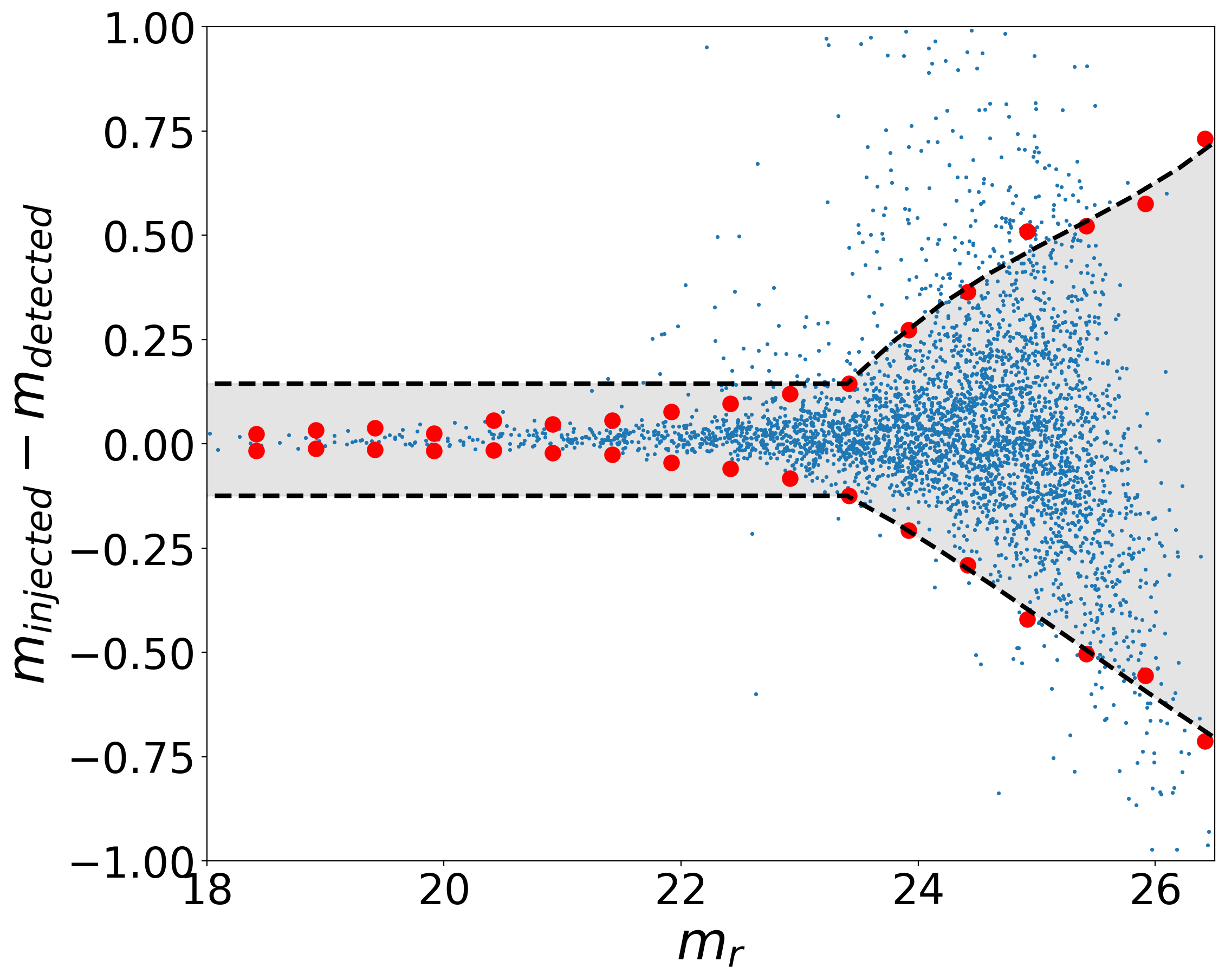}
    \caption{ The adopted selection criteria for the injected sources used to estimate the completeness of the field in the $r$-band.
    Blue points represent the matched detected-injected sources in the field, while red dots indicate the median $\Delta mag$ $\pm 2rms_{MAD}$ within each 0.5 magnitude bin. The black dashed line depicts the function adopted for source selection. All sources within the gray dashed region were selected for estimating the completeness function. }
    \label{fig:deltamag}
\end{figure}

\begin{table*}
    \caption{Fitted $m_{50}$ and interpolated $m_{80}$ magnitude limits provided for both on-galaxy and off-galaxy fields in all four passbands. The last column reports the distance between the center of the off-galaxy regions and NGC\,3640.}
    \small
    \centering
    \begin{tabular}{cccccccccc}
    \toprule
    \toprule
     Region & \multicolumn{2}{c}{$u-band$} & \multicolumn{2}{c}{$g-band$}  & \multicolumn{2}{c}{$r-band$}  & \multicolumn{2}{c}{$i-band$} &$R_{gal}$\\ 
      \cmidrule(lr){2-3} \cmidrule(lr){4-5} \cmidrule(lr){6-7} \cmidrule(lr){8-9}

         & $m_{80}$ &$m_{50}$& $m_{80}$ &$m_{50}$& $m_{80}$ &$m_{50}$& $m_{80}$ &$m_{50}$ &($arcmin$)\\

      \hline

     On-galaxy      &24.50 &24.76 &24.70 &25.73 &24.70&26.09  &24.10 &24.88 &  ...\\
     Off-galaxy \#1 &24.50 &24.79 &25.10 &25.82 &25.30&26.06  &24.30 &24.92 & $\sim$12\\
     Off-galaxy \#2 &24.40 &24.64 &25.10 &25.59 &25.10&25.84  &24.20 &24.86 & $\sim$24\\
     Off-galaxy \#3 &23.90 &24.23 &24.70 &25.07 &24.70&25.43  &23.80 &24.44 & $\sim$45\\

     \bottomrule
    \bottomrule
    \end{tabular}    
    \label{tab:pn_off_gal_comp}
\end{table*}

\subsection{Completeness}
\label{sec:completeness}
Part of the analysis that we perform in the forthcoming sections requires correcting the number counts of faint sources by the detection completeness function, which we derived using the procedure outlined in this section. This function provides the fraction of detected sources over the injected ones ($f\equiv N_{\rm detected}$/$N_{\rm injected}$) as a function of magnitude. We have developed a Python code to estimate the completeness function, in four main steps:
\begin{enumerate}[i)]
    \item Point Spread Function modeling;
    \item Injection of artificial sources;
    \item Detection of sources and matching with the catalog of injected ones;
    \item Estimation of the completeness ratios, and fitting of the completeness functions. 
\end{enumerate}

In more detail, we first derived a PSF model from $\sim$10 bright, compact and isolated sources in the image. A cutout of the selected PSF-model candidates was extracted from the image using the 'extract stars' routine in Astropy. The PSF stars were given as input into the EPSFBuilder module of the photutils package\footnote{\url{https://photutils.readthedocs.io/en/stable/epsf.html}} to generate an effective PSF of 43$\times$43 $pixel^2$ ($\sim$9$\times$9 $arcsec^2$). 

The magnitudes of $\sim$2000 artificial stars to be injected were generated using a probability-density distribution (PDF) that matches with the observed total luminosity function of the sources in the image. The PDF is populated independently for each available passband using a random generator\footnote{One should note that the total luminosity function also contains the non-point sources component. However, we are not interested in the completeness fractions for extended objects, and the impact of ignoring the two components (compact and extended) is negligible for our purposes.}, and the stars are then injected into the image of interest along an equally spaced grid, with a spacing of $\sim 15\farcs0$. The simulations we run are, by construction, independent of the source color. This assumption is made with the intent of keeping the passbands separate, so as to avoid reducing the quality of our data to that of the image with the least depth in the matched sample (see also Sect. \ref{sec:gclf} and footnote \ref{gclg_footnote}).

We run our detection and photometry procedures on the simulated image as described in the previous Sect. \ref{sec:photometry}, adopting the same parameters that were used to generate the catalogs. 

The catalogs of detected and injected sources were matched within 1$\farcs$0 radius, and cleaned for spurious matches. For the cleaning process we inspected the broadening of the $\Delta mag=m_{\rm injected}-m_{\rm detected}$ function versus the detected magnitude. We adopt a magnitude dependent, iterative sigma-clipping approach. Briefly, we first divide our sample into bins of 0.5 mag width  and calculate the median $\Delta mag$ and its $rms_{\rm MAD}$ within each bin. For bright sources we considered as matched the sources within a fixed $\Delta mag$ interval. For faint sources, instead, we consider the sources to be matched when $\Delta mag\leq \pm 2 rms_{\rm MAD}$.
The bright/faint separation cut is set to the point where the $rms_{MAD}\gsim 0.1$ mag. This approach is depicted in Fig. \ref{fig:deltamag} for one of the $r$-band simulations. In the figure, the gray shaded area at magnitudes brighter/fainter than $m_r\sim23$ mag is the region of selection for bright/faint sources. 

Finally, the ratio of the number of detected sources which passed the cleaning stage was normalized to the number of injected sources in the same magnitude bin, $f=N_{detected}$/$N_{injected}$,  providing us with the discrete completeness ratio behaviour as a function of magnitude. 
To reduce sampling effects, we repeated the simulations 150 times, during which the positions and magnitudes of the injected sources varied. Ultimately, the completeness function is obtained as the median of all the completeness functions generated.

We fitted the completeness ratios adopting  a modified Fermi function \citep{alamo13}:

\begin{equation}
    f_F(m)=\frac{1+C\cdot \exp[b(m-m_{50})]}{1+\exp[a(m-m_{50})]}
\end{equation}

\noindent
where $m_{50}$ is the magnitude of the 50\% completeness limit, $a$ regulates the steepness of the cutoff, $b$ (which must be $<$ $a$) influences the point at which the deviation above unity starts, and $C$ determines the amplitude of the deviation.

The completeness function in each band was determined by using the procedure described above, injecting the artificial stars on cutouts of the images with $\sim$ 6$\times$6 $arcmin^{2}$ size (see Fig. \ref{fig:comple_on_gal}). We considered both fields centered on NGC\,3640/NGC\,3641, and off-galaxy fields. The size of the cutouts are big enough to contain $\sim$5 $R_e$ of the galaxies. Table \ref{tab:pn_off_gal_comp} reports the fitted $m_{50}$, as well as the interpolated magnitude of the 80\% completeness limits ($m_{80}$) for the on-galaxy and off-galaxy fields. For the off-galaxy fields, we also give the galactocentric distance, with respect to NGC\,3640 measured from the center of the field. Figure \ref{fig:comple_on_gal} shows the completeness functions derived in three off-galaxy regions and on-galaxy for all four filters. In the figure, we also reported the expected position of the turn-over magnitude of the GCLF  (see Sect. \ref{sec:mag_sel}). In all filters, but the $u$-band, the data are 80\% complete or more at the TOM level. The $u$-band data has 0\% completeness level at the turn-over magnitude, meaning that the $ugri$ matched catalog is limited to the very bright end of the GCLF.
As expected, the off-galaxy regions are more complete than the on-galaxy region at fixed magnitude, with the exception of the region closer to the edge of the image ($R_{\text{gal}}>>30$ $arcmin$, see below).

We also derived the completeness functions for 26 other $6\times 6~arcmin^2$ off-galaxy regions at various galactocentric radii, $R_{gal}$. The resulting 50\% and 80\% magnitude limits for the $r$-band are reported in Table \ref{tab:off_gal_compl}. The full sample of fitted functions and the $m_{80}$ vs. $R_{gal}$ plots are also reported in Fig. \ref{fig:m80_rad}. In the left panel of the figure all functions derived in boxes with $R_{gal}\leq 30'$ are shown with gray dashed curves, while the functions at larger radii are plotted with red dot-dashed lines. The right panel shows the $m_{80}$ vs. $R_{gal}$ behavior.

A small but non negligible depth variation with $R_{gal}$ exists, a result which is expected because of the specific observing strategy adopted, as described in \citet{iodice16fds}. The images were acquired using a step-dither observing strategy. This result will be later used to select the areas suitable for estimating the population of contaminants (mostly foreground MW stars, but also background galaxies) for the purpose of analyzing the GC population (see Sect. \ref{sec:anlaysis_gc}).

\subsection{Testing the PSF}
\label{sec:psf}

To obtain a reliable completeness function, we need to have a well-characterized PSF. To test the robustness of the PSF models used for deriving the completeness functions, we compared the magnitudes of unsaturated, bright and compact sources obtained using PSF photometry\footnote{We used the PSFPhotometry module from the photutils Python package (for more details, see the link \href{https://photutils.readthedocs.io/en/stable/psf.html}{here}.)} to the magnitudes obtained from SExtractor. The median $\Delta phot=mag_{SExtractor}-mag_{PSF}$ and the corresponding $rms_{MAD}$ are reported in Table \ref{tab:psf_tests}. In all cases, but the $u$-band, the
offset is within the errors. Even in the $u$-band, the effect is negligible for the specific purpose of deriving a completeness function, especially if one considers that the $u$ image quality is the worst among the available images. 

 Using the same PSFPhotometry, we  checked the quality of the PSF model using the PSF residuals images, where stars are modeled and subtracted using this task. As an independent proof of the reliability of our PSF-modeling procedure, the PSF subtracted images have generally very flat residuals.

\begin{table}[h!]

   \caption{Comparison between SExtractor aperture corrected and PSF photometry. }
    \centering
    \begin{tabular}{ccc}
    \hline \hline
    
         Filter &\ $\Delta phot$ &\ $rms_{MAD}$ \\
         & (mag) &(mag)\\
\hline
    $u$&-0.03 &0.01  \\ 
    $g$&-0.01 &0.01    \\ 
    $r$&0.00 &0.02   \\ 
    $i$&0.00&0.01    \\ 
    
    \hline
    \end{tabular}
 
    \label{tab:psf_tests}
\end{table}



\section{GC selection}
\label{sec:3}

In this section, we will present the procedures to identify  the GCs from the $ugri$ and $gri$ matched catalogs. For reference, the typical radius of a GC in the MW is $\sim$ 2.5 $pc$ \citep{brodie06}. At the adopted distance for NGC\,3640 (see Table \ref{tab:properties}), one OmegaCAm pixel (0.21 $arcsec$) corresponds to $\sim27.5$ $pc$, meaning that at the image quality of our data, GCs are point-like sources. 
Hence, we choose to identify the GC candidates using their shape (i.e. their morphometry) and their photometric properties. in particular we  select the GCs based on their  magnitude, compactness and color.

\begin{figure*}[h!]
    \sidecaption    \includegraphics[width=\columnwidth]{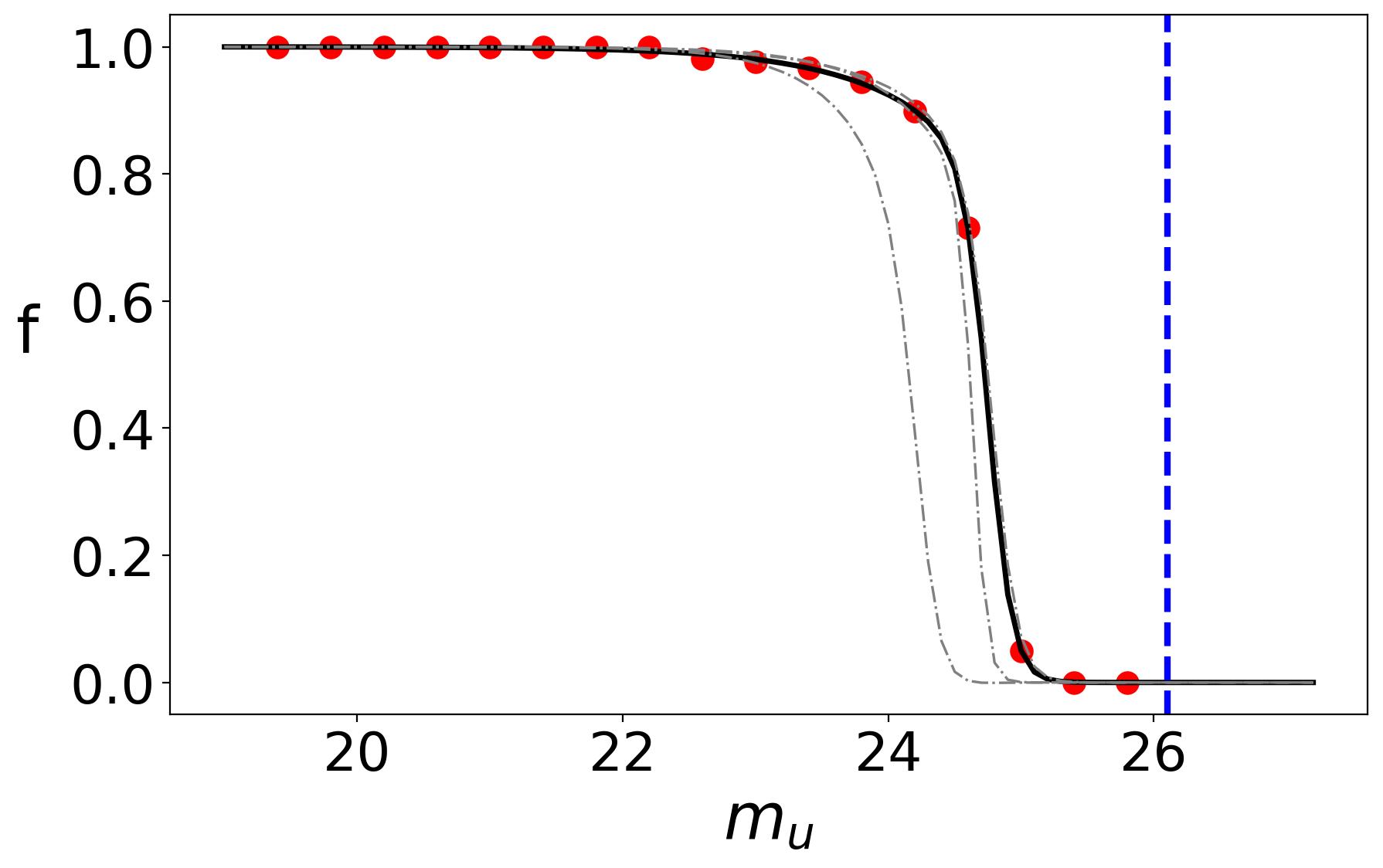}
    \includegraphics[width=\columnwidth]{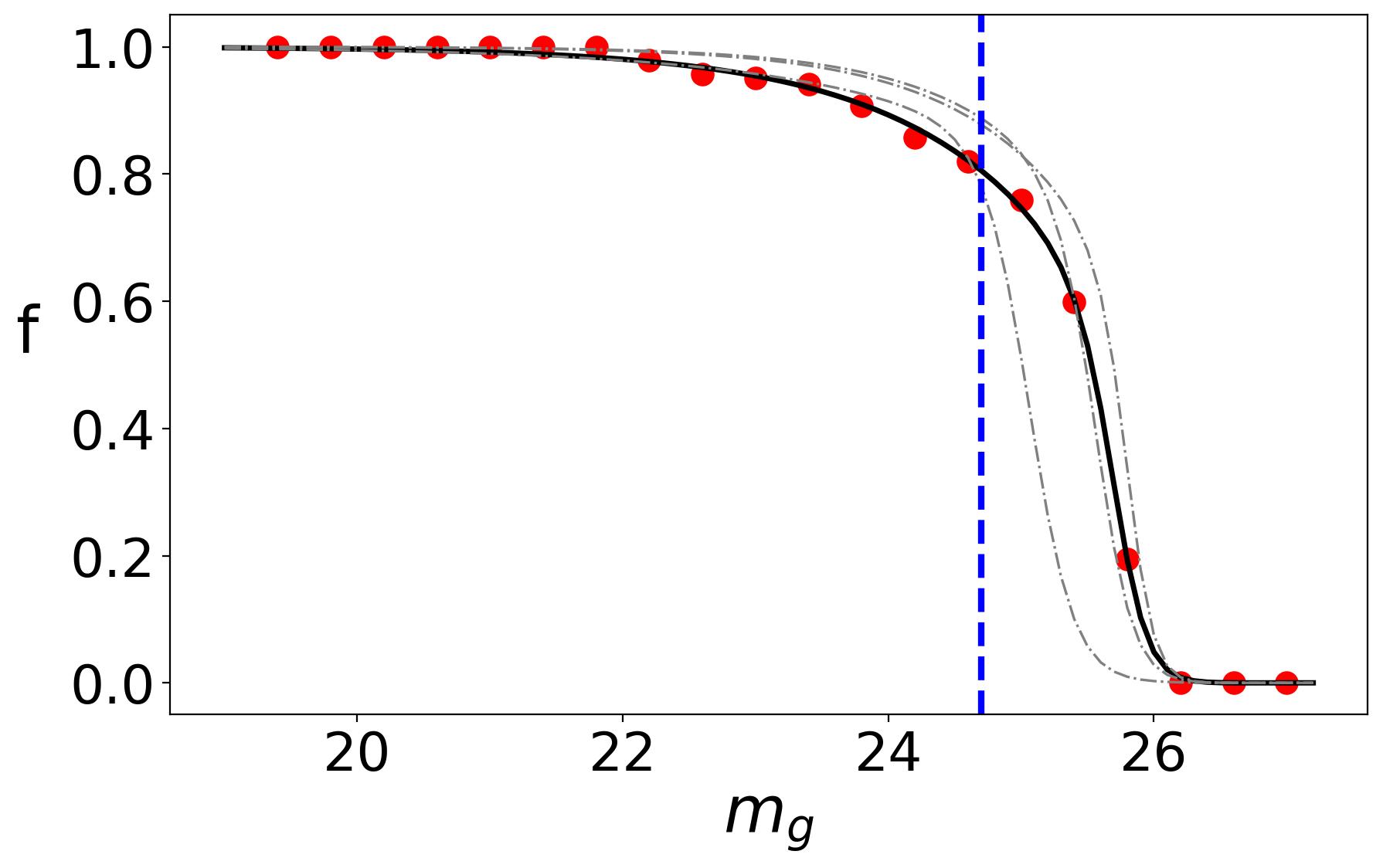}
    \includegraphics[width=\columnwidth]{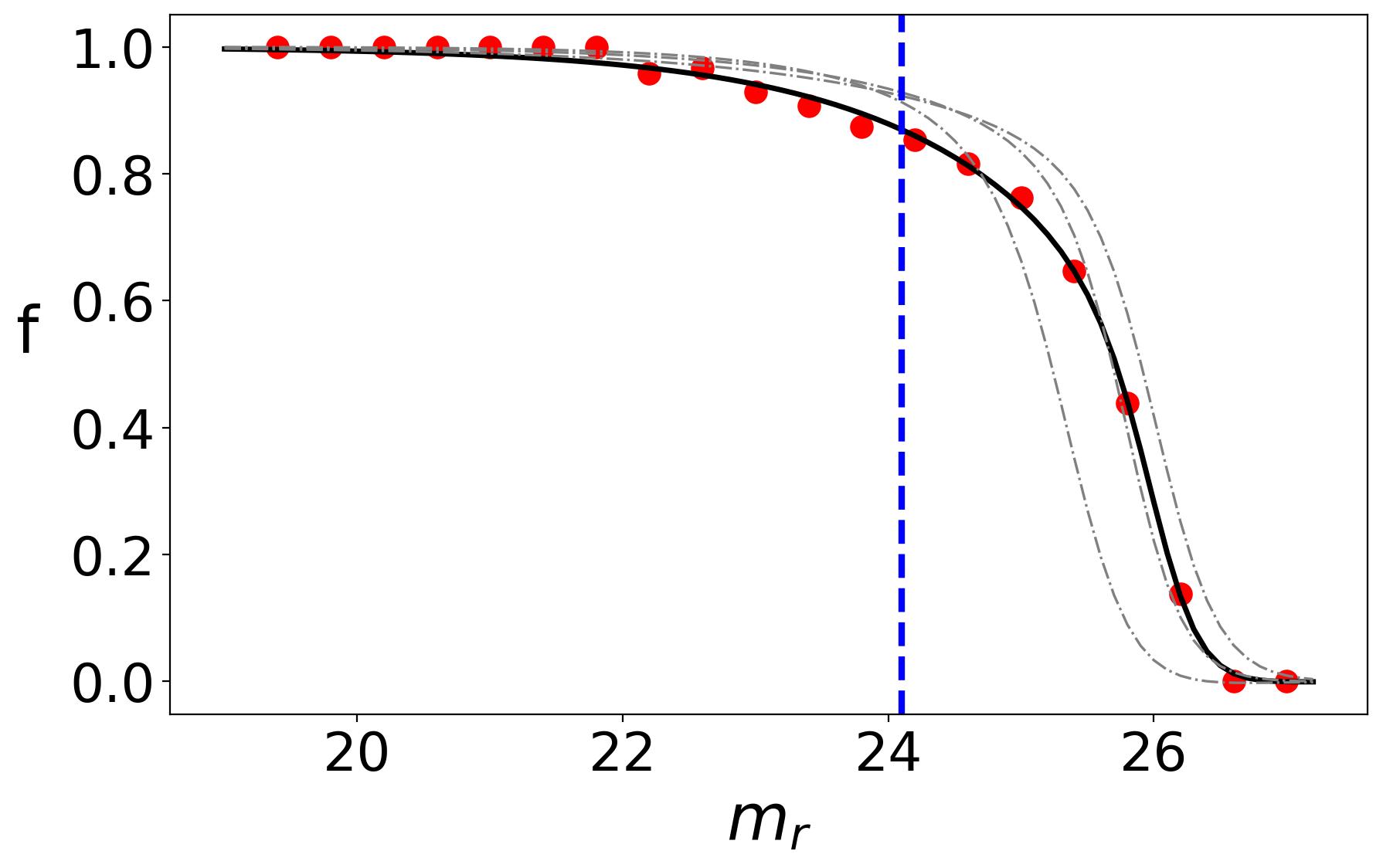}
    \includegraphics[width=\columnwidth]{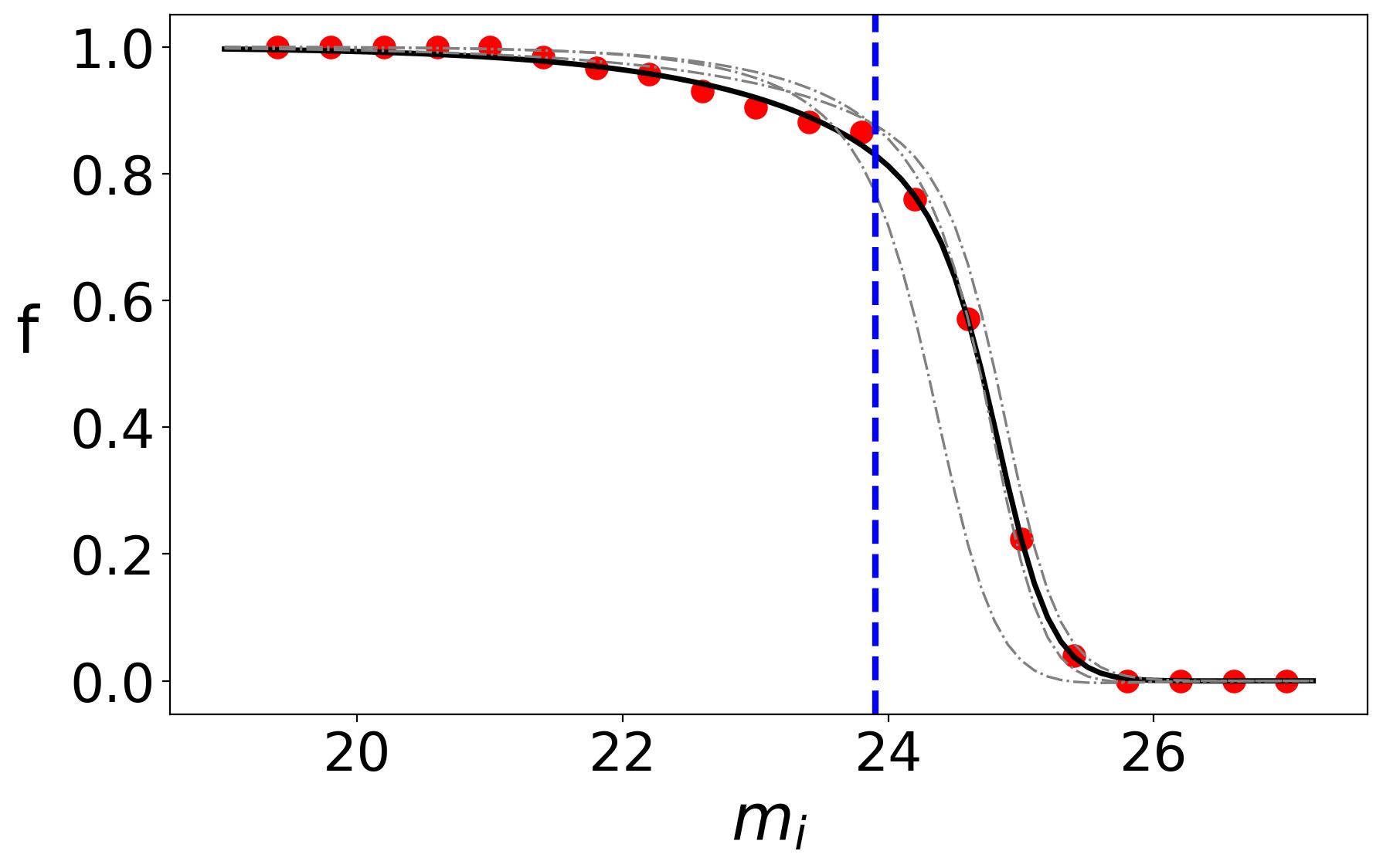}
    \caption{Estimated completeness of the 4 passbands for both on-galaxy, including NGC\,3640 and NGC\,3641, and three off-galaxy regions in a box with a size of $\sim$ 6$\times$6 $arcmin^{2}$. Dark solid lines and gray dashed lines are the fitted Fermi function \citep{alamo13} for the on-galaxy and off-galaxy, respectively. Red dots are completeness measurements per magnitude bin for the on-galaxy region. Blue vertical dashed lines are the expected turn-over magnitudes in each passband (see Sect. \ref{sec:mag_sel}). $Upper$ $row$: estimated completeness of the $u$- and $g$-band, respectively. $Lower$ $row$: estimated completeness of the $r$- and $i$-band, respectively.  }
    \label{fig:comple_on_gal}
\end{figure*}
\begin{figure*}[h!]
    \centering
    \includegraphics[width=\columnwidth]{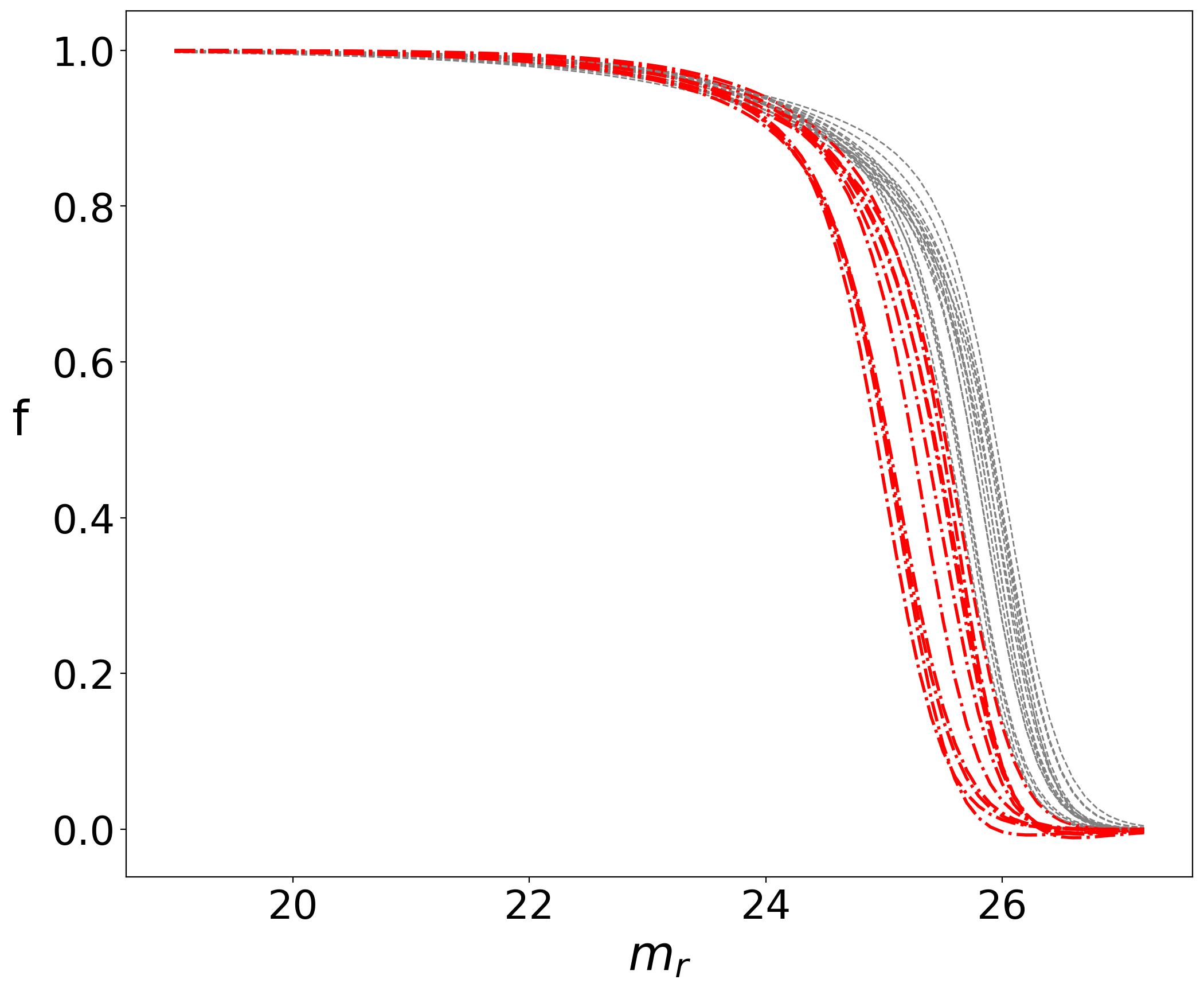}
    \includegraphics[width=\columnwidth]{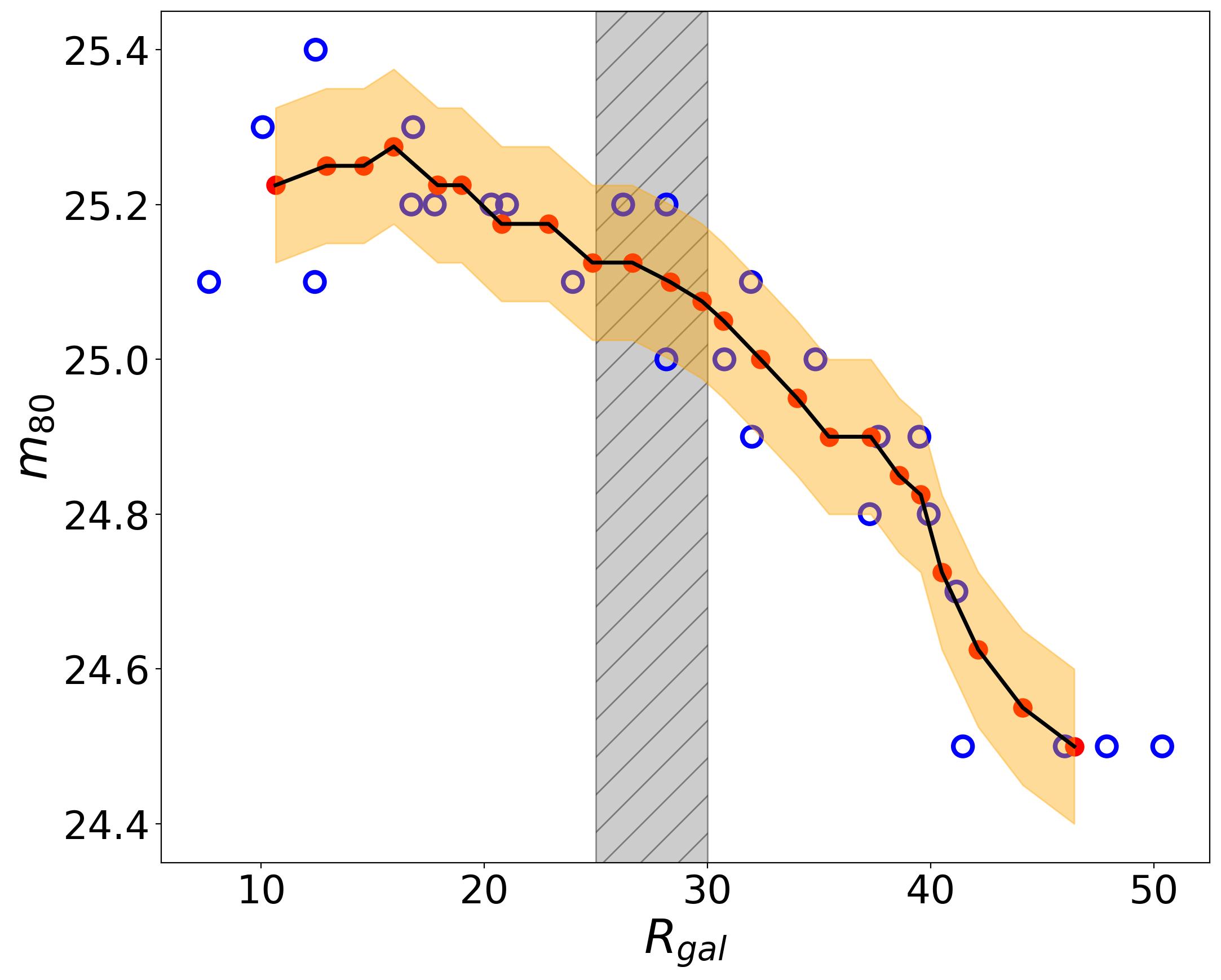}
    \caption{$Left$ $panel$: Completeness functions for 26 off-galaxy regions at various galactocentric radii estimated in a region of 6$\times$6 $arcmin^2$ for the $r$-band. In red, all the regions with $R_{gal}\geq30'$, while functions at smaller galactocentric radii are plotted with gray lines. 
    $Right$ $panel$: radial trend of the interpolated 80\% magnitude limit for each function. Blue open dots are the interpolated 80\% magnitude limit for each region, while red dots are the values obtained through a running mean procedure with a window width of 4. The yellow region is $\pm$ 0.1 mag around the red points.}
    \label{fig:m80_rad}
\end{figure*}

\subsection{Magnitude selection}
\label{sec:mag_sel}

The GC system in a galaxy exhibits a universal Gaussian GCLF, which can be used to define the magnitude range expected for GCs in the target galaxy, once the distance to the host is known. The width of the GCLF scales with the total galaxy luminosity. 
Using Eq. (5) from \citet{villegas10}, and adopting a total galaxy magnitude $M_{z,gal} \sim {-}22.4$ mag (see Table \ref{tab:properties}), we evaluated $\sigma_{g}^{GCLF}= 1.1 \pm$ 0.2 mag.
Adopting a turn-over magnitude of $M_{g}^{TOM} \sim$ -7.5 mag\footnote{Derived assuming a mean TOM for Fornax galaxies of $\sim24.0$ mag from \citet{villegas10}, and a distance modulus to this cluster of $m{-}M=31.51\pm0.15$ from \citep{blakeslee09}.}, and the distance modulus to NGC\,3640 of 32.14 from \citet{tonry01}, the expected GCLF peak would be $m_{g}^{TOM} \sim$ 24.7 mag. Assuming GC median color $(g{-}r)\sim$ 0.6 mag, $(g{-}i)\sim$ 1.0 mag and $(u{-}r)\sim$ 2.0 mag from \citet{cantiello20}, we obtained the TOM in the other observed filters:  $m_{u}^{TOM} \sim$ 26.1 mag,  $m_{r}^{TOM} \sim$ 24.1 mag, $m_{i}^{TOM} \sim$ 23.7 mag. Additionally, assuming identical GCLF width in the $r$ and $g$ bands\footnote{As reported in Table \ref{tab:properties} the GCLF width does not change much ($\Delta\sigma=0.02\pm0.3 mag$) from the $g$- to the $z-$band, allowing us to assume the value of $\sigma_g$ in the $r$-band as well.}, we select all sources in the matched catalog within $\pm3\sigma^{GCLF}$ around the $m_r^{TOM}$, namely $20.8\ mag\leq m_r\leq 27.4\ mag$, as GC candidates. As reported in Table \ref{tab:sel_par_1}, we adopted a more generous cut at the bright limit, to avoid any possible bias from the adopted distance to the galaxy. The faint detection limit is, in practice, much brighter than the expected $m_{TOM}+3\sigma$ because of the incompleteness. We applied the magnitude cut only from the $r$-band, as we consider this as our reference passband, both because of its fainter depth relative to the TOM (see Fig. \ref{fig:comple_on_gal}) and for the compactness of its point-like sources (see Table \ref{tab:mean_par} and Sect. \ref{sec:morpho_sel}).

\begin{table}[t]
\caption{Photometric and morphometric parameters limits adopted for GCs selections. The $r$-band measurements are adopted as reference.} 
\centering        
\begin{tabular}{c c c }    
\hline
\hline    

 Parameter & \ Min &  \ Max  \\   
\hline
\\             
FWHM (arcsec) & .... & 2 \\
 Flux radius (arcsec) &....& 1.0\\
 Elongation & ....   &  2\\

 $m_r$ (mag)&  20.4& ....\\

\hline     
\end{tabular} 

\label{tab:sel_par_1}
\end{table}

\subsection{Morphometric selection}
\label{sec:morpho_sel}

\begin{figure}[ht!]
    \centering
    \includegraphics[width=\columnwidth]{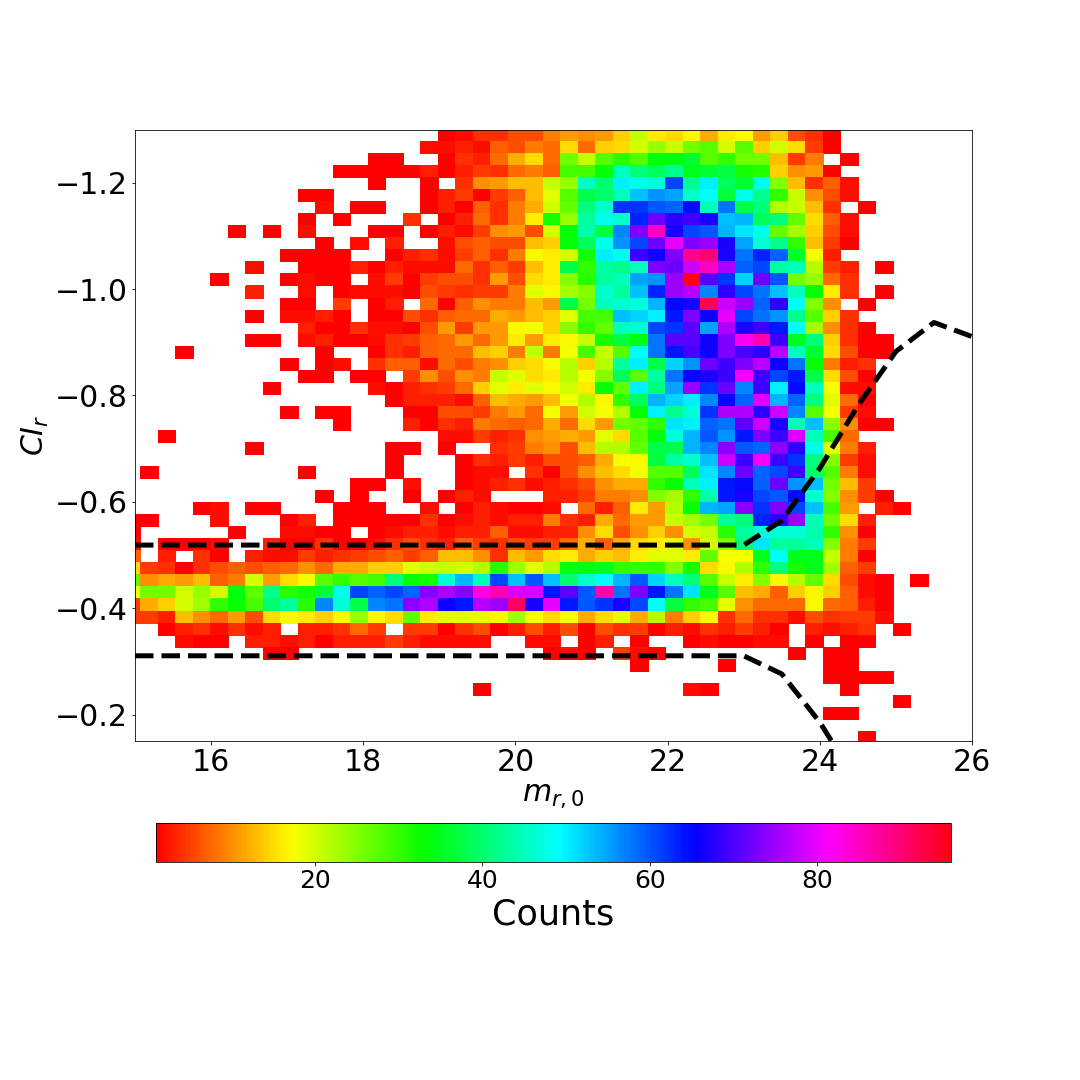}
    \caption{Hess diagram of the CI measured for the sources into the full $ugri$ matched catalog. The dashed black lines define the adopted intervals for the selection of compact sources, derived from numerical simulations.}
    \label{fig:CI_sel}
\end{figure}

To identify GC candidates, we also adopt the FWHM, flux radius, and elongation (major-to-minor axis ratio) properties, all derived using SExtractor. We decided not to use the star classifier provided by SExtractor due to its lower accuracy in classifying stars and galaxies in the fainter regime. Although there is some degree of redundancy between these indicators, this approach helps in culling the catalog more robustly than by using just one indicator. In addition, for each source, we measured the magnitude concentration index, CI, defined as the difference in magnitude measured at two radial apertures \citep{peng11}. This parameter, has proven to be very effective in separating compact sources from extended ones. After  several tests, we adopted  as reference the  8- and 4-pixel aperture radii: CI=$m_{8 pix}-m_{4 pix}$. For compact sources, after aperture correction is applied to the magnitudes at both radii, the CI should be statistically consistent with zero. However, we chose to not apply any aperture correction to the magnitudes used to estimate the CI, hence our CI sequence of compact sources would simply be offset with respect to zero, by an amount that corresponds to the AC difference for magnitudes at 8 and 4 pixel. Figure \ref{fig:CI_sel} shows the CI versus magnitude plot in $r$,  which is our reference selection passband thanks to its image quality. This provides the best separation diagnostic between point sources and slightly resolved background ones.

To define selection ranges for identifying compact sources, we independently estimated the mean properties of point-like, bright sources in each passband using the matched $ugri$ catalog. The results, along with the $rms_{MAD}$, are presented in Table \ref{tab:mean_par}.
  
When dealing with the selection of faint sources, two factors require consideration: the increase in photometric error, and the blending with brighter sources. The former introduces broadening in the distribution of the selection parameter considered as the magnitude gets fainter, while the latter results in fainter sources exhibiting larger radii due to the contamination from neighbouring sources which increases the local background. In order to take into account these effects and taking CI as reference parameter, we applied a selection procedure similar to the one adopted for the cleaning of the injected-to-detected catalog for the study of the completeness (Sect. \ref{sec:completeness}). The mean locus of the accepted sources is fixed to the median CI of the bright sample, while the broadening is taken to be $\pm 5 \sigma_{mag.\ bright}$ for sources brighter than $m_r\sim 23$ mag, and it is set to $\pm 3\sigma_{mag}$ for sources fainter than $m_r=23$ mag. Here, $\sigma_{mag.\ bright}$ is the $rms$ of the CI$_r$ reported in Table \ref{tab:mean_par}, while $\sigma_{mag}$ is the $rms$  determined in the each magnitude bin fainter than $m_r=23$ mag.
The results from this approach are shown in Fig. \ref{fig:CI_sel}, where the $r$-band CI for the matched $ugri$ catalog is reported, together with the curves 
\begin{table}[t]
\caption{Median and standard deviation for compact and bright sources estimate independently in the four passbands using the $ugri$ matched catalog.} 
\centering   
\begin{tabular}{c c ccccc }    
\hline
\hline    
 Band & FWHM  & CI  & FLUX RADIUS  \\  
    &   ($arcsec$)&  ($mag$) & ($arcsec$)\\
\hline
\\             
$u$& 0.95 $\pm$ 0.03 & -0.87$\pm$  0.02 & 0.63$\pm$0.01\\
$g$& 0.90 $\pm$  0.03& -0.79$\pm$  0.02& 0.61 $\pm$ 0.01 \\ 
$r$& 0.56 $\pm$ 0.04&  -0.41$\pm$  0.02 & 0.40$\pm$ 0.01 \\
$i$& 0.70 $\pm$  0.06& -0.56 $\pm$  0.02&0.49 $\pm$ 0.01 \\

\hline     
\end{tabular}

\label{tab:mean_par}
\end{table} 


\noindent
adopted for culling the sample: the accepted compact sources are within black dashed lines.

\begin{figure*}[h!]
    \centering
    \includegraphics[width=\hsize]{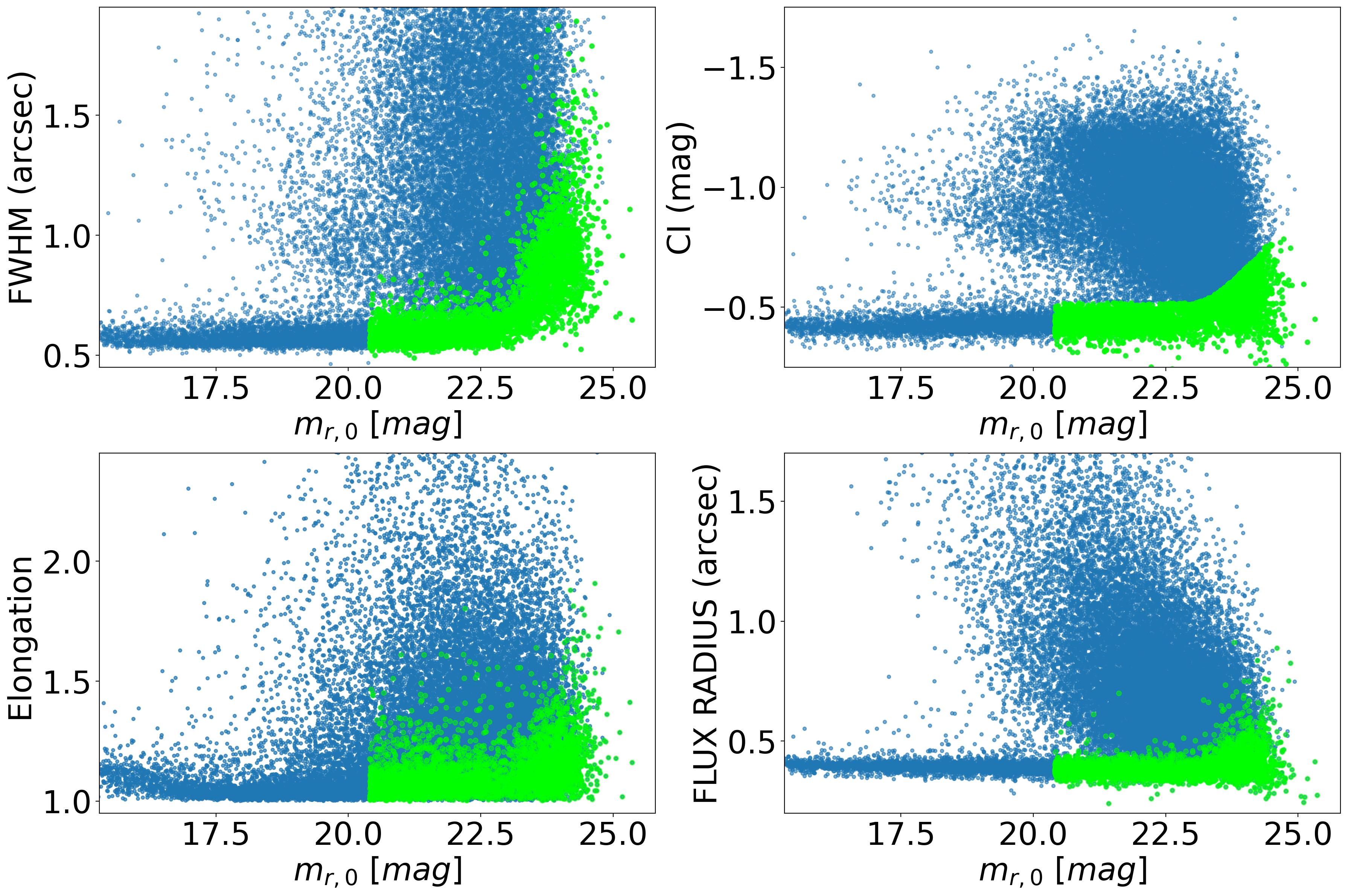}
    \caption{Morphometric parameters used for the selection of GC candidates. Blue dots show are all the sources in the $ugri$ matched catalog; green filled circles are the sources selected based on the CI, magnitude and the intervals reported in Table \ref{tab:sel_par_1}. }
    \label{fig:param_sel}
\end{figure*}

After the $CI_r$ cleaning, we also inspected the other morphometric parameters, and further clean the sample from possible outliers, as shown in Fig. \ref{fig:param_sel}. For the FWHM, elongation and FLUX\_RADIUS parameters, we decided to adopt upper/lower limits to identify bona-fide GCs. Table \ref{tab:sel_par_1} contains the parameters values we adopted for the final catalog of GC candidates. To choose these limits,  we made several tests to balance the effects of completeness and contamination, allowing the selection parameters vary from narrower intervals (lower completeness and contamination) to broader (higher completeness and contamination). For these tests, which were run iteratively with the color-color selection (see section \ref{sec:col_sel}) and included a comparison of the GC population properties  (i.e. radial profile, color distribution and density map, see Sect. \ref{sec:anlaysis_gc}) for each adopted selection range, we observed consistency in the results over a wide range of adopted selection intervals. This result suggest that, globally, the selection is driven by the CI parameter. In conclusion, we decided to adopt the wide  selection ranges of the FWHM, elongation and flux radius as reported in Table \ref{tab:sel_par_1}.

\subsection{Color selection}
\label{sec:col_sel}

The selection of $bona-fide$ GC candidates can be improved taking advantage of the multi-band coverage of the present dataset. With an approach similar to the one presented in \citet{cantiello20} and relying on the same GC master catalog adopted there (composed mainly of spectroscopically confirmed GC in Fornax), we used color-color plots to further refine the catalog of pre-selected GC candidates.

\begin{figure*}[h!]
    \sidecaption
    \includegraphics[width=\columnwidth]{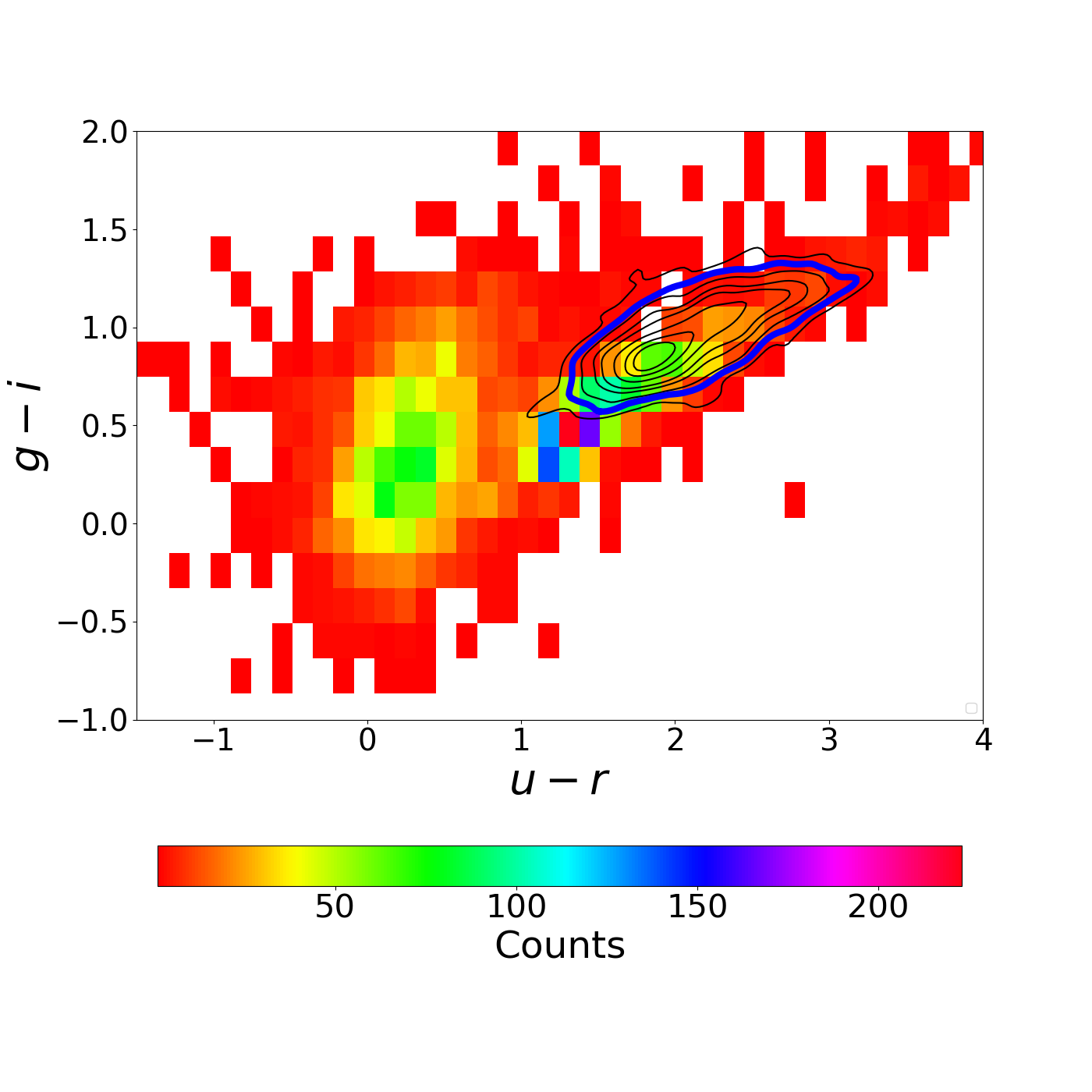}
    \includegraphics[width=\columnwidth]{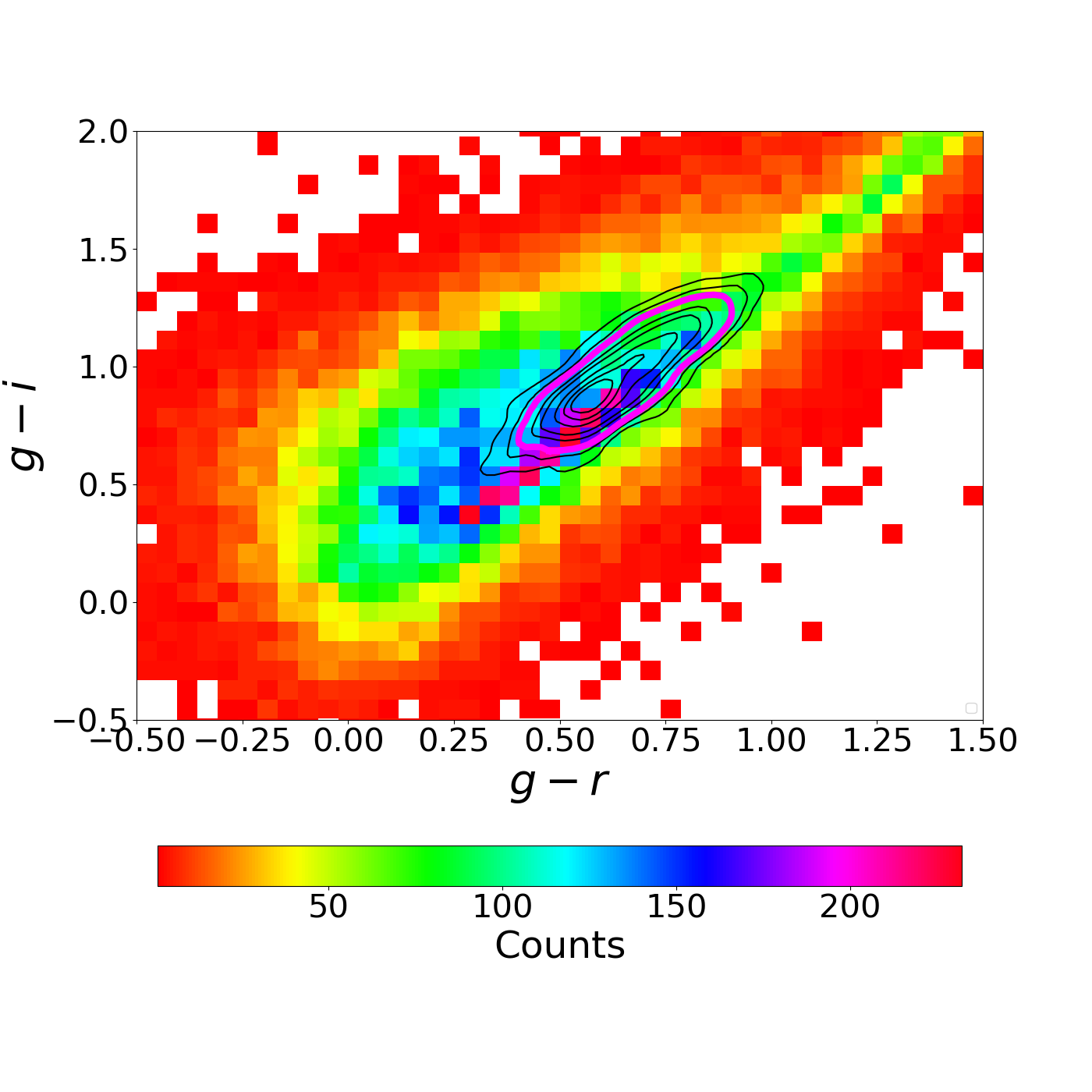}
    \caption{Color-color Hess diagrams of the pre-selected  GCs candidates in the $ugri$ (left panel) and $gri$ (right) catalogs. Black solid lines are the isodensity contours of GCs from the master catalog used in
\citet{cantiello20}. Each isodensity contour contains a certain percentage of GCs from the master catalog. In particular, going from the outer contour (lower GCs density) to the inner one (higher GCs density) the contours contain: 90\%, 85\%, 80\%, 70\%, 60\%, 30\%, 20\% 10\% of GCs in the master catalog. The 85\% (80\%) isodensity contour used for the selection are  highlighted in blue (left) and magenta (right).}
    \label{fig:color_sel}
\end{figure*}

The color-color plots for the $ugri$ and $gri$ catalogs are reported in Fig. \ref{fig:color_sel}, as Hess diagrams. 
Additionally, we included isodensity contours of the GCs from the master catalog obtained from the Fornax Deep Survey \citep{iodice16fds}. The contour level containing 85\% (80\%) of the sources in the catalog is highlighted with a thick blue (magenta) line for the $ugri$ ($gri$) catalog and serves as our reference contour\footnote{The isodensity levels of the master catalog are obtained using a kernel density estimator\footnote{\url{https://docs.scipy.org/doc/scipy/reference/generated/scipy.stats.gaussian_kde.html}} (KDE). The adopted parameters were:  bw\_method='scott' and bw\_adjust=0.7.}.

Ideally, one would like to have a GC candidate sample with low contamination and high completeness. However in order to choose which contour to use for color-color selection, we need to balance two effects. Narrow color-color density contours imply low contamination but also a fairly incomplete GC sample; wide  contours imply higher completeness, but also high contamination. To search for the contour level that is the best compromise between the two effects we made several tests. In each test, we adopted different contour levels, and inspected the properties of the resulting catalog (2D-map, radial and color distributions, etc.). Finally, we chose the contour levels containing the 85\%\footnote{The choice to use two different contour levels is consequence of the higher contamination presented in the $gri$ catalog.} (80\%) of  the sources of the master catalog as our reference for the $ugri$ ($gri$) matched catalog because narrower contours resulted in too sparse final sample of GC candidates, which  is prone to statistical sampling effects, while larger contours were too much contaminated by the background sources and spurious over/under densities in the regions of the mosaic characterised by higher/lower S/N.

The color-color selection algorithm was independently run on the $ugri$  and on the $gri$ catalog. The former is expected to provide a final GC-sample with lower contamination compared to the $gri$, thanks to the use of two independent colors, $u{-}r$ and $g{-}i$, and the wider wavelength range covered (more details will be given in Sect. \ref{sec:anlaysis_gc}) . However, it is also much less complete than the final $gri$ GC-sample, because of the much shallower $u$-band depth compared to the other passbands (Sect. \ref{sec:completeness}).

\subsection{Final catalog}

The final sample of GC candidates derived from the $ugri$ ($gri$) catalog contains 769 (4944) sources. The basic parameters of this sample are reported in Table \ref{tab:gc_catalog}. 
The  catalog contains: $i)$ the source ID and its RA and Dec coordinates; $(ii)$ the automatically estimated magnitudes by SExtractor, the aperture corrected AB magnitudes and their error in all available bands;  $(iii)$ the morphometric parameters (FWHM, Concentration Index, Flux Radius, elongation) in the reference passband (i.e. $r$-band), and the $E_{B-V}$ reddening at the position of the source.

This catalog still contains point source contaminants matching 
the properties of the GC population, especially at the faint magnitude level. Nevertheless, thanks to the
large covered area by VST images, and assuming that any population of contaminants
is uniform over the inspected area, the GCs in the region can be analyzed using background subtraction methods  \citep[][see Sect. \ref{sec:anlaysis_gc}]{cantiello15}.

\newpage

\newpage

\begin{figure*}[h!]
    \centering
    \includegraphics[width=\columnwidth]{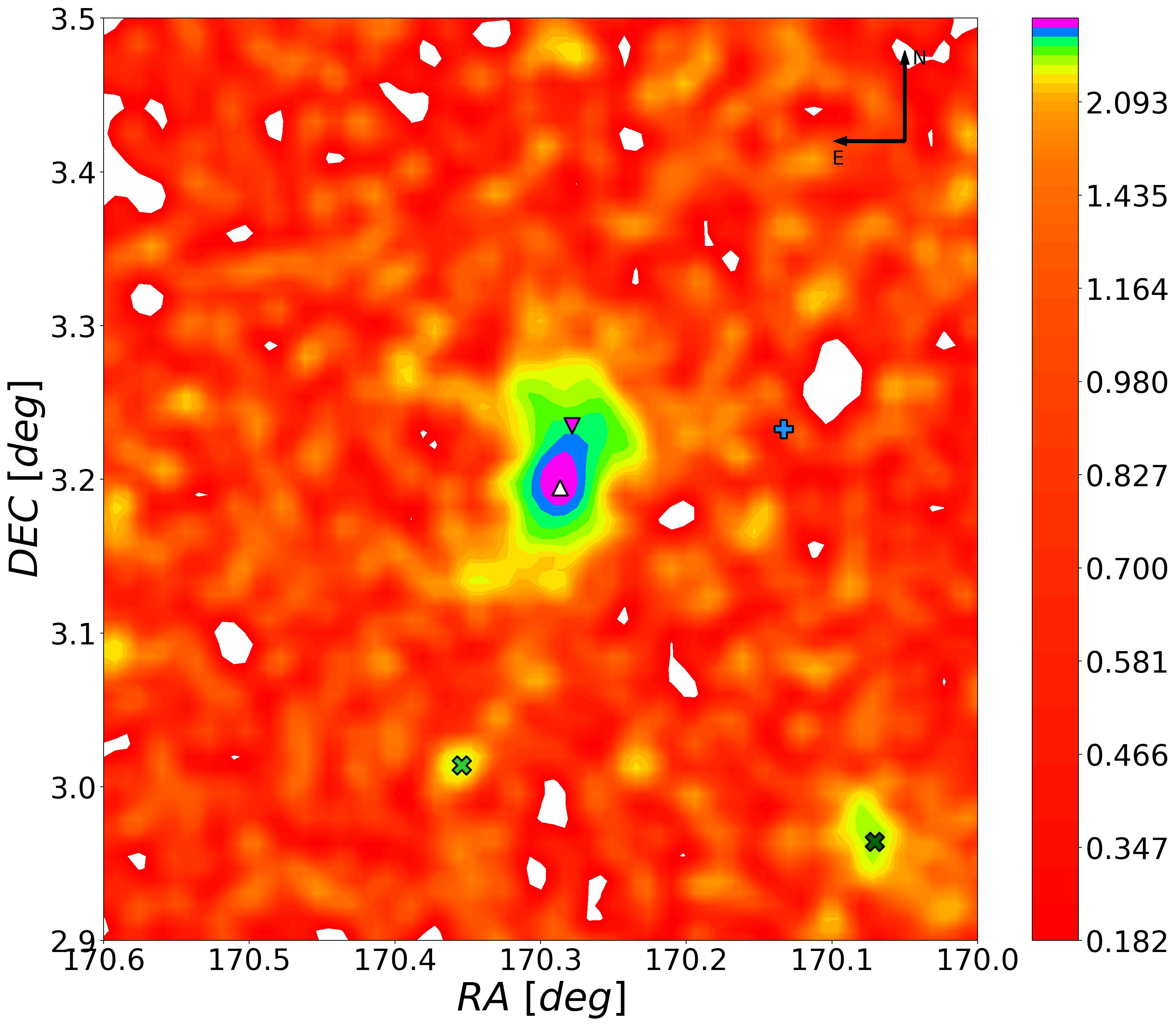}
    \includegraphics[width=\columnwidth]{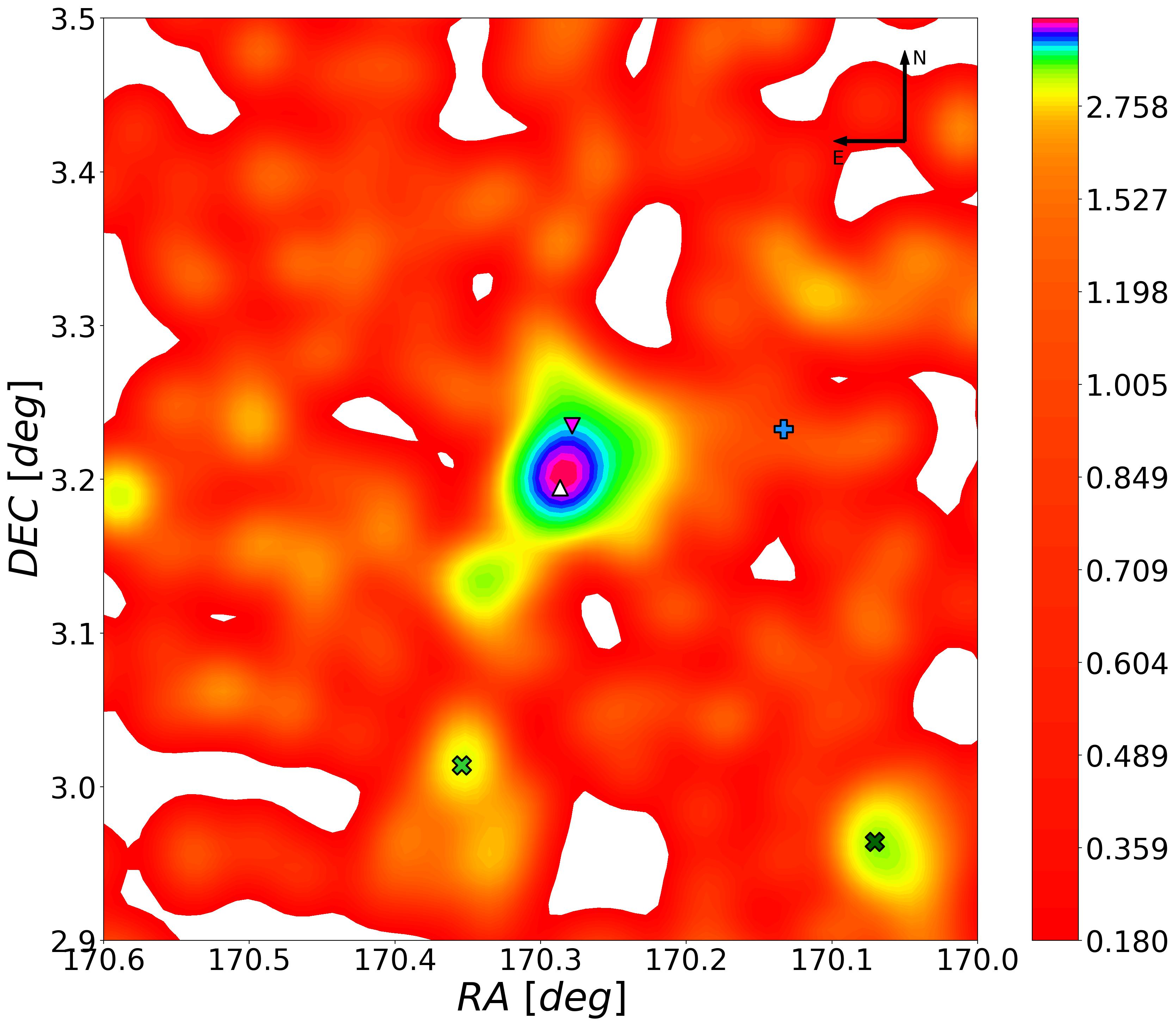}

    \caption{2D surface distribution of GCs candidates within 0.6$\times$0.6 $deg^2$ around NGC\,3640 and NGC\,3641. The positions of NGC\,3640 and NGC\,3641 are marked with the magenta and white triangles, respectively. The blue-filled plus symbol shows the location of the Dwarf\,12. The lime and darkgreen crosses show the position of NGC\,3643 and NGC\,3630, respectively. $Left$ $panel$: 2D density map of GCs candidates selected from the $gri$ matched catalog. $Right$ $panel$: the same plot but with GCs candidates selected from the $ugri$ matched catalog.} 
    \label{fig:2d_maps}
\end{figure*}

\begin{figure*}[h!]
\centering
     \includegraphics[width=\columnwidth+3cm]{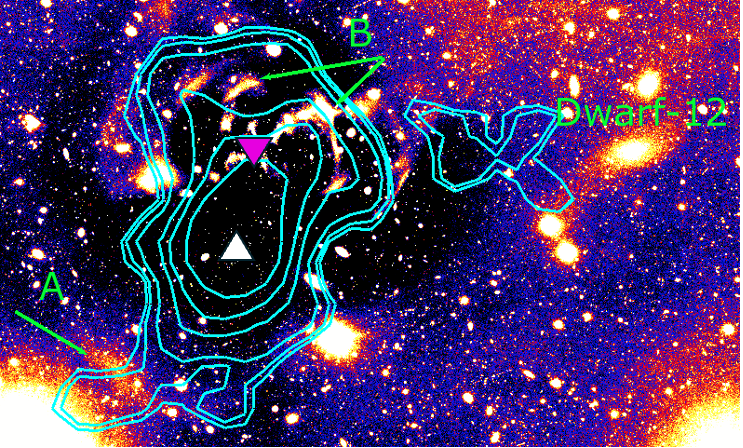}
     
     \includegraphics[width=\columnwidth-3cm]{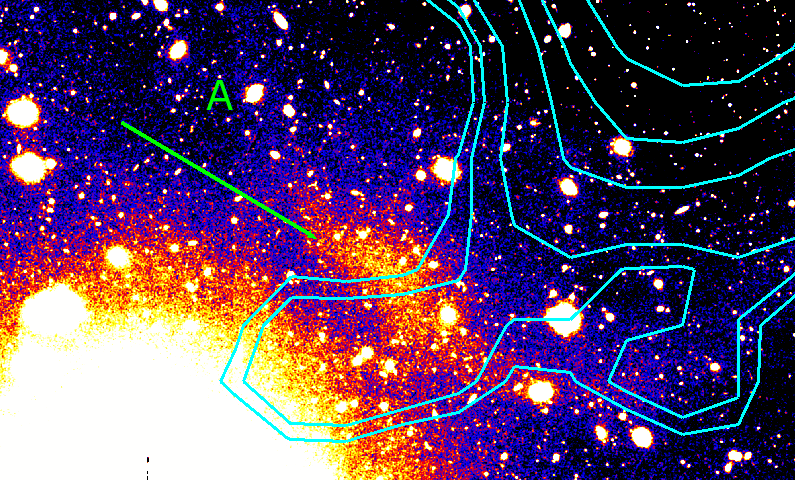}
     \includegraphics[width=\columnwidth-3cm]{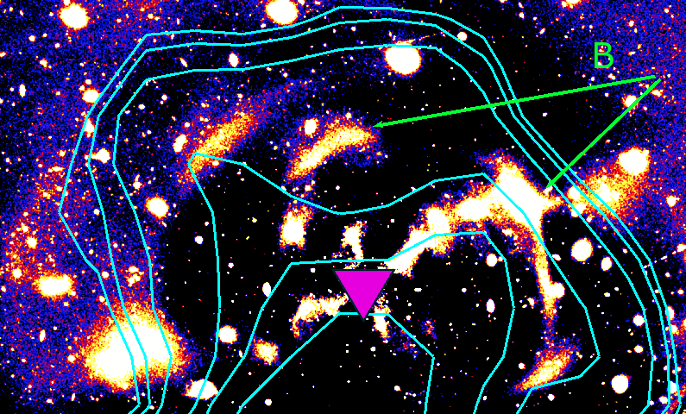}
     \includegraphics[width=\columnwidth-3cm]{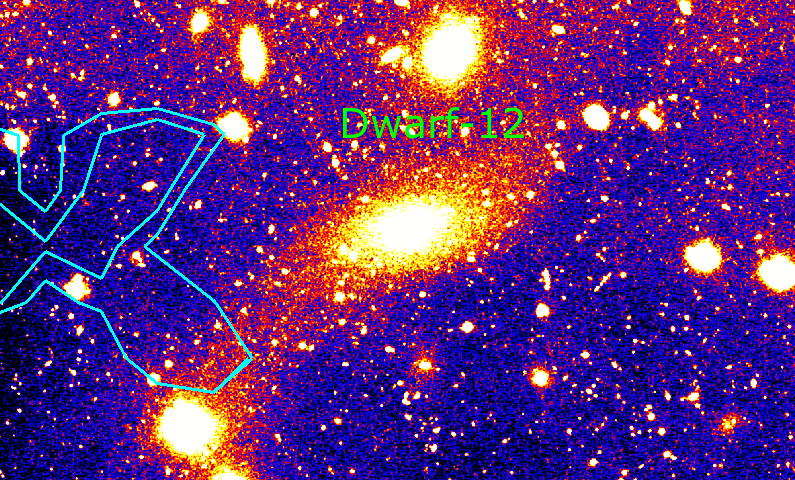}
     \caption{$Upper$ $row$: Cutout of 0.9$\times$0.7 sq. degrees of the galaxy-subtracted image in the $r$-band. Light-blue contours are the isodensity contours extracted from the surface density map (see Fig. \ref{fig:2d_maps}, left panel). Magenta and white triangles represent the center of NGC\,3640 and NGC\,3641, respectively. Green labels indicate the connection regions between GC population and LSB features described in Sect. \ref{sec:density2d}. Label A points towards an LSB tail South-East of NGC3640; label B points to shell structures North-West of NGC3640.
     $Lower$ $row$: Zoom-in on the the three labeled regions.}
    \label{fig:contour_over}
\end{figure*}


\newpage
\clearpage

\section{GC population: sample analysis }
\label{sec:anlaysis_gc}
In this section, we study the spatial, color and luminosity distribution of the selected GC samples. We expect that the final GC-list still contains some contamination from fore- and back-ground sources. With the only exception of the 2D surface density maps, to reduce the effect of such contamination we take advantage of the large observed field, and use a statistical background decontamination method, as described in our previous works \citep{cantiello15,dabrusco16}. This technique involves statistically characterizing the GC contamination levels by inspecting regions far from the galaxies, where no GCs are expected to be. The underlying assumption is that the background sources identified as contaminants are evenly distributed across the field. Therefore, by using the final catalogs of GC candidates, the on-galaxy population is decontaminated from spurious sources using the catalog of off-galaxy candidates as a reference.

\begin{figure}
    \centering
    \includegraphics[width=\columnwidth]{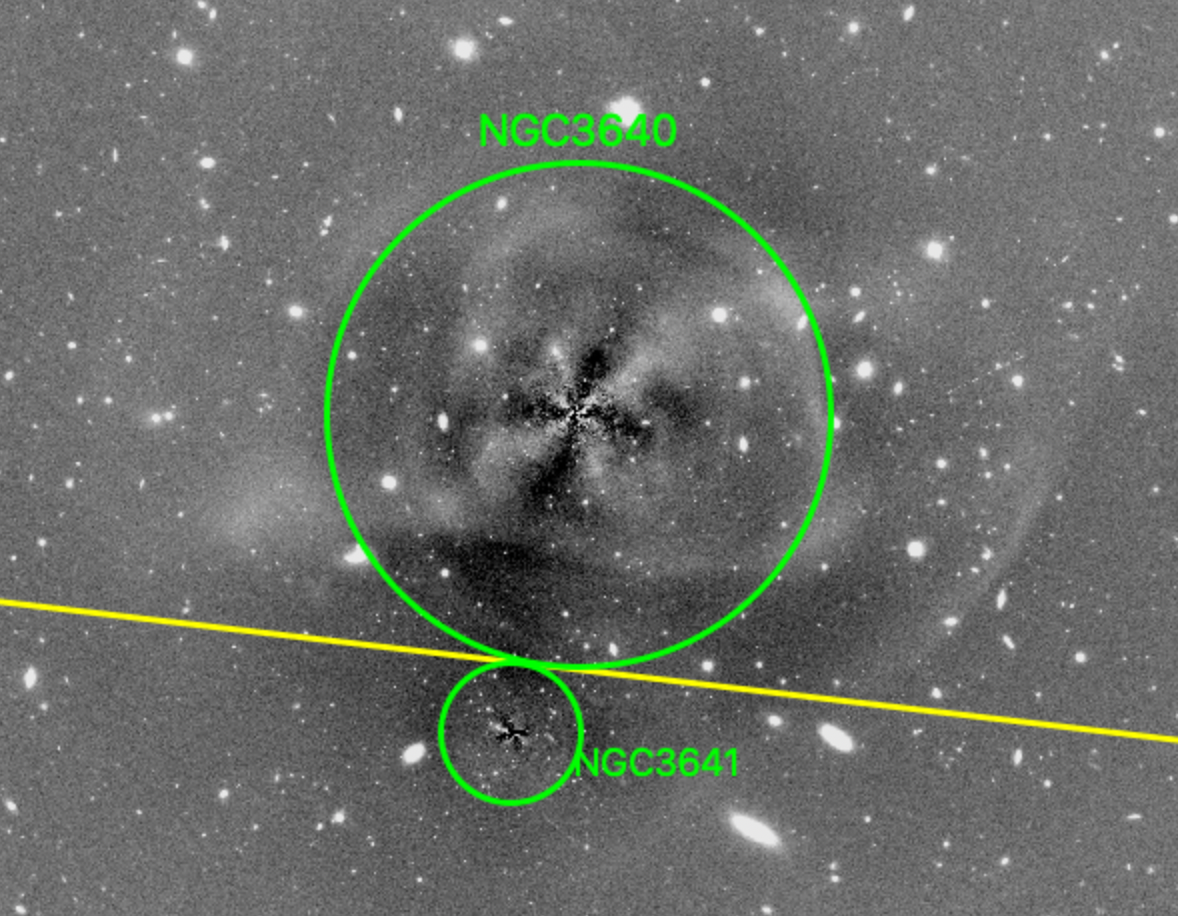}
    \caption{Adopted Gc separation method. Green circles represent 3.9$R_e$ circles centered on each galaxy. The yellow line is the tangent common to the two circles, and it is used to split the GCs population of NGC\,3640 from the one over NGC\,3641. The cutout size is 9 $\times$ 7 arcmin$^2$.}
    \label{fig:sepration}
\end{figure}

\subsection{Surface density map of GCs over NGC\,3640 field}
\label{sec:density2d}

Thanks to the large format of the VST images, under the assumption that the level of contamination is roughly constant across the observed field, we can study the 2D map of the GC-candidates, without the need of subtracting the background. We inspected the GC distribution maps centered on NGC\,3640, applying a kernel density estimator technique\footnote{We used the $kdeplot$ packaged of seaborn v.11.2 (link \href{https://seaborn.pydata.org/generated/seaborn.kdeplot.html}{here}) with the following parameters: bw\_method='scott' and bw\_adjust=0.1. }. The distribution is inspected using both the $ugri$ and the $gri$ catalogs. A zoom on the central region, over a galactocentric radius of $\sim0.3$ degrees, is shown in Fig. \ref{fig:2d_maps}. 


In the figure, the global GC population, centered on the pair of galaxies NGC\,3640 and NGC\,3641, stretches from southeast to northwest with an inclination nearly identical
to the line connecting the two galaxies. A key result from our analysis is that among the two galaxies, NGC\,3641, which is also  fainter/less massive, is the one showing the largest central density of GCs. Indeed, the peak of the density appears on NGC\,3641.
In the 2D map, an over-density of GC candidates on the other two bright galaxies in the group, NGC\,3630 and NGC\,3643, is also observed.

To further investigate the 2D distribution of GCs, the iso-density contours from the  surface density map in Fig. \ref{fig:2d_maps} are shown in Fig. \ref{fig:contour_over} overlaid on the galaxy-subtracted image. 
The southeastern tail of the contour levels appears to be spatially connected to an LSB feature labeled "A" in Fig. \ref{fig:contour_over}. In the northeast extension of the density contours, we note two such connections. First, the GC contours follow the shape of what appears to be a complex structure of streams and merging features north on NGC\,3640 (label B in the figure). Secondly, a bridge of GCs seems to connect the distribution on the galaxy pair along the direction of a disturbed dwarf in the field (Dwarf\,12, see Sect. \ref{sec:lsb}). We inspected the contour levels in the background region (between 25 and 30 $arcmin$, see next section) to ensure that the bridge is a real feature and not an artifact. Observations of the background contour levels reveal more rounded structures, with no evidence of any distinct patterns. Inspecting the contour levels in the background regions, we find them to host a lower density population and to be less extended compared to the asymmetric structure of the GC bridge and the other substructures mentioned here.

As a sanity check, we also inspected the maps from the $ugri$ catalog. The general behaviour of the features just described (south-east/north-west stretch, highest density peak on NGC\,3641, broader distribution in the region of complex streams, etc.) is preserved, as seen in Fig. \ref{fig:2d_maps} (right panel).
  
We also investigated the overdensity of sources with color $0.15\leq(g-i)\leq0.65$ and $1\leq(u-r)\leq 1.6$, present in the left panel in Fig. \ref{fig:color_sel}, to understand if these sources are foreground MW stars or young GCs. We inspected the 2D density map to understand the nature of these sources. We observed a U-shape overdenstity features extending from south-est to south-west respect to the galaxy pair. This overdensity might be associated to a  intermediate GC population formed after merging event of NGC\,3640. However, spectroscopic datasets are required to confirm or rule out this scenario.

\subsection{Radial GC number density profile}
\label{sec:radial}

To study the radial distributions of the GC systems of NGC\,3640 and NGC\,3641, we investigated the azimuthally averaged surface density profiles centered on each galaxy. Because of the small angular separation between the two galaxies,
we need to adopt some criterion to separate our sample of GC candidates into two subgroups: one belonging to NGC\,3640 and another one to NGC\,3641\footnote{Ideally one would require spectroscopy, which would be very effective in this system thanks to the radial velocity offset between the two (see Table \ref{tab:properties}).}.

Figure \ref{fig:sepration} shows the criterion we adopted. The separation criterion relies on the effective radius of the galaxies, which is directly linked to the radial extent of the two GC populations  \citep{alamo21}. We split the sample using the tangent line between the two circles shown in Fig. \ref{fig:sepration}. These circles are centered on the galaxies and have a radius of $\alpha$ effective radii, $\alpha R_e$. The value of $\alpha$ that makes the two circles tangent is $\alpha\sim3.9$. 

\begin{figure}
    \centering
    \includegraphics[width=\columnwidth]{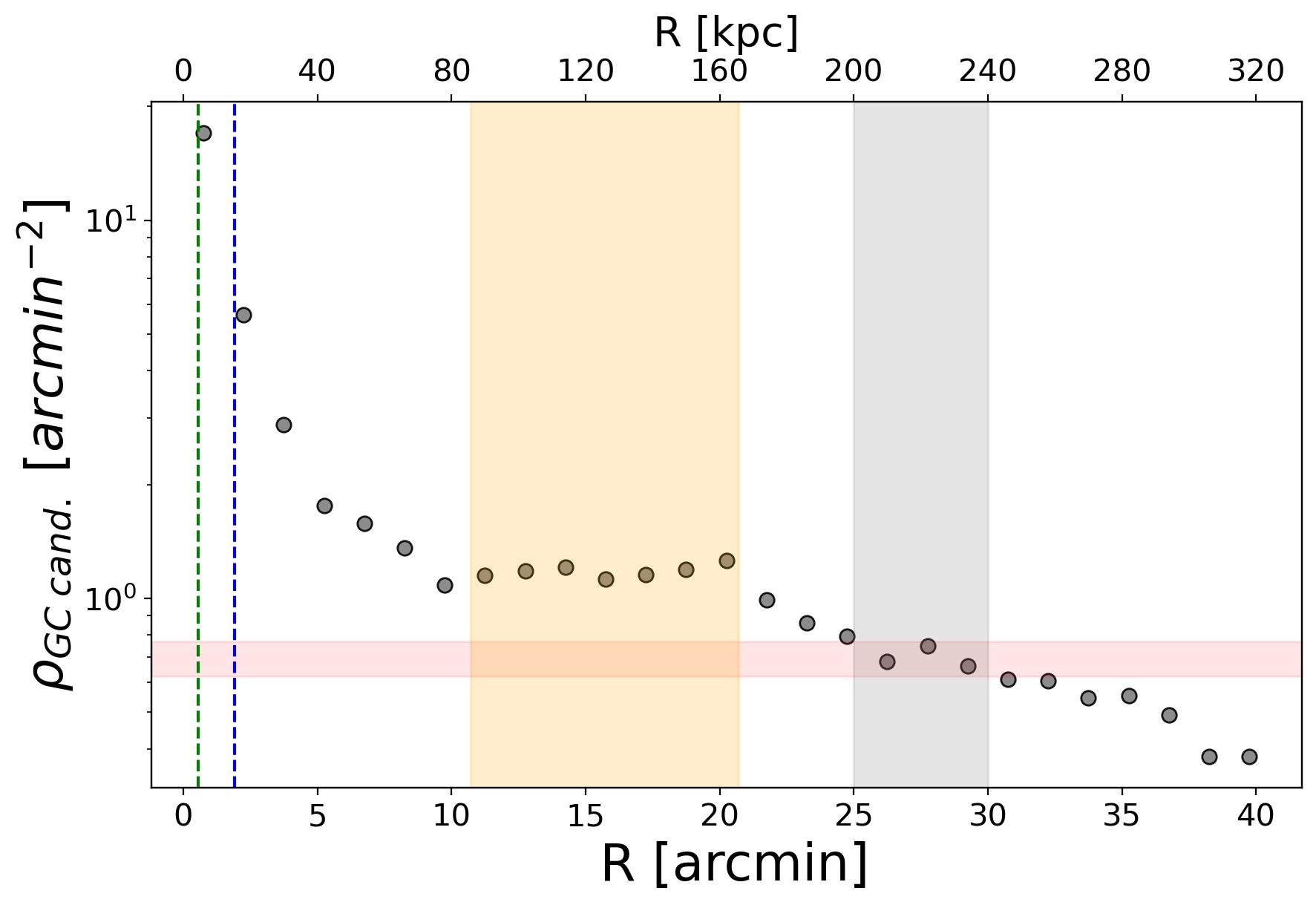}
    \caption{Radial distribution of the global GCs population plotted as surface density vs. projected radial distance from NGC\,3640 from the tangent point in Fig. \ref{fig:sepration}. The $y$-axis is in $logarithmic$ scale. The green and blue dashed line represent the location of NGC\,3641 and NGC\,3640, respectively. The gray shaded region shows the radial range used to estimate the residual contamination level. The red horizontal line is the 1$\sigma$ background level. The yellow shaded region identifies the flat profile region likely associated with the intra-group GCs.}
    \label{fig:radial}
\end{figure}

Then, all GC candidates  above the tangent (yellow line  in Fig. \ref{fig:sepration}) are associated to NGC\,3640, the ones below the tangent are associated to NGC\,3641. For the sake of clarity, it is essential to emphasize that the described criterion has been fine-tuned specifically for this case. We also inspected the surface brightness (SB) level of the two galaxies, and found that the point where the two SB profiles reach equality, is at about the same location as the tangent point (within a few $arcsec$). For the motivation detailed below, we decided to use the criterion based on the effective radii.

Before inspecting the radial density profile on the single galaxies, we analysed the radial surface density profile for the global GCs population (i.e. the combined GCs systems of NGC\,3640 and NGC\,3641 without any distinction), assuming the center to be the tangent point where the two circles meet in Fig. \ref{fig:sepration}.

\begin{figure*}
    \centering
    \includegraphics[width=\columnwidth]{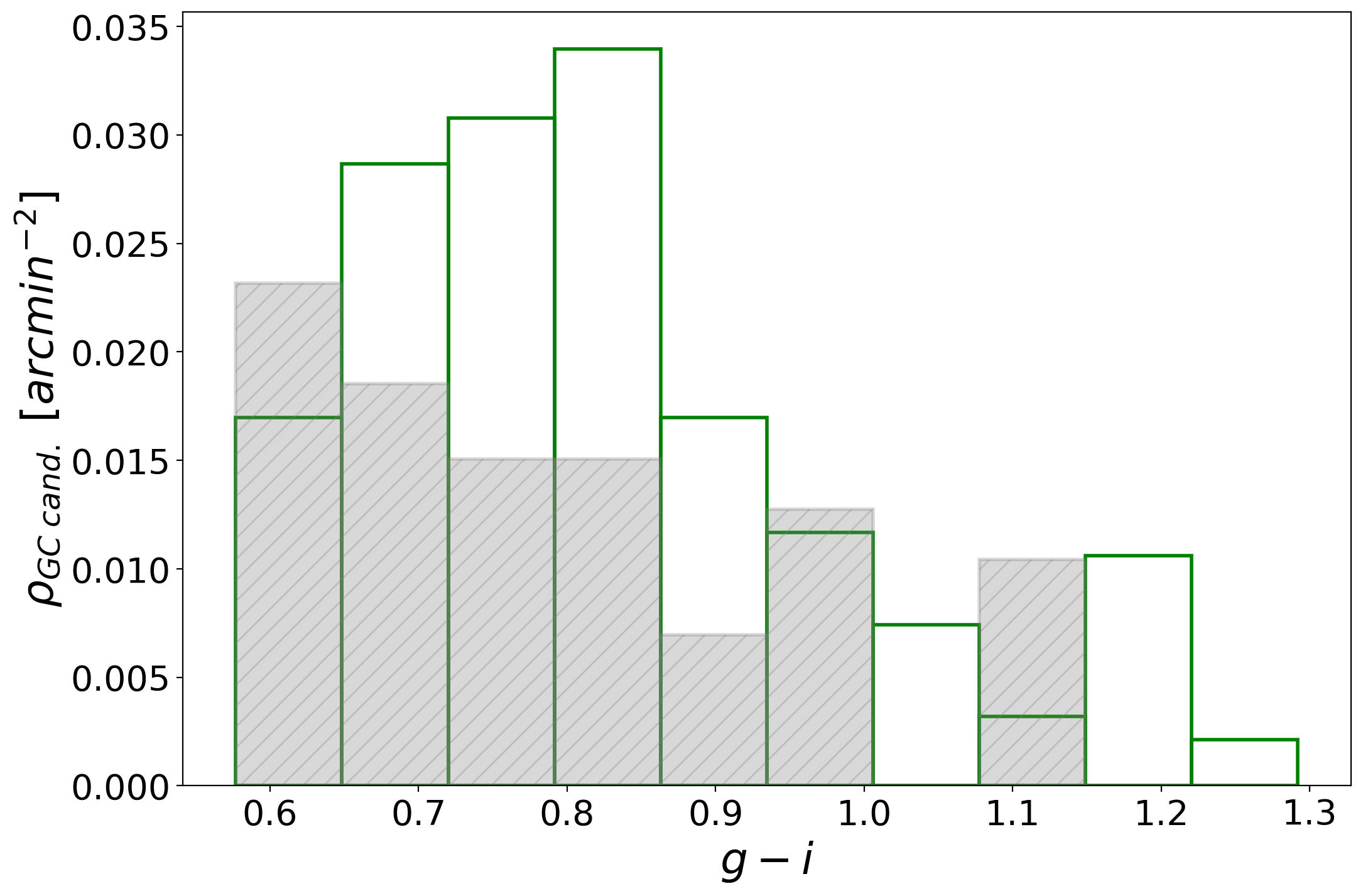}
    \includegraphics[width=\columnwidth]{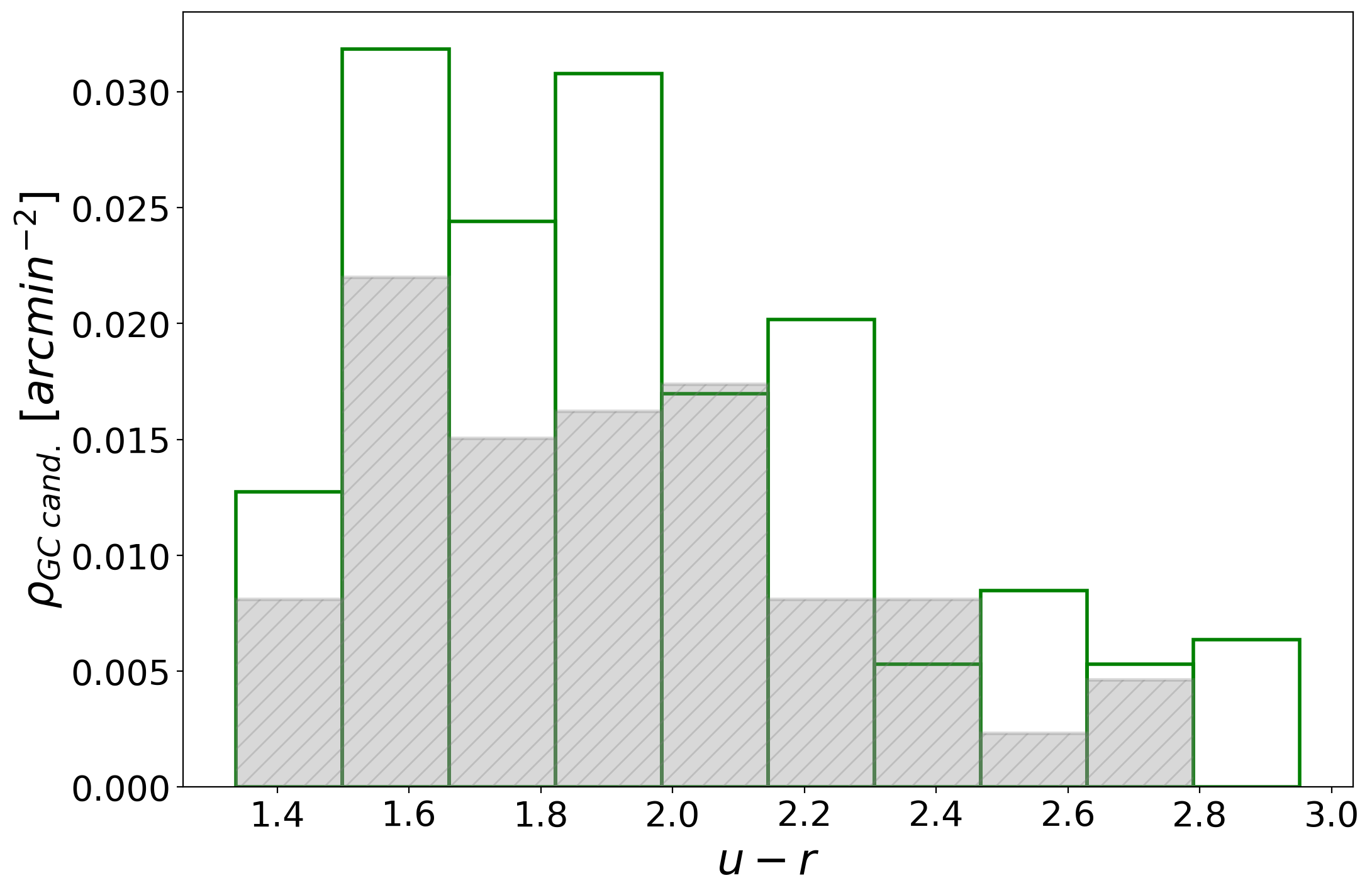}
    \caption{Color distribution of the GC candidates in the background region (25-30 $arcmin$, grey shaded bars) and in the plateau region (10-20 $arcmin$, green lines). The $(g-i)$ distributions are shown in the left panel, while the $(u-r)$ distributions are shown in the right panel.}
    \label{fig:palt_back_conf}
\end{figure*}

We considered concentric circular annuli, 1.5 $arcmin$ wide, and counted the GCs candidates within each annulus. For this profile, we masked the regions around saturated pixels, bright stars, image artefacts, etc. The area of the annuli is corrected for such masking.
Figure \ref{fig:radial} shows the radial density distribution of the bulk GCs population using the $gri$ catalog.
We mainly used this global GC profile to identify the region where the background density becomes constant. Since this catalog is richer than the ones we get splitting the sample in two, it allows to constrain more robustly the background contamination level. The gray-shaded region in Fig. \ref{fig:radial} indicates the spatial regime, between 25 and 30 $arcmin$, that we adopted to estimate the background level (shown as a red-shaded region in the figure). We assumed that all GC candidates  in this region are interlopers (either background galaxies or foreground MW stars) that pass our photometric and morphological selections. We chose this region because interlopers should have a uniform distribution around the group, leading to a flat radial distribution. The region also needs to be away from the outer image area, which typically has lower signal-to-noise ratio because of the observational strategy. At group-centric radii beyond 30 $arcmin$ the distribution approaches to the edges of the images, resulting in regions with lower completeness fractions (Sect. \ref{sec:completeness}). This generates the drop seen in the density profile at $R>30$ $arcmin$.
The profile in Fig. \ref{fig:radial} shows a monotonic decrease within $R\leq10$ $arcmin$, followed by a flat density up to 20 $arcmin$. We rejected the flat region between 10 and 20 $arcmin$ as a possible background area because it has a GC candidate density a factor of $\sim$2 higher than the 25-30 $arcmin$ region, which cannot be attributed solely to differences in image depth (see Fig. \ref{fig:m80_rad}, right). Additionally, the color distribution of sources in this region shows a blue-density peak compared to the distribution within 25-30 $arcmin$ (see Fig. \ref{fig:palt_back_conf}), as expected in the case of a residual GC component. As discussed below, we associate this flat region with a possible population of intra-group GC candidates, which drops off after $\sim25$ arcmin.


We inspected the same radial profile for the $ugri$ catalog and, although it has many fewer GC candidates, the general behaviour (in terms of trend and background identification) is confirmed.

\afterpage{
\begin{figure*}
    \centering
    \includegraphics[width=\columnwidth]{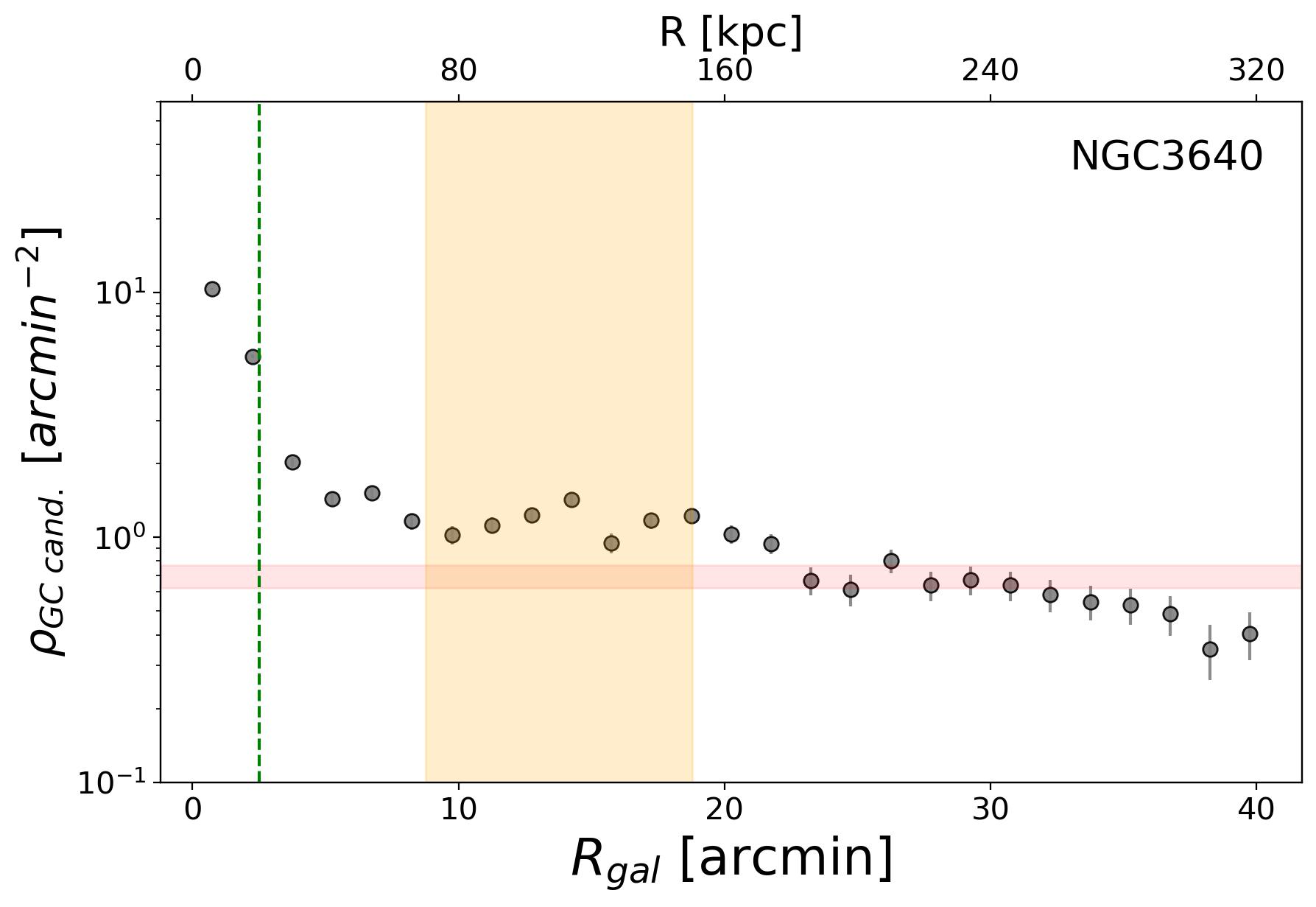}
    \includegraphics[width=\columnwidth]{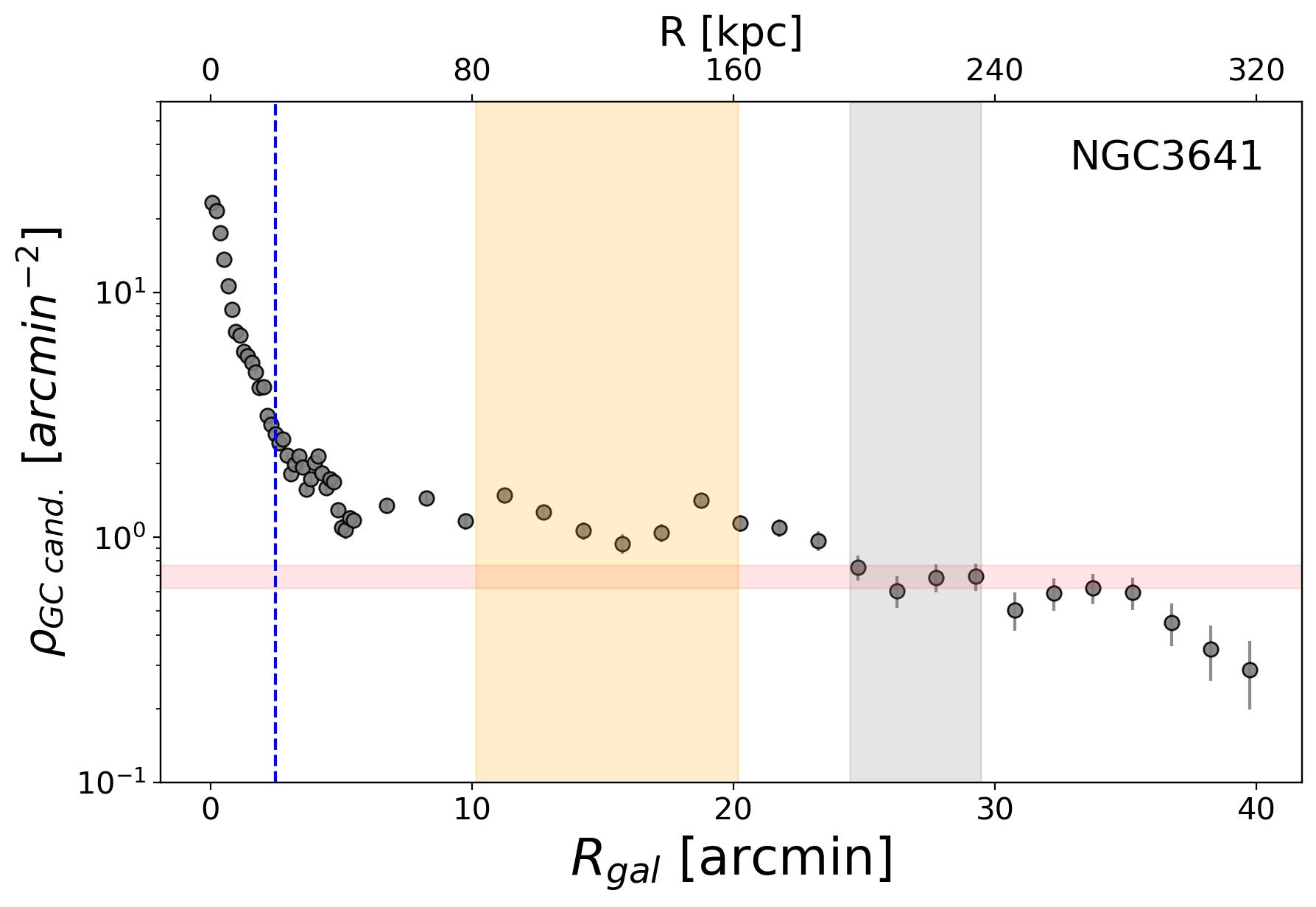}

    \caption{Radial distributions of GCs associated with NGC\,3640 (left panel) and NGC\,3641 (right). Only GCs associated with each galaxy, split as described in Sect. \ref{sec:radial}, are used for the radial density profiles. Symbols and colors are the same as in Fig. \ref{fig:radial}.}
    \label{fig:radial_lin_both}
\end{figure*} 

\begin{figure*}    
    \centering
  \includegraphics[width=\columnwidth]{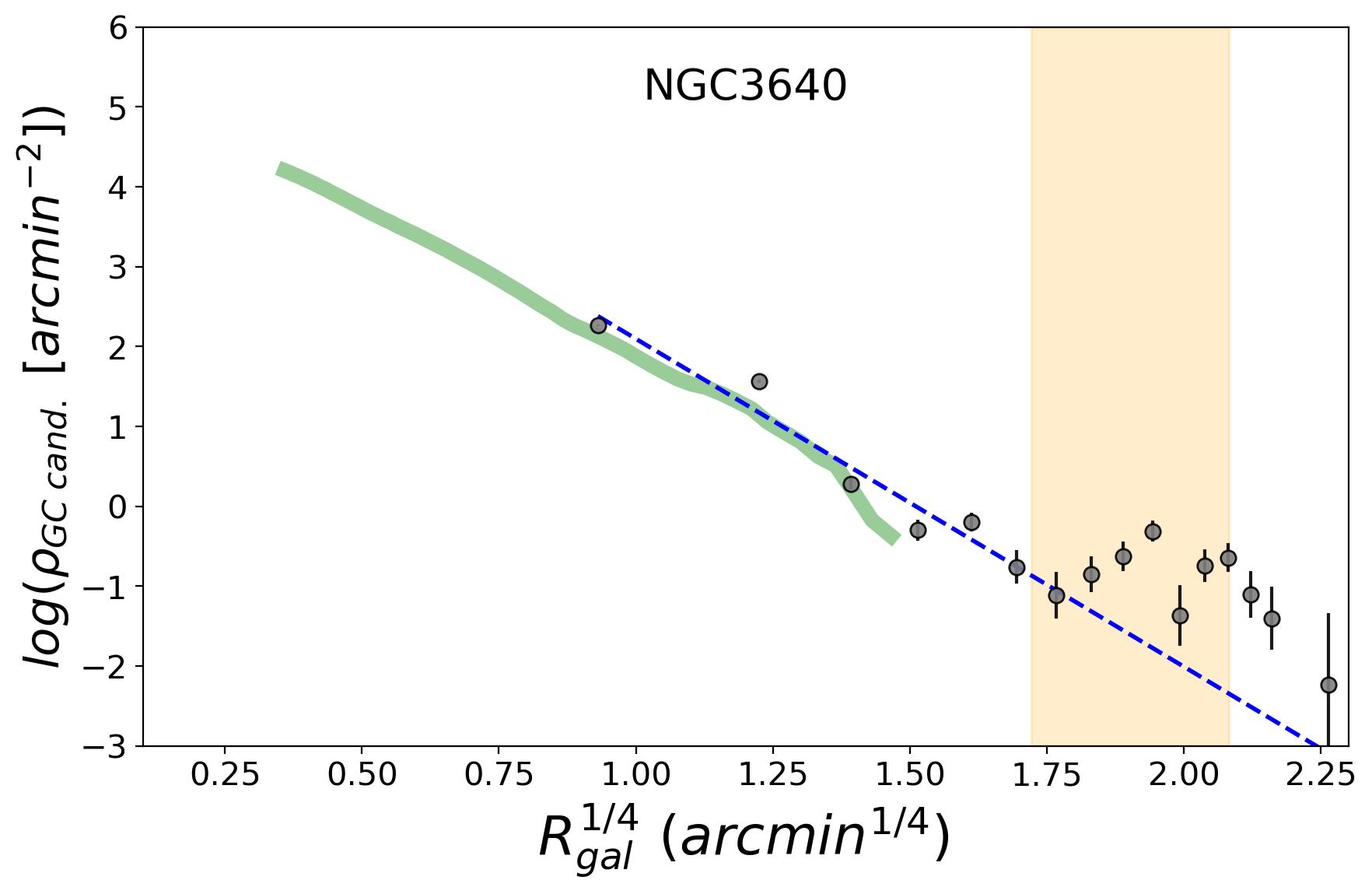}
    \includegraphics[width=\columnwidth]{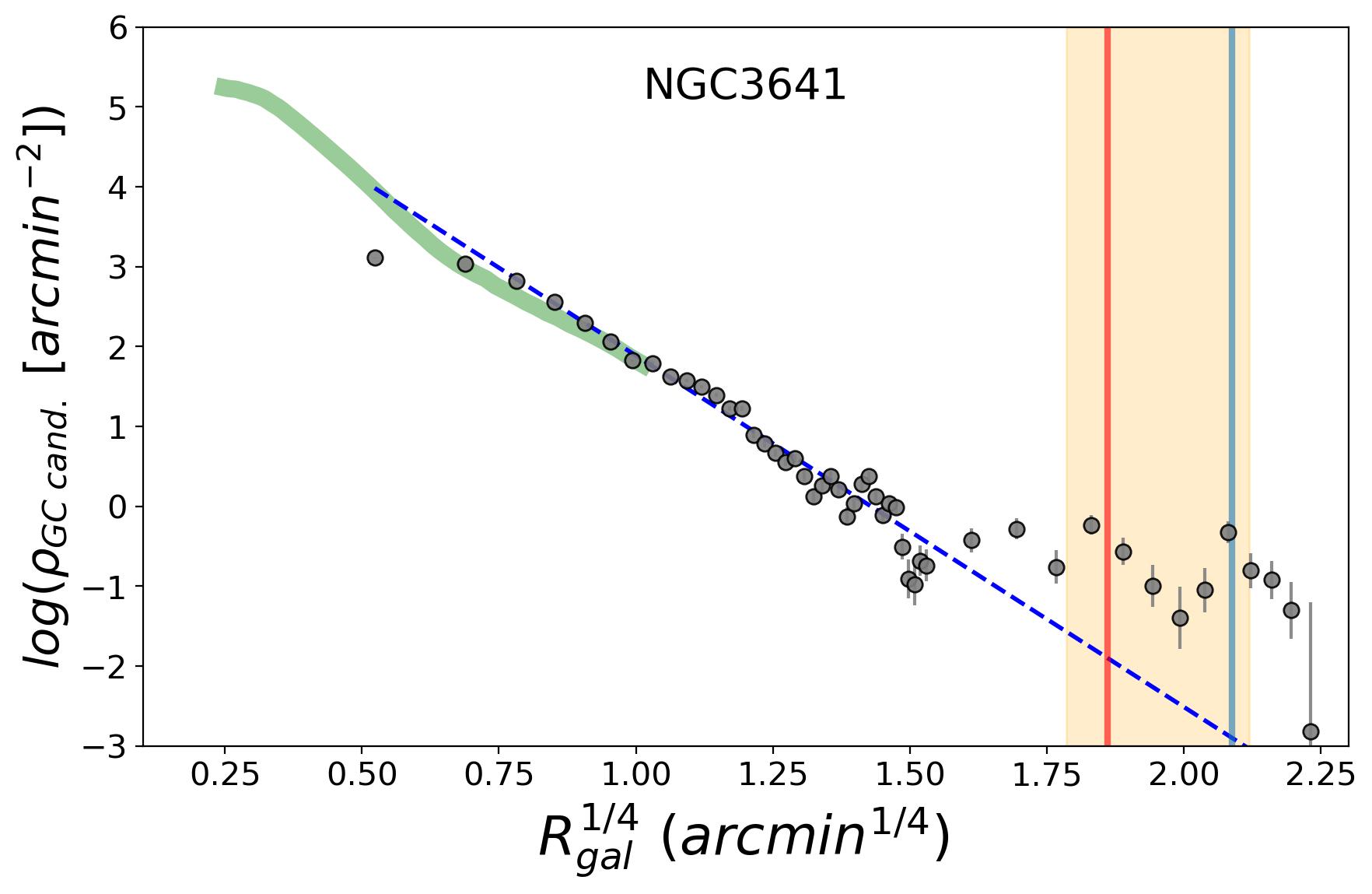}
   
    \caption{Background-subtracted radial distributions of GCs associated with NGC\,3640 (left panel) and NGC\,3641 (right panel), fitted with a $R^{1/4}$ de Vaucouleurs profile. On the y-axis, the logarithm of the background-subtracted density profile is plotted against the fourth root of the galactocentric distance from each galaxy photocenter. The green shaded lines represent the surface brightness profile of the galaxy in the $g$-band with an arbitrary shift. The yellow shaded areas delineate a region in the density profiles likely associated with intra-group GCs. Light-blue and red vertical solid lines mark the distances where NGC\,3630 and NGC\,3643 are located, respectively.}
    \label{fig:radial_fit_both}
\end{figure*} 

}

\subsection{Individual GC Radial profile}
\label{sec:radial_ind}

As anticipated, we also analyzed the GCs radial density profiles considering all GC candidates above the separation limit (tangent yellow line in Fig. \ref{fig:sepration}) associated with NGC\,3640, and the ones in the remaining area that are associated with NGC\,3641. We used circular concentric annuli, adopting a constant annular width of 1.5 $arcmin$ for NGC\,3640. For NGC 3641, we used radial binning with a width of 0.15 $arcmin$ up to 6 $arcmin$, followed by an annular width of 1.5 $arcmin$. This adjustment was necessary to provide a more precise characterization of the inner region of the galaxy, where the density of GCs is relatively high.

Figure \ref{fig:radial_lin_both} shows the  radial profiles for NGC\,3640 (left panel) and NGC\,3641 (right). The light-red horizontal line is the background level from the global profile, estimated as shown in Fig. \ref{fig:radial} with a thickness corresponding to  1$\sigma$ level. Both profiles show decreasing trend outward from the galaxy center until $R_{gal}\sim$10 $arcmin$ for NGC\,3640 and to $\sim$5 $arcmin$ for NGC\,3641. The density reaches a plateau at  $\sim$20 $arcmin$ before dropping to the background level. The background levels for both galaxies agree within errors with the global value. Table \ref{tab:back_value} reports the estimated  mean and the $rms$ of the global background (Fig. \ref{fig:radial}, between 25-30 $arcmin$), along with the values estimated for the individual galaxies  within the gray shaded regions in Fig. \ref{fig:radial}. 

The two radial profiles in Fig. \ref{fig:radial_lin_both}, highlight a feature already discussed in the previous section: NGC\,3641 shows substantially larger GC density in the inner regions compared to NGC\,3640, despite it being $\sim$ 3 mag fainter (see Table \ref{tab:properties}).


\begin{table}

    \caption{Measured and fitted properties of the radial GC density profile.   }
    \tiny
    
    \centering
    \begin{tabular}{ccccc}
        \hline
    \hline
        Galaxy & Background level &$R_{t}$& \multicolumn{2}{c}{de Vaucouleurs Law} \\ 
\cmidrule(lr){4-5} 

& ($arcmin^{-2}$)& ($arcmin$)& $a_0$ &$a_1$ \\
&(1)&(2)&(3)&(4)\\

            \hline
    
         Global     & 0.69 $\pm$ 0.04 & $\sim $11 &... &...\\
         NGC\,3640& 0.68 $\pm$ 0.07  & $\sim$ 9& $6.2\pm 0.8$ &$-4.1\pm0.6$  \\
         NGC\,3641 & 0.69 $\pm$  0.05 &  $\sim$ 5 &$6.3\pm 0.2$ & $-4.4\pm0.1$ \\
             \hline
    \hline
    \end{tabular}

    \begin{justify}
        \textbf{Notes:} Col. (1): Estimated background level from each radial profile; Col. (2): transition radius between the central radial trend and the plateau region in the radial profile; Col. (3-4): fitted parameters of the de Vaucouleurs profile: log($\rho_{GC}$)=$a_0$ + $a_1r^{1/4}$.
    \end{justify}
    \label{tab:back_value}
\end{table}

We also inspected the background subtracted radial profiles, shown in Figure \ref{fig:radial_fit_both}, which we fitted with a $R^{1/4}$ de Vaucouleurs profile, log($\rho_{GC}$)=$a_0$ + $a_1r^{1/4}$, out to the transition radius between the central radial trend and the plateau ($R_t$$\sim$9 $arcmin$ for NGC\,3640, abd $R_t$$\sim$5 $arcmin$ for NGC\,3641). In the figure we also show the galaxy SB profiles in the $g$-band (with an arbitrary shift; R. Ragusa, priv. communication). The preliminary galaxy light profile reaches a depth of $\mu_g\sim 28\ mag/arcsec^2$ at a distance of $\sim$ 5 $arcmin$ from NGC\,3640 and $\sim$ 1 $arcmin$ from NGC\,3641. 

For both galaxies we observe a good match between the SB profiles and the radial GC densities with the GCs profile reaching larger radii, a result consistent with expectations \citep{harris01,alamo12,kartha13,cantiello15}. 

In a previous work from the VEGAS series, \citet{ragusa2023A&A...670L..20R}, taking advantage of the same wide-field used here, analyzed the LSB features in the intra-group space within this field. 
Focusing on the intra-group light (IGL), the authors found that the light fraction of this component compared to the total luminosity of the pair is $\sim$8$\%$ in both $g$ and $r$ bands, with average colors $g-r$ $\sim$ 0.7 mag and $g-i$ $\sim$ 0.9 mag, which are in full agreement with theoretical predictions \citep[e.g.][]{Contini2019}. These results were achieved using the well-tested method in the literature of multi-component decomposition of the averaged azimuthally surface brightness profile of the brightest central galaxy in the group or cluster, applied in this case to NGC\,3640. The results of this analysis (Ragusa et al. , in prep.) indicated a transition radius of $\sim$ 1.5 $arcmin$. In the context of IGL/ICL studies, the transition radius refers to the separation radius between the brighter and inner component of the galaxy, gravitationally bound to it (which, in the case of NGC\,3640, follows a Sersic profile), and the outer exponential component, representing the diffuse stellar envelope of the galaxy plus the IGL component. This transition radius, related to the behavior of the core/envelope stars in the galaxy, should not be confused with the GC transition radius described above, which is about a factor of 7 times larger. Although we expect that in a virialized system the two transition radii would be correlated with each other, here the galaxy system is actively merging. It will be of interest to analyze this correlation using a comprehensive study based on multiple systems --both interacting and non-interacting-- and on numerical simulations of galaxies that include proper GC systems modeling.

The radial density profile of NGC\,3640 shown in Fig. \ref{fig:radial_fit_both} (left panel) shows a plateau of GCs density between 10$\leq R_{gal} [arcmin] \leq 20$ which may be associated with an extended population of intra-group GCs, possibly generated or enriched by the recent merging activity that characterizes this group. For the profile of NGC\,3641 (right panel) at the galactocentric distance of NGC\,3630 and NGC\,3643, we notice two relatively small peaks in the distribution over the plateau over-density, corresponding to the galactocentric radii of the GC systems in these two galaxies (also visible in the 2D maps Fig. \ref{fig:2d_maps}).

We also inspected the radial trends of the red and blue GC sub-populations in both the $ugri$ and $gri$ catalog for NGC\,3641. We identified the blue population as all the sources with $(g-i)\leq0.95$ and the red ones with $(g-i)>0.95$ (see Sect. \ref{sec:color_dist}). In both cases galaxies the red population is more concentrated than the blue one and contributes less to the region where the intra-group population is located.
 
All these results are also confirmed from the analysis of the GC candidates using the $ugri$ catalog. Additionally, these results are also confirmed when adopting a different criterion for radial binning. As a test, we made radial bins chosen to have an equal number of GC candidates (thus the same statistical error) this leads to the same results illustrated above.

\subsection{Color distribution}
\label{sec:color_dist}

As discussed previously, studying the colors of GCs provides valuable information that enables us to trace mainly the metallicity distribution of the GC population, thereby revealing insights into the formation history of the host galaxy.

We examine the color distribution of the GC candidates associated with NGC\,3640 and NGC\,3641 using both the $ugri$ and $gri$ GCs candidates catalogs. We take the first catalog as reference in this section, as the analysis of colors benefits from the broader wavelength coverage (and lower contamination) of the $ugri$ sample, despite its shallower depth due to the brighter magnitude cut of the $u$-band data.

As discussed in Sect. \ref{sec:radial}, we initially associate the GC candidates based on their galactocentric distance from the galaxies. Here, we further narrowed down the candidates analysed by imposing that their distances is: $R\leq R_t$ (see Sect. \ref{sec:radial_ind} and Table \ref{tab:back_value}).

\begin{figure*}[h!]
    \centering

    \includegraphics[width=\columnwidth]{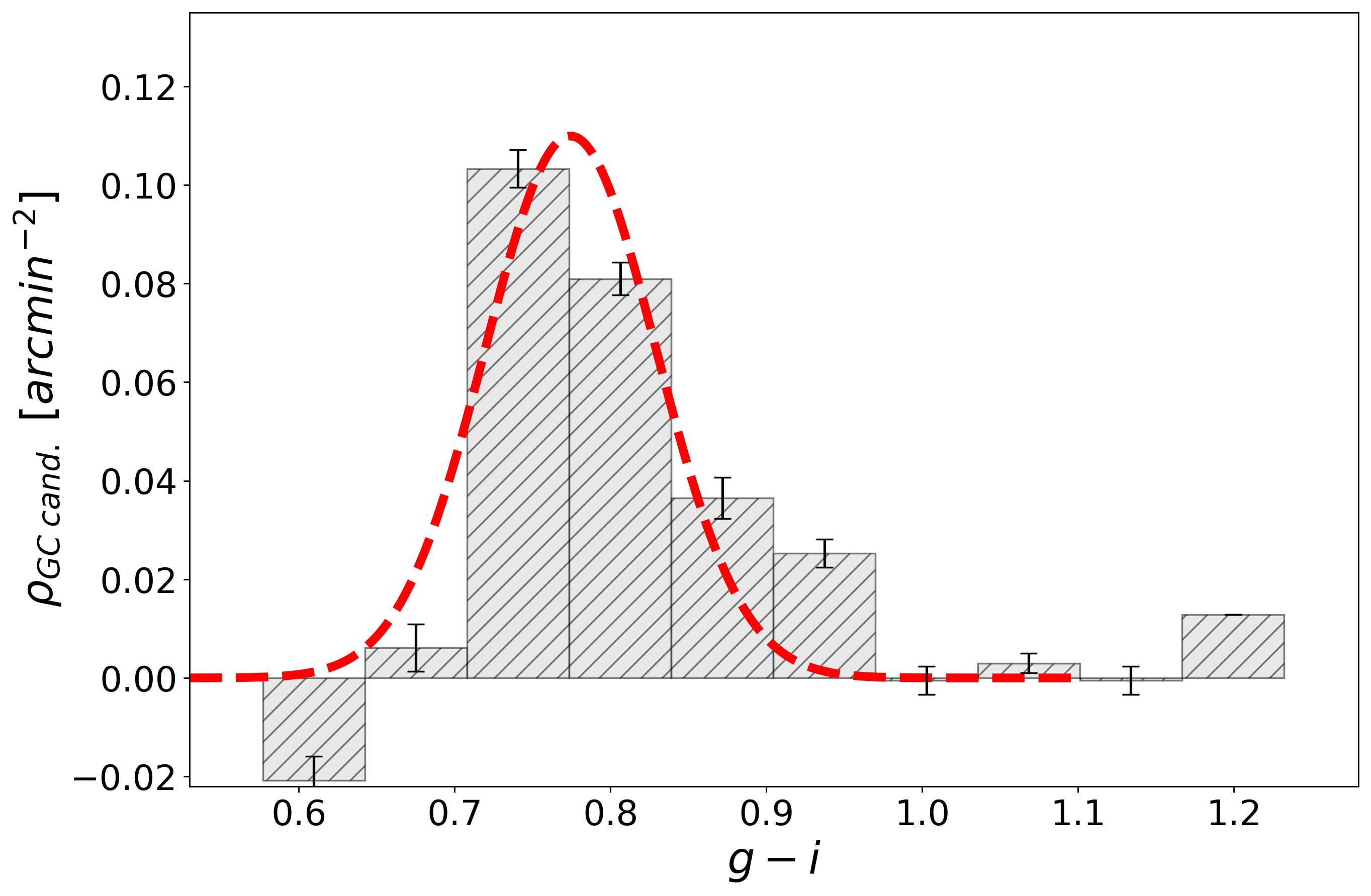}
    \includegraphics[width=\columnwidth]{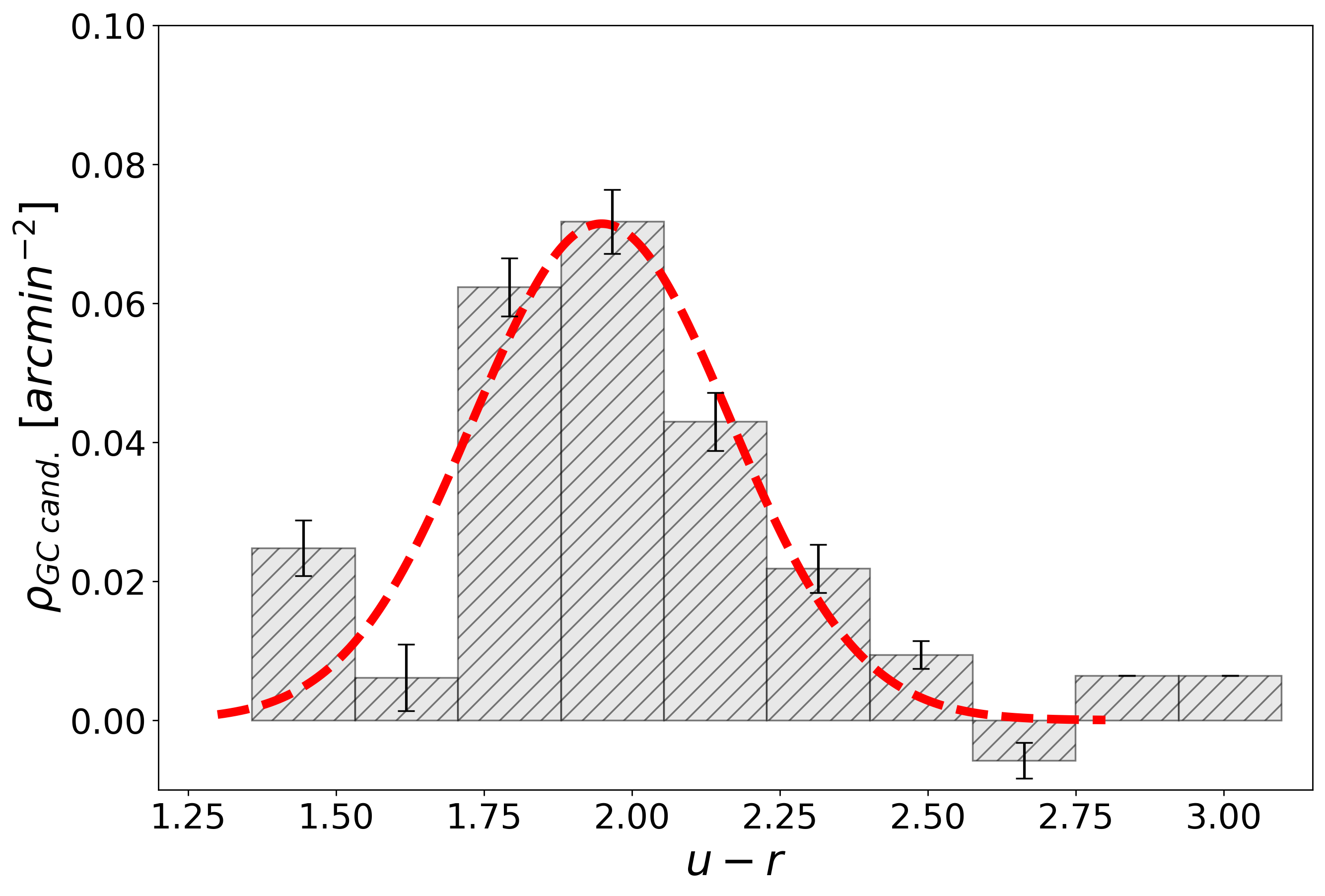}
    \caption{Background-subtracted $g-i$ (left) and $u-r$ (right) color distributions for the GCs population associated to NGC\,3640. The red dashed line is the fitted Gaussian color distribution.}
    \label{fig:color_3640}
\end{figure*}

\begin{figure*}[h!]
    \centering

    \includegraphics[width=\columnwidth]{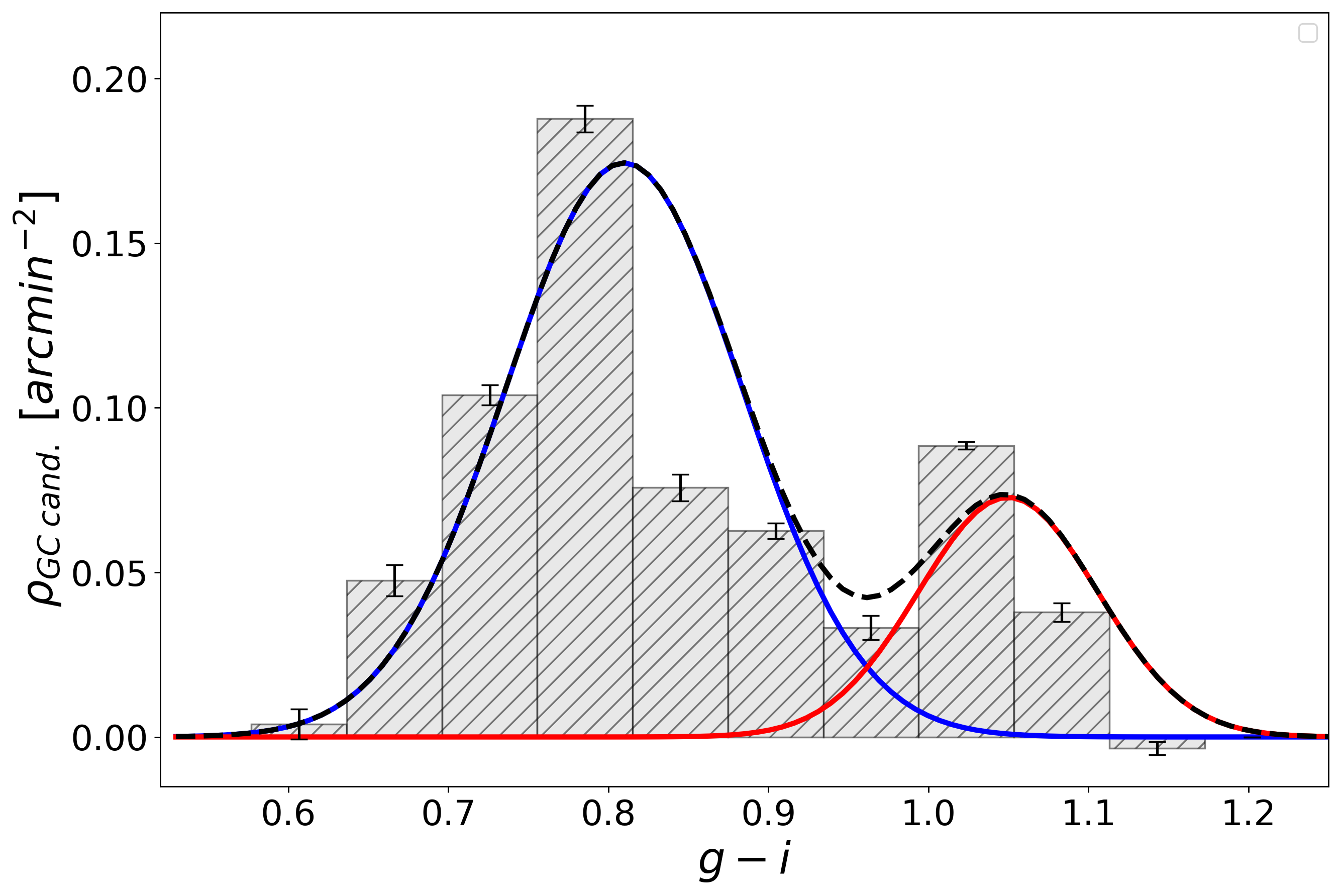}
    \includegraphics[width=\columnwidth]{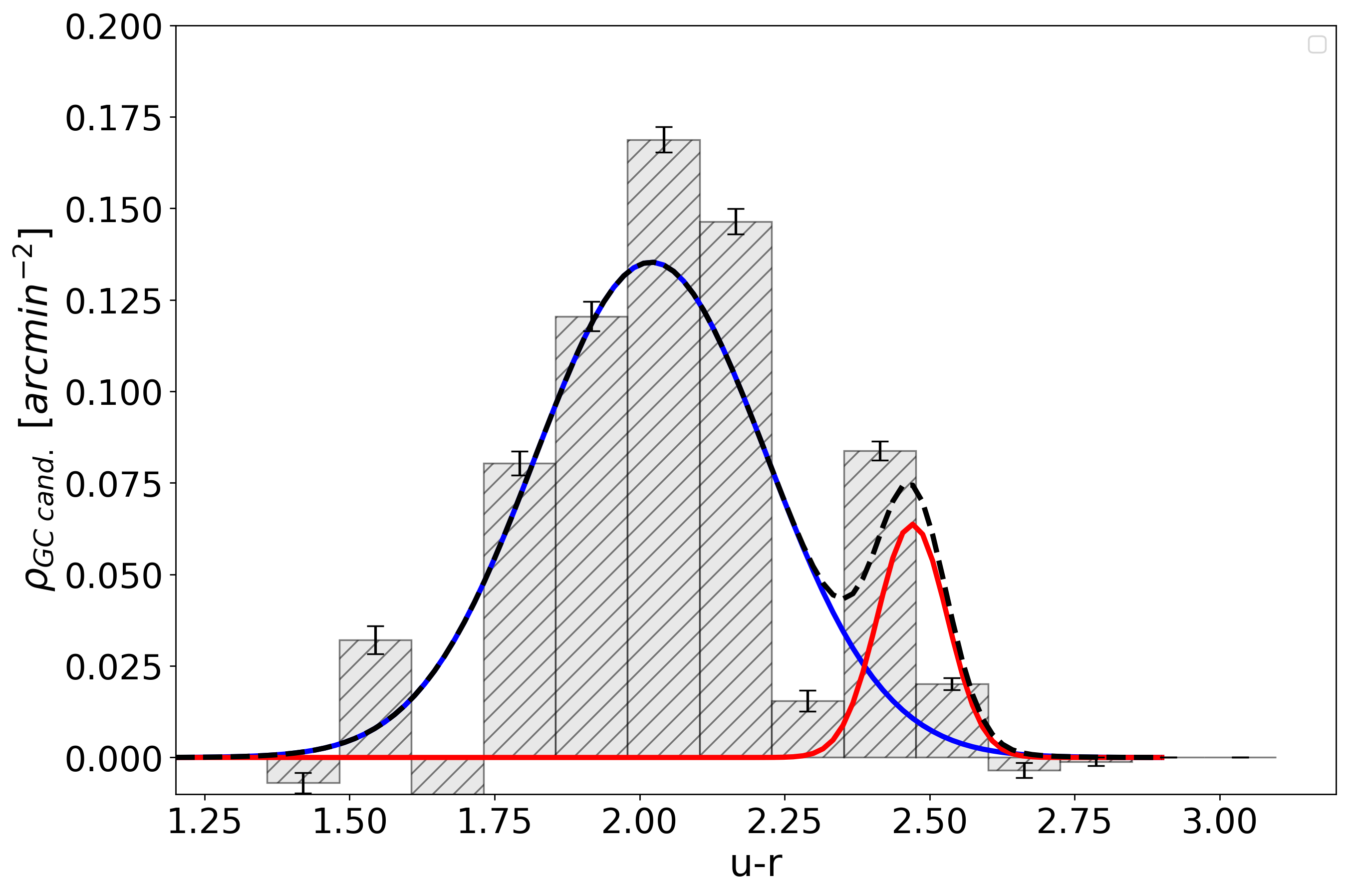}
    \caption{ Background-subtracted $g-i$ (left) and $u-r$ (right) color distributions for the GCs population associated to NGC\,3641. The blue and red solid lines show the fitted Gaussians associated to the two sub-populations, while the black dashed line shows the sum of the two.}
    \label{fig:color_3641}
\end{figure*}

To correct for background contamination, we  used the background region defined in Sect. \ref{sec:radial}, to characterize the color distribution of the background sources. 
Figs. \ref{fig:color_3640}-\ref{fig:color_3641} show the color histograms for $(g-i)$ and $(u-r)$. The gray-filled bars represent the corrected color distribution (total minus background distribution) for the sample of GC candidates in NGC\,3640 and NGC\,3641, respectively. The error bars are computed assuming Poisson errors in the number of contaminating sources in each color bin.

\begin{table*}[h!]
\caption{Color distribution fit parameters for the  GC population of NGC\,3640 and NGC\,3641 using the $ugri$ and $gri$ catalogs, along with the blue/red fraction (f), the kurtosis (k), and the separation value (D).}
    \small
    \centering
    \begin{tabular}{ccccccccccc}
    \hline
    \hline
     Galaxy & Color & $\langle \mu_1 \rangle$&  $\langle \sigma_1\rangle$  & $\langle $$\mu_2\rangle$& $\langle\sigma_2\rangle$ & $\langle f_b(\%)\rangle$ & $\langle f_r(\%)\rangle$ & $\langle\kappa\rangle$& $\langle D\rangle$ & Catalog \\
      & & (mag)&(mag)&(mag)&(mag)& &&&&\\
      \hline

     \multirow{2}{*}{NGC\,3640}  &  $g-i$&0.77$\pm$0.01 &0.05$\pm$ 0.01&...&...&...&...&...&...& $ugri$\\
         &$u-r$&1.95 $\pm$ 0.03 &0.22$\pm$0.03&...&...&...&...&...&...&$ugri$\\
         \hline
         
     \multirow{2}{*}{NGC\,3641} &   $g-i$& 0.780$\pm$ 0.004 & 0.074$\pm$ 0.002&1.02$\pm$0.01& 0.056$\pm$ 0.005 &74 &26&-0.75$\pm$0.04&3.5$\pm$0.1 & $ugri$\\
        & $u-r$&   2.03$\pm$0.01 & 0.20$\pm$ 0.01& 2.47$\pm$0.01&0.06$\pm$ 0.01&91&9&  -0.13$\pm$0.09 &2.9$\pm$0.3 & $ugri$ \\
         \hline
    
   NGC\,3640  &  $g-i$&0.82$\pm$0.03 &0.22$\pm$ 0.04&...&...&...&...&...&...&$gri$\\
     
   NGC\,3641     &$g-i$& 0.82$\pm$ 0.01 & 0.089$\pm$ 0.052&1.03$\pm$0.03& 0.11$\pm$ 0.01 &60           
                                   &40&-0.39$\pm$0.07&2.1$\pm$0.3 & $gri$\\
         
         \hline
         \hline
    \end{tabular}
    
    \label{tab:gmm_all}
\end{table*}

To study the detailed properties of the color distribution (e.g the presence of bimodality), the probability distribution functions obtained from the background-subtracted color distributions were randomly populated with $\sim$ 300 points. The probability function was derived by subtracting the normalized color distribution of sources within the background region (see Sect. \ref{sec:radial}) from the corresponding profile of on-galaxy sources, which was obtained using all GC candidates within $R_t$. The negative density bins ($\rho<0$) were set to zero. Then, the resulting distribution was inspected using Gaussian Mixture Models \citep[GMM,][]{muratov10}, which uses a likelihood-ratio test to compare the goodness of fit for double (or multiple) Gaussian versus a single Gaussian. GMM outputs the best-fit Gaussian parameters:  mean ($\mu$), standard deviation ($\sigma$) and the fraction of sources attributed to each Gaussian. 
Additionally, it also provides statistical tests to assess the confidence level of the multi-modal fit. These tests include the kurtosis, which measures the skewness of the distribution, and $D$, a parameter that is based on the separation of the means
relative to their widths expected to be $>2$ for bimodal distributions. A unimodal distribution would exhibit a strong peak, resulting in positive kurtosis value and $D<2$. For NGC\,3640 GC population, GMM confirms the preference for a unimodal distribution; whereas for NGC 3641, a color bimodality is preferred in both colors.

The numerical experiment of repopulating the distribution functions was repeated ten times. Table \ref{tab:gmm_all} reports the median and $rms_{MAD}$ of the parameters from the Gaussian best fit to the distributions for NGC\,3640 and NGC\,3641. Figures \ref{fig:color_3640}, \ref{fig:color_3641} show the background-subtracted distribution with the fitted Gaussians. 

In order to verify if our results are independent from the adopted choice we conducted several tests, including: $i)$ varying the distance up to which we considered the GC candidates belonging to the galaxies\footnote{Among the several distances we tested, we also inspect the GC color distribution adopting the sources with $R_{gal}\leq3.9R_e$ (i.e. all the sources within the green circle in Fig. \ref{fig:sepration}). Even with only a few tens of objects we confirm the observed behavior.}; $ii)$  adopting the GC candidates associated with the plateau as background region. All the tests yielded results consistent with our reference GC selection. The GC population on NGC\,3640 exhibited a unimodal color distribution, while NGC\,3641 consistently showed a bimodal distribution. However, the GMM runs provide statistically different fractions 
for the red population, $\sim$ 26\% ($\sim$9\%) of the total background-corrected GC population for $g-i$ ($u-r$) color. The two colors have different sensitivities to metallicity, with $(u-r)$ being the more sensitive of the two, and this difference might be an expression of this effect. To further inspect this behavior, we made the same analysis for the $(g{-}i)$ color distribution, using the more contaminated $gri$ matched catalog. The analysis with the $gri$ confirms that NGC\,3640 has a preference for a unimodal skewed blue distribution, while NGC\,3641 confirms its bimodality with a fraction of red GCs of the order of 40\%. Table \ref{tab:gmm_all} reports the GMM parameters for the $(g-i)$ color for NGC\,3641.
With the aim of inspecting the mean positions of red/blue GC-candidates around NGC\,3641, we divided the GC sample into two sub-populations: a blue one with $(g-i)\leq0.95$ mag, and a red one with  $(g-i)>0.95$ mag. The adopted separation value is determined as the point where the two color distributions intersect shown in Fig. \ref{fig:color_3641}. The mean galactocentric distance for red GCs is $\sim$1 $arcmin$, while for the blue GCs we find $\sim$2 $arcmin$. The higher central concentration of the red population relative to the blue one is a feature already observed in massive galaxies \citep{cantiello18fds,Hazra2022}.  We obtain consistent results even if we assume the blue/red cut using $(u{-}r)\approx2.25$ mag, or if the $gri$ catalog is used. Actually, using the $(u{-}r)$ color cut, we observe a red sub-population slightly more concentrated ($\sim$0.6 $arcmin$) toward the central region than the one observed using the $(g-i)$ cut.

\begin{table}[ht]
\caption{Adopted selection parameters and corresponding ranges for identifying GC candidates in the $g$- and $r$-band to study the GCLF.}
\centering

\begin{tabular}{@{}ccccccp{\linewidth}@{}}
\toprule 
Parameters& \multicolumn{2}{c}{$g-band$} & \multicolumn{2}{c}{$r-band$} \\ 
\cmidrule(lr){2-3} \cmidrule(lr){4-5} 
 & min & max& min& max \\ \midrule
FWHM (")&0.84 &2.34 & 0.50 & 2.0\\
Flux Radius (")& ...& 1.21&... &1.0\\
Elongation& ... & 2.0 &...&2.0\\ 
Magnitude & 21 &...&20.4&...\\

\bottomrule 
\end{tabular}

\label{tab:sel_gclf}
\end{table}

\subsection{GC Luminosity function and specific frequency}
\label{sec:gclf}

In this section, we focus on the analysis of the GCLF for both NGC\,3640 and NGC\,3641 to estimate the total number of GCs ($N_{GC}$) and constrain the relative distance between the two galaxies by comparing their GCLFs. We adopted a single-band data, the $g$- and $r$-bands, discarding the  $u$- and $i$-bands. The $g$ and $r$ catalogs are derived independently from the respective images as described in Sect. \ref{sec:photometry}. The $u$-band data are too shallow, failing to reach the turnover magnitude of the GCLF (see Fig. \ref{fig:comple_on_gal}). The $i$-band, instead, is characterized by a high fraction of spurious detections close to the galaxy core, making them difficult to reject based solely on a single-band analysis. 
The decision to adopt a single-band approach arises from the lack of a completeness function estimated for the multi-band matched catalog, which requires a procedure that accounts for the color of the sources\footnote{\label{gclg_footnote}To determine the completeness fraction of a GC-candidates catalog derived from multi-band cross-matches, we would need to simulate source injection using the expected GCLF, color distribution, and galaxy background contamination in all used bands. The function itself would then be derived from a selection based on the same criteria adopted for identifying GC-candidates. However, such a procedure would be excessive for the present work. Moreover, the expected outcome would, in most cases, be dominated by the worst passband in the matched sample. Ultimately, this approach would not benefit from the advantage of the large format of VST data.}.

\begin{figure*}[h!]

    \includegraphics[width=\columnwidth]{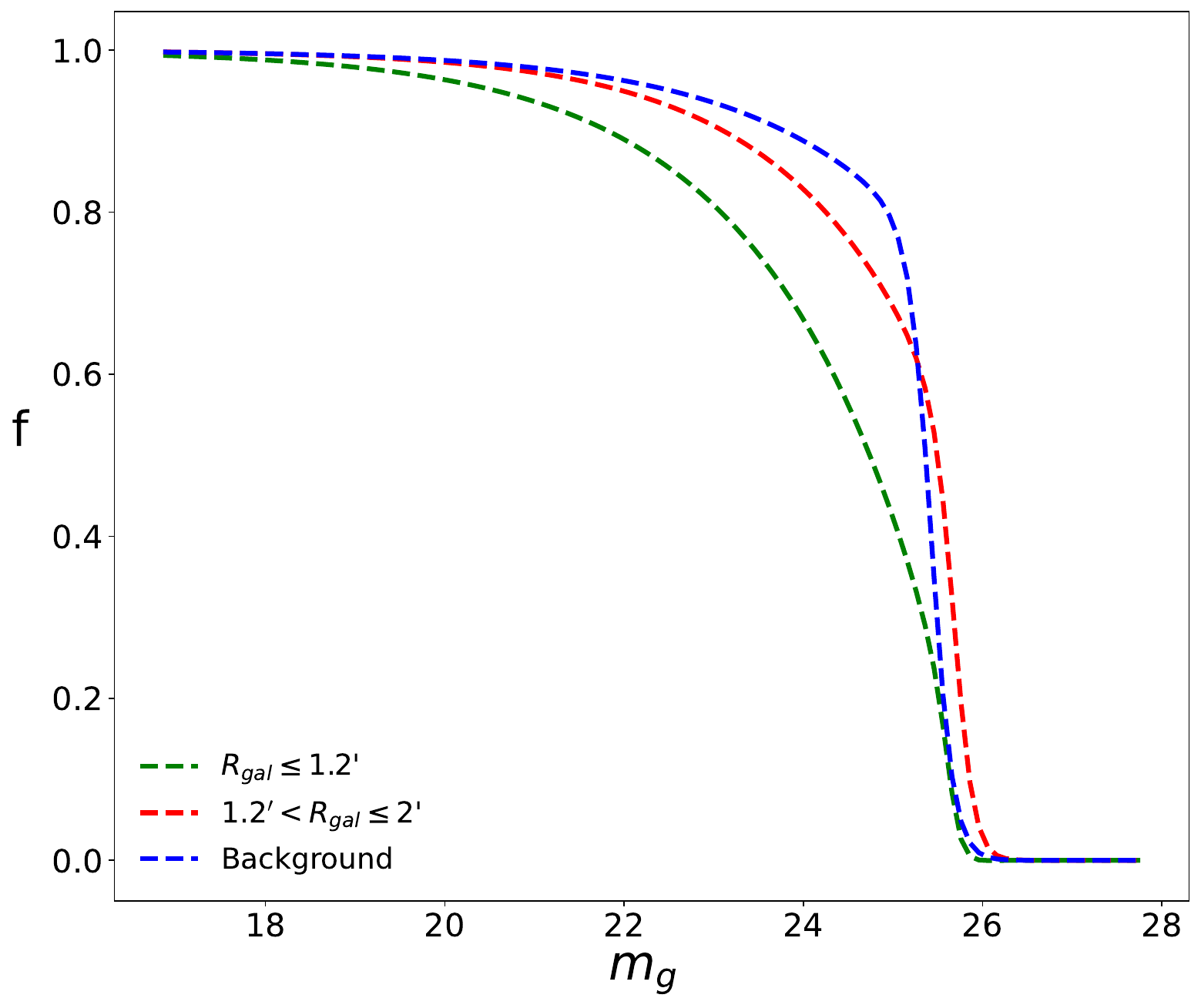 }
    \includegraphics[width=\columnwidth]{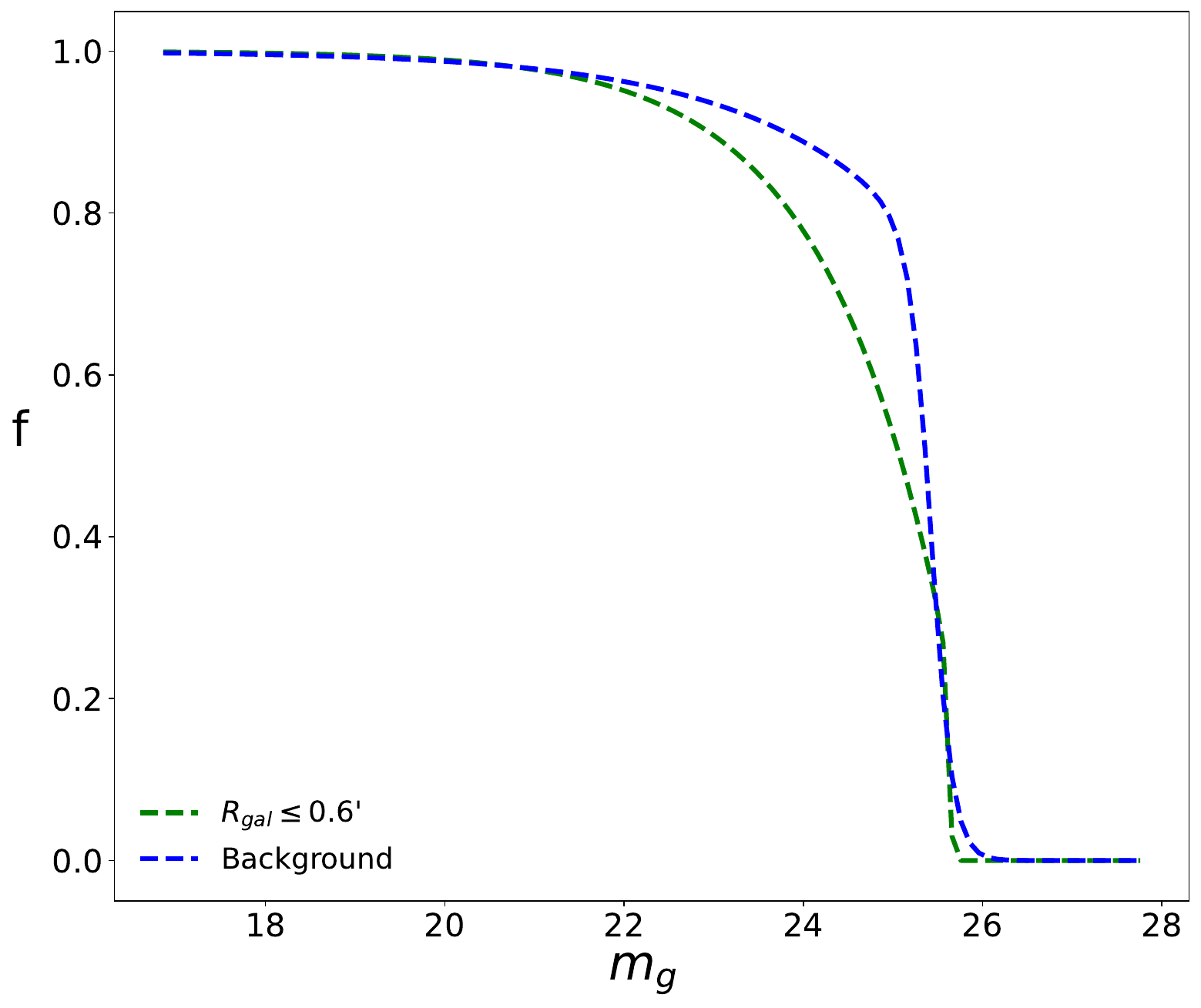}

    \caption{Radial completeness fraction in the $g$-band for both the on-galaxy and off-galaxy background fields. The fraction was determined by following the procedure described in Sect. \ref{sec:completeness} and also applying the selection criteria described in Sect. \ref{sec:gclf}. $Left\ panel:$ Radial completeness over NGC\,3640. The green dashed line is the completeness function obtained within 1.2 $arcmin$ from the galaxy center. The red dashed line is the completeness function between 1.2 and 2 $arcmin$ from the galaxy center. The blue dashed line represents the mean completeness of the four background regions inspected along the annulus with an inner radius of 25 $arcmin$ and an outer radius of 30 $arcmin$.  $Right\ panel:$ Radial completeness function over NGC\,3641. The green dashed line is the completeness function within 0.6 $arcmin$ from the center of the galaxy.}
    \label{fig:comple_gclf_}

\end{figure*}

\begin{figure*}[h!]

    \includegraphics[width=\columnwidth]{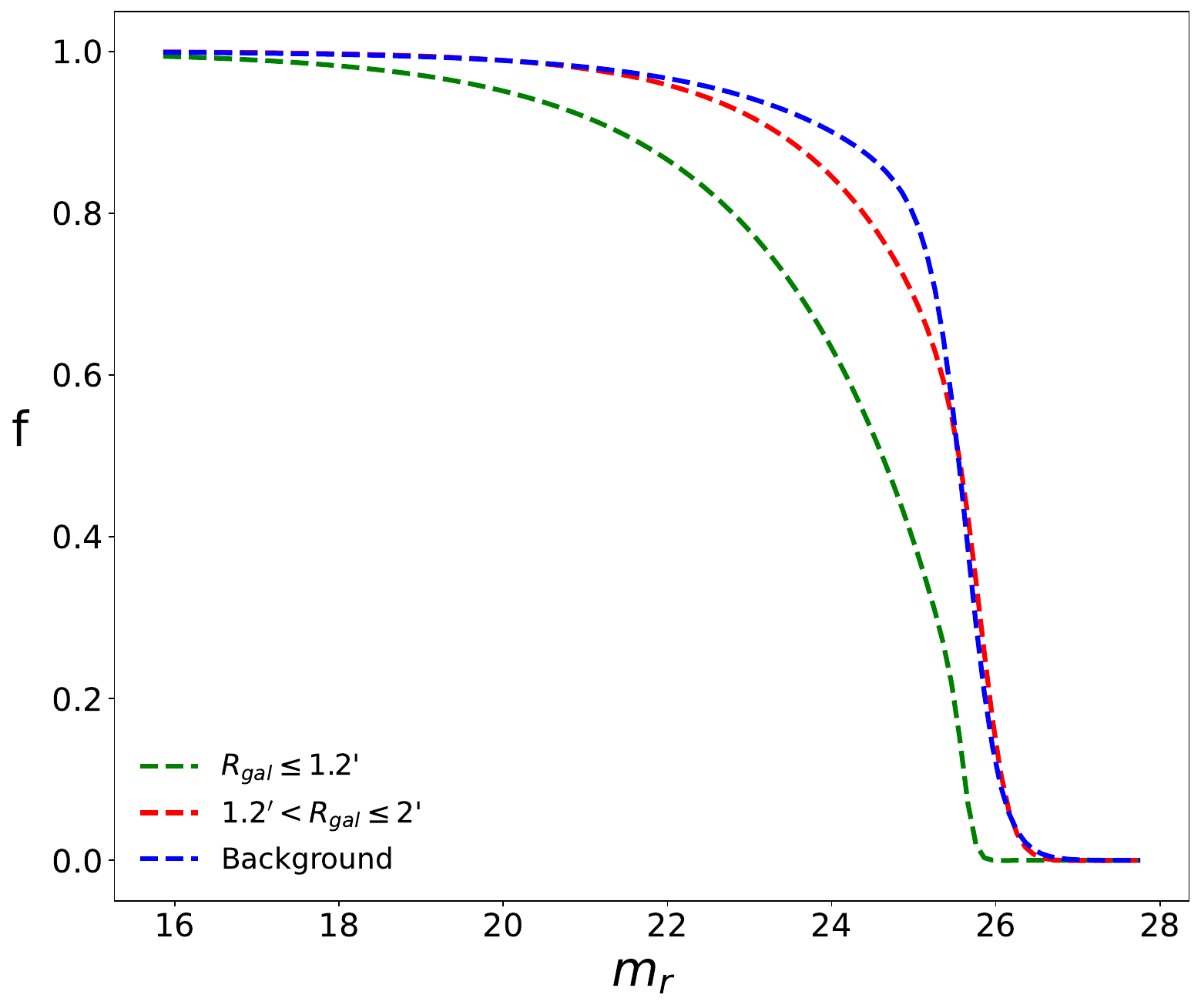 }
    \includegraphics[width=\columnwidth]{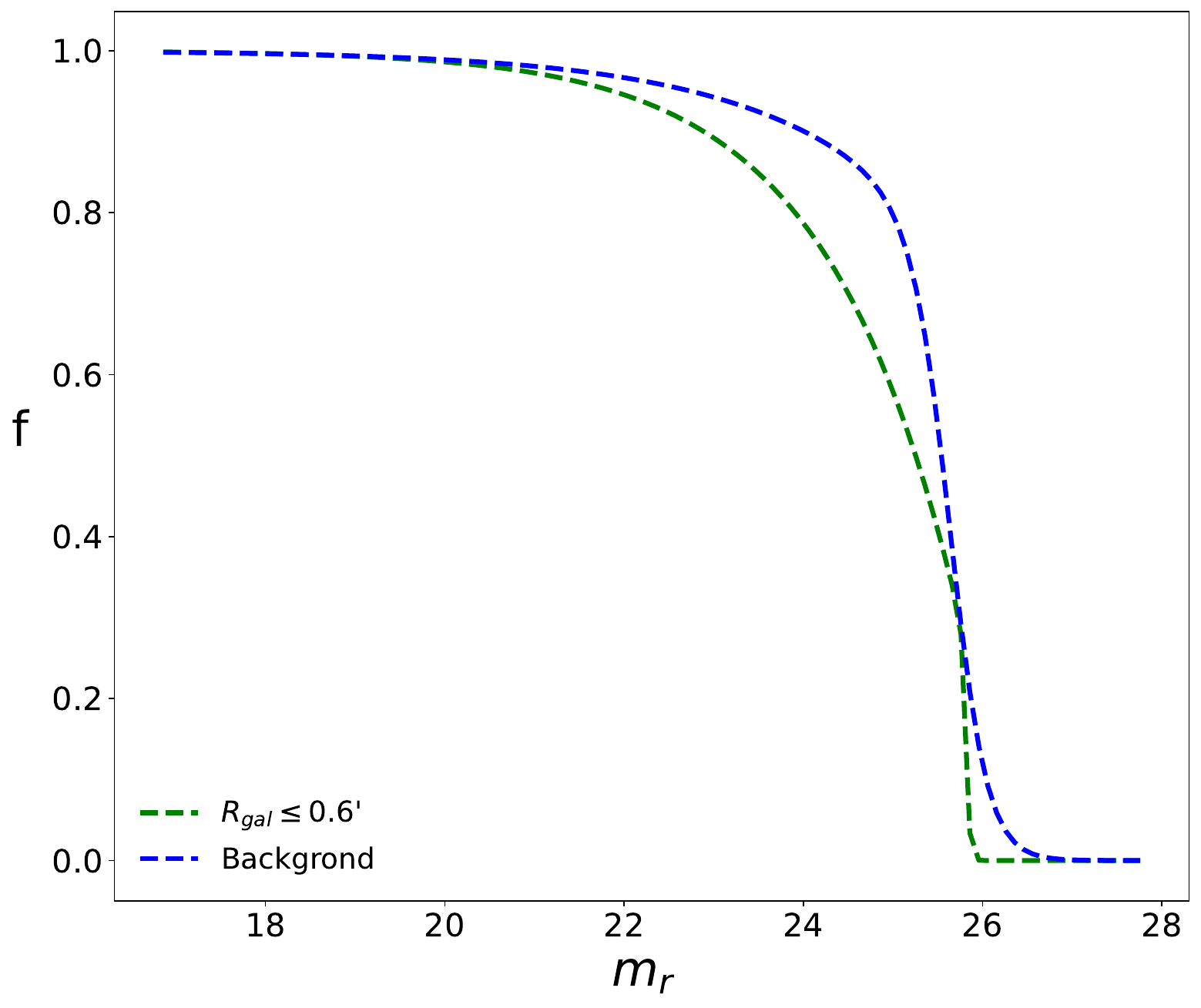}

    \caption{Same as in Fig. \ref{fig:comple_gclf_} but for the $r$-band.}
    \label{fig:comple_gclf_r}

\end{figure*}

\begin{figure*}[h!]
    \sidecaption
    \includegraphics[width=\columnwidth]{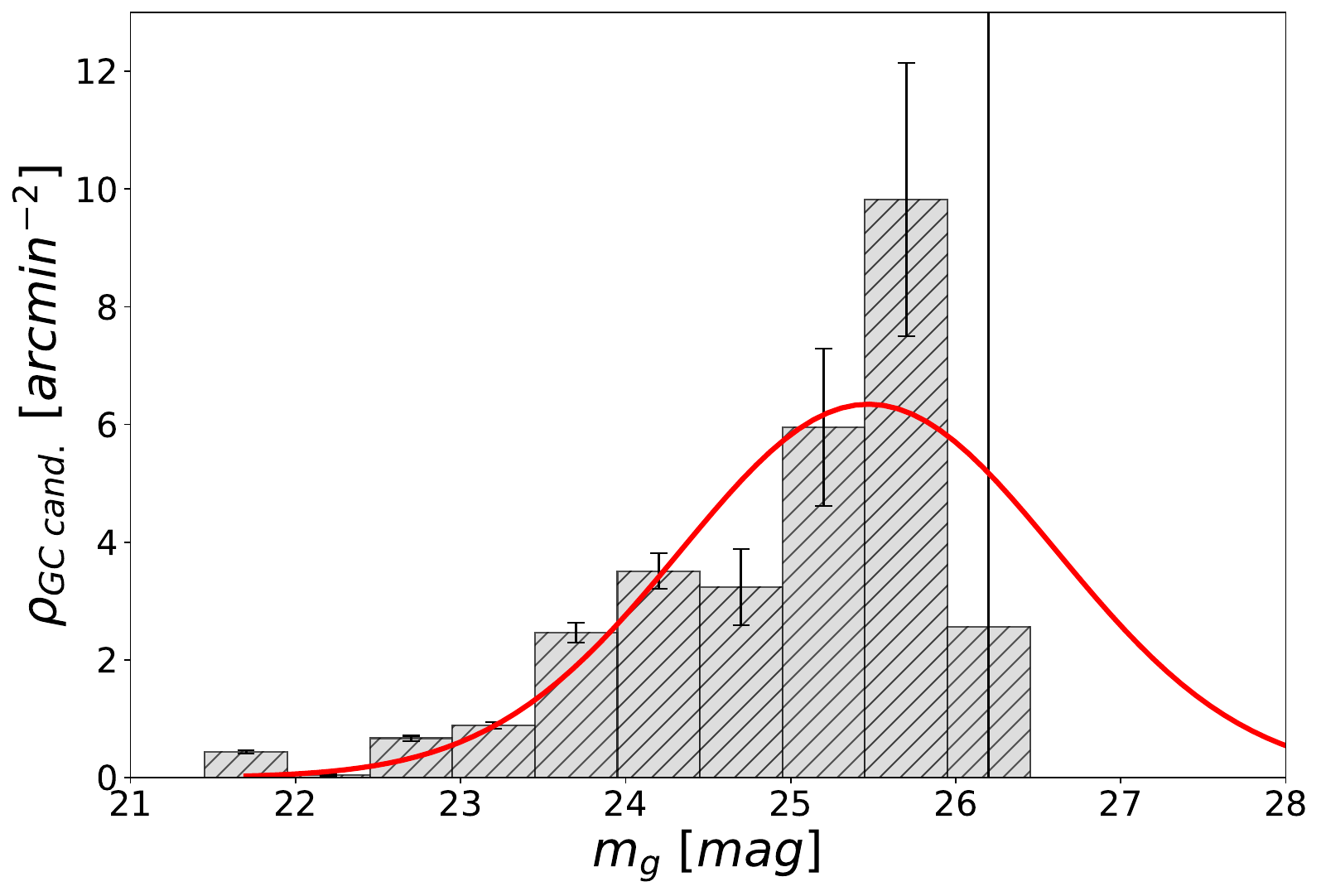}
    \includegraphics[width=\columnwidth]{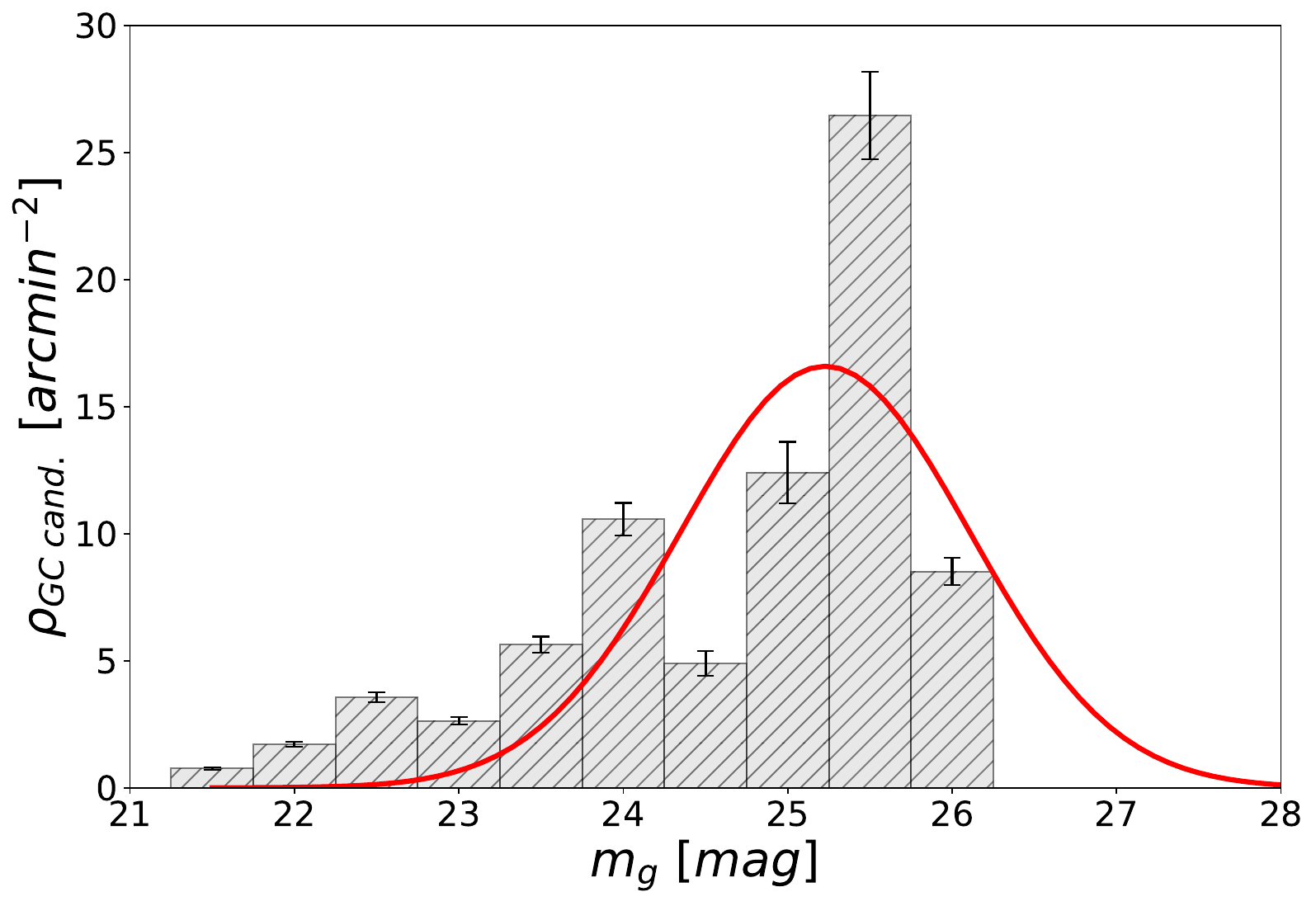}
    \caption{Globular cluster luminosity function for NGC\,3640 (left panel) and NGC\,3641 (right panel) in the $g$-band. The gray bars represent the completeness-background corrected GCLF, while the red solid line depicts the fitted Gaussian curve. }
    \label{fig:gclf_g}
\end{figure*}

\begin{figure*}[h!]
    \centering

    \includegraphics[width=\columnwidth]{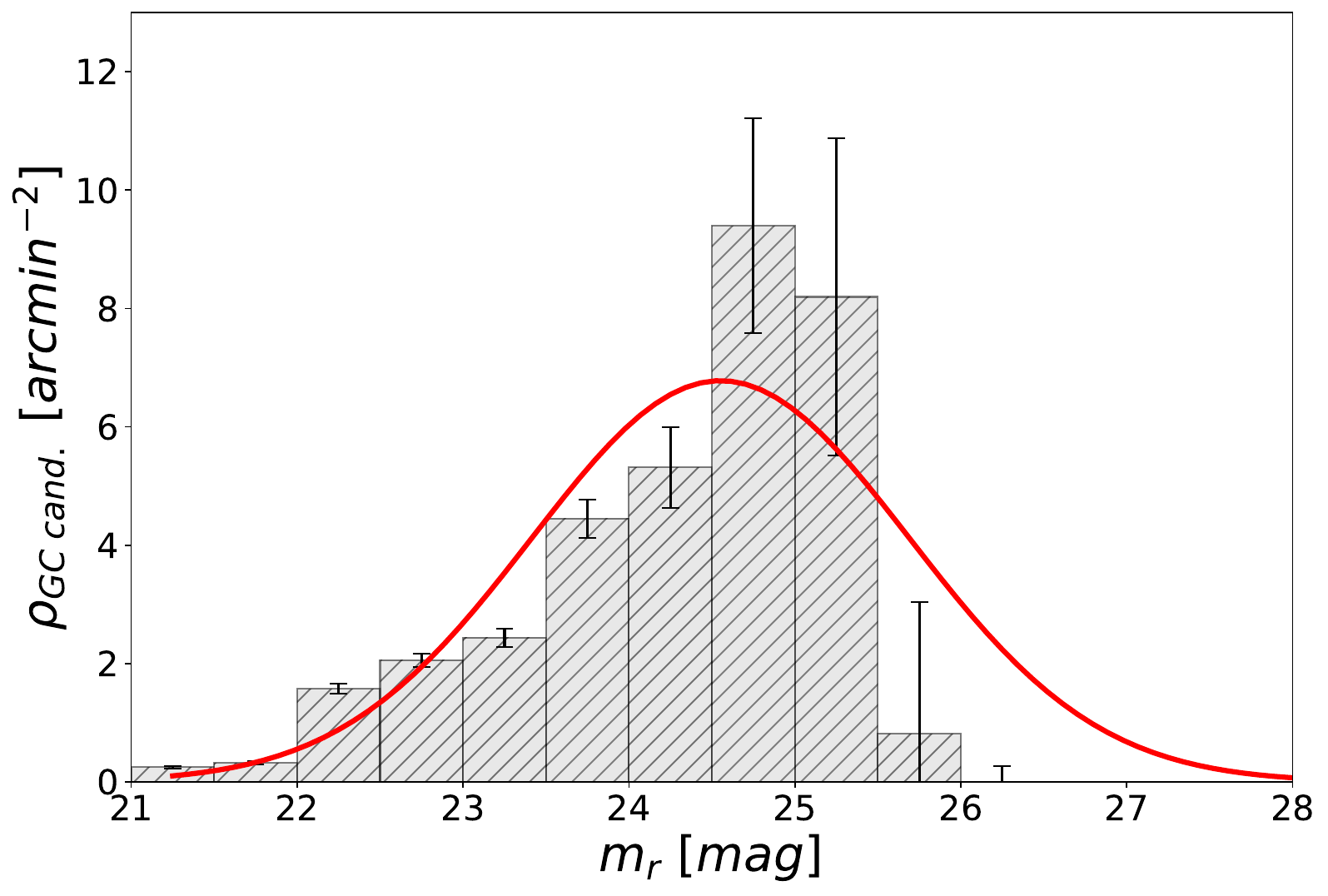}
        \includegraphics[width=\columnwidth]{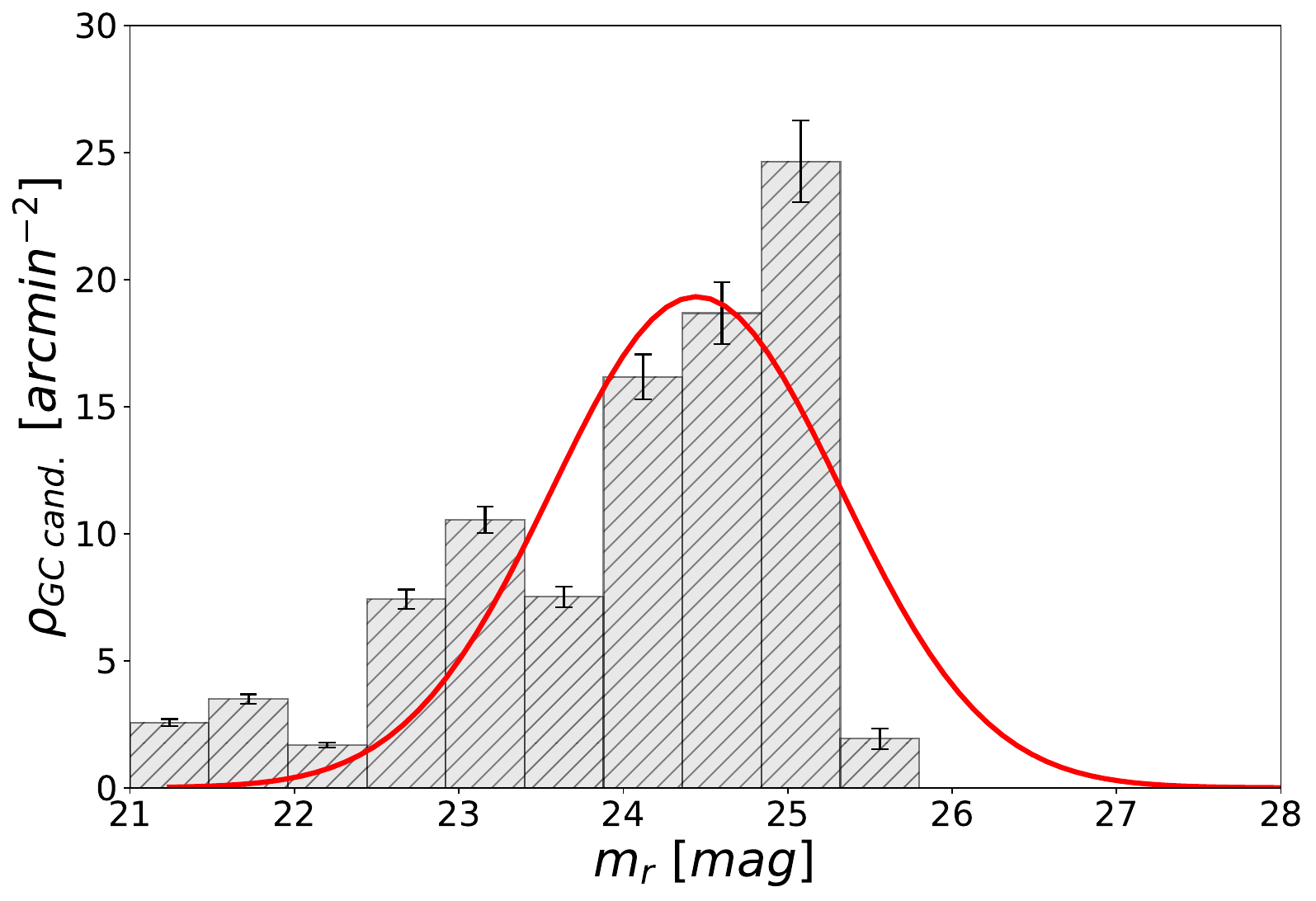}

    \caption{Same as in Fig. \ref{fig:gclf_g} but in $r$-band.}
    \label{fig:gclf_r}
\end{figure*}

Single-band catalogs exhibit higher contamination but also greater completeness compared to the multi-band ones. As also discussed in Section \ref{sec:completeness}, we take advantage of the wide sky coverage of the VST imager to efficiently characterize the population of fore/back-ground interlopers. This allows us to clean the contaminants from the luminosity function by subtracting the density function derived in background areas from the on-galaxy luminosity function, thereby revealing the luminosity function of GCs.
For the $r$-band, which is our reference image for morphometry  (FWHM, CI, Flux radius, elongation), we adopted the same magnitude, morphometric and CI selection strategy described in Sect. \ref{sec:morpho_sel}. 
For the $g$-band data, where selection parameters were not required except for colors in the multi-band selection, we obtained selection parameters independently from the $r$ data.  We applied the CI cuts using the same basic procedure outlined in in Sect. \ref{sec:morpho_sel}.

Additionally, for both bands, we applied a lower threshold to the FWHM, set at the median FWHM value minus 3$\times\sigma_{FWHM}$, to exclude sources with a FWHM much smaller than the value expected for point-like sources \footnote{The additional cut was not necessary for the $ugri$ and $gri$ catalogs because, thanks to the matching procedure, we were able to discard these detected sources --likely fake detections-- from our catalogs.}. The final selection parameters are given in Table \ref{tab:sel_gclf}.

\begin{table}[htbp]
\caption{Position of the peak of the GCLF for NGC\,3640 and NGC\,3641, the peak separation ($\Delta\mu$), and the total number of GCs obtained by accounting for the missing faint-end of the GCLF and spatial incompleteness.}
\small

\begin{tabularx}{\columnwidth}{@{}c *{4}{>{\centering\arraybackslash}X} c@{}}
\toprule
\toprule
Passband & \multicolumn{2}{c}{NGC3640} & \multicolumn{2}{c}{NGC3641} & $\Delta\mu$ \\ 
\cmidrule(lr){2-3} \cmidrule(lr){4-5} 
 & $\mu$ & $N_{GC}$ & $\mu$ & $N_{GC}$ \\ 
\midrule

$g$ & 25.2$\pm$0.2 & 906$\pm$272 & 25.0$\pm$0.1 & 168$\pm$50 & 0.2$\pm$0.2 \\
$r$ & 24.5$\pm$0.2 & 1048$\pm$314 & 24.46$\pm$0.05 & 200$\pm$60 & 0.0$\pm$0.2 \\
\bottomrule
\bottomrule
\end{tabularx}

\label{tab:gclf}
\end{table}

To characterize the GCLF, all sources within 3.9 $R_e$ from the galaxy center were included in the analysis (see Sect. \ref{sec:radial} and Fig. \ref{fig:sepration}). For the background region, we adopted the same area as described in Sect. \ref{sec:radial}, specifically, group-centric radii between 25 and 30 $arcmin$ (measured from the tangency point of the two circles in Fig. \ref{fig:sepration}). Then, for the on-galaxy and off-galaxy regions, we derived the LFs.
To ensure a reliable GCLF, completeness corrections for undetected sources are mandatory, for both the LF on-galaxy and the background LFs.

We calculate the radial completeness fraction within the adopted region, using the same strategy outlined in Sect. \ref{sec:completeness}. For NGC\,3640, we retrieved the completeness function in two regions: one within 1.2 $arcmin$ ($\sim2R_e$) and another one between 1.2 and 2 $arcmin$ ($\sim3.9R_e$) from the center of the galaxy. For NGC\,3641, we only retrieved the completeness function within the 3.9$R_e$, as analyzing a smaller region would suffer from sampling effects. 
However, in both cases, we also take in to account the shape selection  criteria in deriving the completeness (see Table \ref{tab:sel_gclf}). For the background region, the completeness is obtained from the average functions estimated in 4 different regions across the adopted radial annulus\footnote{The completeness is indeed very stable across the background annulus, with an $rms$ of 0.01 (1\%) across the four different regions.}. Figure \ref{fig:comple_gclf_} and \ref{fig:comple_gclf_r} present the completeness functions derived in the $g$- and $r$-band, respectively. Once more highlighting the expected observation that completeness on-galaxy is lower than that of the background/off-galaxy.

The gray shaded bars in Figs. \ref{fig:gclf_g}-\ref{fig:gclf_r} show the completeness corrected and background subtracted LF for both the galaxies in the $g$- and $r$-bands, respectively. The uncertainty on each magnitude bin was calculated assuming Poisson errors on the number of contaminating objects, and a conservative 5\% error on the completeness corrections.

Each distribution was then fitted with a Gaussian function. Initially, we allowed the fit to estimate the dispersion of the distribution, and obtained values consistent within the errors with the expected $\sigma^{GCLF}$ (see Table \ref{tab:properties}). However, given the very large uncertainties on $\sigma_{GCLF}$ we chose to fix the value of the GCLF width to the empirical expectation \citep{villegas10}. We repeated this procedure $\sim$10 times, varying the binning width and the phasing of the histogram. The final adopted values for the peaks and uncertainties are the mean and the $rms$ of the obtained fitting parameters.

The positions of the peaks of the Gaussian functions shown in Figs. \ref{fig:gclf_g}-\ref{fig:gclf_r} (red lines) are reported in Table \ref{tab:gclf}. 
In this table we also report the $\Delta\mu$ between the GCLF peaks. Taking advantage of the GCLF as a distance indicator, we notice that in both bands the two galaxies are consistent with being basically at the same distance. Possibly, the $g$-band estimates reveal NGC\,3640 being $\sim10\%$ farther than its fainter companion. However, given the higher quality of our $r$-band, we prefer the scenario in which the two galaxies are roughly at the same distance which is also the result from SBF distances of the two \citep{tonry01}.

Taking advantage of the fitted Gaussian parameters,  we estimated the total number of GCs, $N_{GC}$, as follows. First, we integrated the GCLF up to its peak and then double the GC number obtained. This provides the total number of GCs within 3.9$R_e$. 
Recent literature quotes a range of 3-5 $R_e$ as the half-number radius for the GC spatial distribution \citep{alamo21, forbes17, lim24}. Adopting these results and considering our area coverage, to estimate $N_{GC}$ we doubled once more the GC number from the GCLF integration. 

To estimate the uncertainty of $N_{GC}$, we derived alternative values for $N_{GC}$ by varying the selection parameters for GC identification and the half-number radius within reasonable intervals. As a result, we observed that $N_{GC}$ changes by $\sim$30\%, which we assume as uncertainty. 
The final $N_{GC}$ values are reported in Table \ref{tab:gclf}. Despite the fact that the $g$- and $r$-band $N_{GC}$ estimates are derived using independent catalogs, with non negligible difference in terms of image quality, we note that the total population of GCs in the two passbands is in good agreement. 

For NGC\,3640, the only value available in the literature is $N_{GC}\sim50$ from \citet{gebhardt99}. However, this value represents only a lower limit, as it was obtained by counting the selected sources over the HST area.

Adopting as $M_V$ the value reported in Table \ref{tab:properties}, we evaluated the specific frequency ($S_{\rm N}$) of the two galaxies, taking as reference the weighted average between the two $N_{GC}$ estimates. 
The specific frequency provides an estimate of the number of GCs per unit of $V$-band luminosity \citep{harris81,hamuy91,harris01} and is derived as:
\begin{equation}
    S_{\rm N}=N_{GC}\times10^{0.4 \times (M_V+15)}
\end{equation}

The $S_{\rm N}$ values we obtain are: $S_{\rm N}$ =2.0$\pm$0.6 for NGC\,3640, and $S_{\rm N}$ = 4.5$\pm$1.6 for NGC\,3641.

\begin{figure*}[h!]
    \centering
    \includegraphics[width=\textwidth]{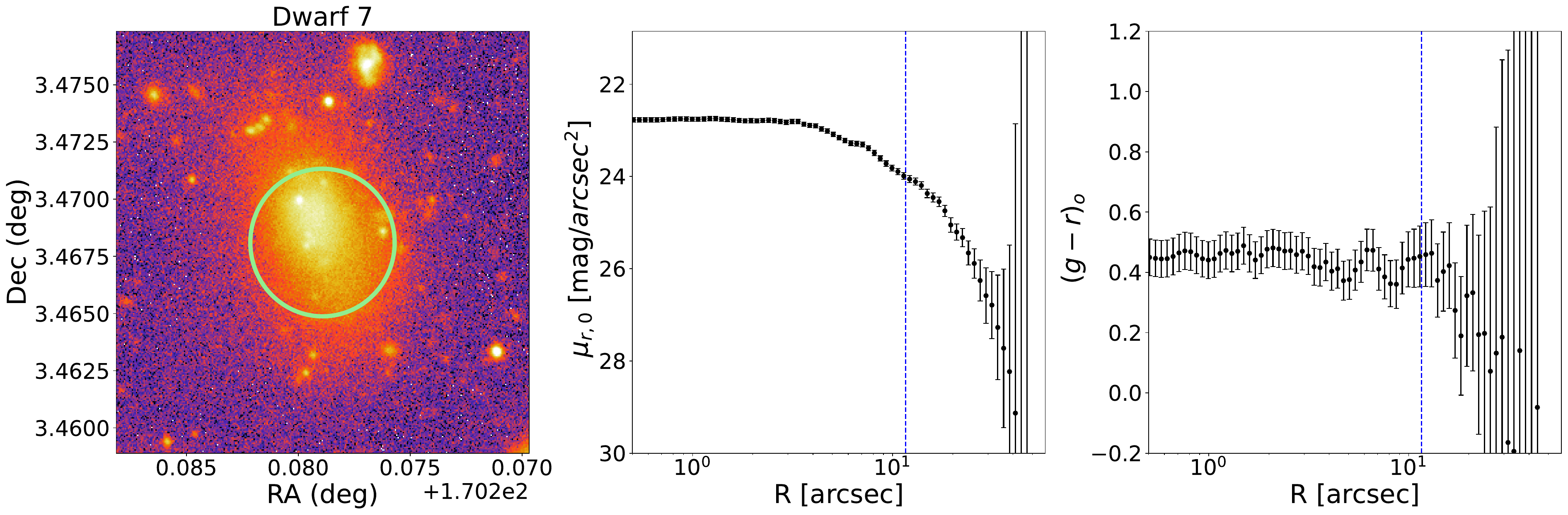}
    \includegraphics[width=\textwidth]{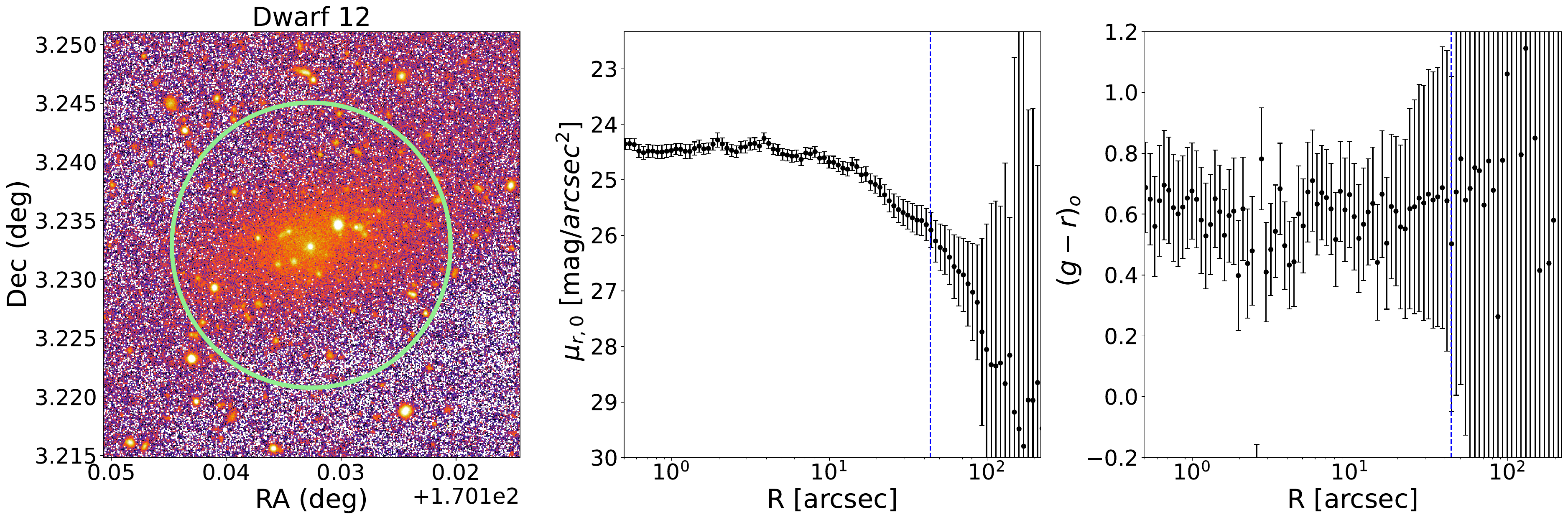}
    \caption{Cutout around the the identified Dwarf galaxies in the $r$-band (left panel), the surface brightness profile (middle panel) and the color profile for two dwarf candidates in our sample (right panel). Green circles and the blue dashed line represent $R_e$ in the $r-band$ reported in Table \ref{tab:lsb}.}
    \label{fig:lsb_ex}
\end{figure*}

\section{Dwarf galaxies in the field and their GCs content}
\label{sec:lsb}

We used the wide-field coverage of the VST data to investigate the GC population in dwarf galaxy candidates across the observed field. The investigation of the GC populations in dwarf galaxies is currently a strongly discussed topic with a number of studies focusing on the universality of the GCLF and their dark matter content \citep[e.g.][]{miller07,georgiev09,georgiev10, rejkuba12,vandoku16,vandoku19,shen21,mueller21, battaglia22}, because these low-mass galaxies provide some of the most critical constraints for cosmology \citep[e.g.][]{bullock17,sales22}.

As a first steps we visually inspected the $g$ and $r$ images independently, and then cross-matched the identified dwarf candidates from both fields (the visual inspection was independently made by the first and second authors of this work). We searched for objects that showed a diffuse brightness profile, with no evidence of the typical features found in bright galaxies (spiral arms, shells, bars), with a number of point-like GC candidates centered around the dwarf (although we did not find any GC-rich dwarf). Then, after inspecting the brightness profiles, we also rejected objects with bright central cores. This yielded to a final sample of 27 galaxies, all listed in Table \ref{tab:lsb}. Among the selected dwarfs, we identify as Dwarf\,8 (upper middle panel in Fig. \ref{fig:color_comp})  the edge-on dwarf spiral galaxy observed by \citet{Schweizer1992} to exhibit HI emission. Indeed, it is one of the bluest galaxies in the sample. Figure \ref{fig:color_comp} shows the position of the dwarfs identified across the observed field. 
 
We also checked the literature and found a sample of dwarfs identified in the MATLAS Survey \citep{duc15,habas20}. In \citet{Poulain21}, they characterized the properties of 19 dwarfs found over $\sim$ 1$\times$1 sq. degrees around NGC\,3640, 10 of which are in common with our list of 27 identified dwarfs. \citet{Marleau21} classified the Dwarf\,12 as an Ultra Diffuse Galaxy (UDG, with $\mu_{0,g}\geq$ 24mag/$arcsec^2$ and $R_e\geq1.5kpc$), a result confirmed by our analysis. \citet{Heesters23} analysed MUSE data of a sample of dwarf galaxies in the MATLAS sample, including three dwarfs identified around NGC\,3640. They confirmed the membership of these dwarfs to the galaxy group. Additionally, from the MATLAS website none of the dwarfs identified in the field around NGC\,3640 show evidence of GCs overdensity, with some exceptions discussed below. In Table \ref{tab:lsb}, we add a column to include information regarding the dwarfs we have in common with MATLAS. In conclusion, adding up our sample of 17 newly detected dwarfs with the total sample identified by the MATLAS survey, brings the total number of dwarf galaxies in the $\sim$1.5$\times$1.5 $deg^2$ area around NGC\,3640 to 26.

To analyze the properties of the galaxies in this work, we developed a code that integrates the AutoProf pipeline \citep{stone21}. In brief, AutoProf automatically fits elliptical isophotes to galaxy images and extracts accurate flux measurements along them. It also estimates the local background value for each object and masks out any sources (e.g., stars, background galaxies, GCs) that could affect surface photometry measurements. In our code, we enhanced the masking procedure by incorporating additional sources to mask --particularly focusing on the inner region of the galaxy-- and added a module to derive the Sersic parameters \citep{sersic68} to the isophotal profiles produced by AutoProf. The procedure was run on cutout images with size of $\sim9\times9\ arcmin^2$ around each of our candidate dwarfs, both in the $g$- and $r$-bands, adopting as reference image the $g$, and then running forced photometry on the $r$-band frame. For each dwarf candidate we derived both structural (effective radius $R_e$, and Sersic index $n$) and photometric (e.g., effective surface brightness $\mu_{eff}$) properties. The resulting parameters are reported in Table \ref{tab:lsb}, together with their uncertainties, which we decided to conservatively round-off to 0.1 on all the fitted parameters. Due to their extremely faint surface brightness, structural and photometric properties could not be retrieved in four of the 27 identified dwarfs. 
Along with these properties, we also estimated the total integrated magnitude and the $(g-r)$ color within one effective radius, adopting the mean  $R_e$ of the $g$- and $r$-bands as radius. All magnitudes and colors are corrected for Galactic extinction using values from \citet{sf11}. Then, we inspected the color-magnitude diagram and compare the properties of the dwarfs selected in this work with the relation retrieved by \citet{misgeld08}: $(g-r) = -0.0314\times M_r + 0.145, \sigma=0.15.$ We found that the 80\% of our sample is within 1$\sigma$ from the relation and the full catalog is included within 2$\sigma$, suggesting that the identified dwarfs are located at roughly the same distance of the group.
To derive the total magnitude, $m_{T}$, we decided to adopt the following procedure. First, we derived $m_{X,T}$ (with X being $g$ or $r$) by using the asymptotic value of the growth curve derived from the fitted isophotes. We also derive $m_T$ using the equation $m_{eq.,T}=\mu_e-2.5log_{10}(2\pi R_{eff}^2)$ from \citet{caon94} and \citet{grahm97}, where $\mu_e$  is the effective surface brightness and $R_{eff}$ is the effective radius. Finally, as a third option, we integrate the light profile up to the radius corresponding to a surface brightness value of $\sim 27\ mag/arcsec^2$ which we considered as our photometry limiting radius, beyond which we are contaminated by the background. The reason for having three different approaches is due to the peculiarities of the profiles for some dwarfs, which did not allow using the simple curve of growth approach straightforwardly.

For galaxies exhibiting a flat asymptotic curve of growth, we assumed the asymptotic value as the total magnitude $m_{T}$ (Case A). In cases where the curve of growth was affected by the presence of bright contaminants (stars or galaxies), we  assumed an upper limit to the magnitude  $m_{eq.,T}$ from the equation above (Case B). Finally, in cases where the behavior of the curve of growth was ambiguous, and the results from $m_{eq.,T}$ were substantially different from the expected magnitudes based on the broad asymptotic value of the curve of growth (e.g. due to  the presence of a plateau region immediately followed by an increase of the flux) we opted for the magnitude estimated using the third method (Case C). For Case A galaxies the uncertainty was determined by taking the semi-difference between the maximum and minimum magnitude estimates among the three methods used. For Case B, we only have upper limits, while for Case C we conservatively assumed an uncertainty of 0.5 mag.

In Figs. \ref{fig:lsb_ex}-\ref{fig:lsb_plot}, we show the cutout from the VST $r-band$ image centered on the dwarf galaxies (left panel), the surface brightness profile in the $r$-band (middle panel) and the color profile for each of the galaxies (right panel).

We also analyzed the GC content in the identified galaxies by examining both the $g$- and $r$-catalogs, and applying to the same selection criteria outlined in Sect. \ref{sec:gclf}. 

Additionally, we selected sources based on their distance from the dwarf galaxy candidate. We considered sources associated with the galaxy to be those located within $4R_e$ \citep{forbes17,lim24}.

For background estimation, we assessed the local background of GC-candidates surrounding each galaxy using three annuli, with radii that are multiples of the galaxy's $R_e$. Specifically, the adopted radii were: 6$R_e$, 7$R_e$, 8$R_e$, and 9$R_e$. For each galaxy, the adopted radius was the mean $R_e$ calculated between the $g$- and $r$-band estimates.
We estimated the number of GCs associated with each galaxy using the formula $N_{GC}^{dw}= (\rho_{gal}-\rho_{background})\times A_{gal}$. 
Here, $\rho_{gal}$ is the on-galaxy GC-candidates density and $\rho_{background}$ is the background density, both measured in number of sources per square $arcmin$. The circular area containing the sources associated with the galaxies ($\pi R^2$, with $R=4 R_e$) was used as $A_{gal}$.  
We repeated this process for all three background regions and adopted the middle value among the three estimates as the final result. None of the galaxies we selected showed a statistically significant overdensity of sources: they either had no GC-candidates or had a number consistent with zero within the uncertainty. Upon a visual inspection, however, we note that the dwarfs with ID=7  and 12 in Table \ref{tab:lsb} exhibit a central concentration of compact sources, consistent with a non-zero GC population. These results are basically consistent with what is observed by the MATLAS team, who also find no GC population in ten of the dwarfs we have in common. Additionally, MATLAS data also suggest that Dwarf\,12 hosts a population of GCs \citep[MATLAS-951,][]{Marleau21}, as well as Dwarf\,13 \citep[MATLAS-976,][]{Marleau24b} which in our data appears to be nucleated (see Appendix \ref{sec:appendix_dw} and Fig. \ref{fig:lsb_plot}).

Regarding Dwarf\,12, this galaxy distinctly exhibits a tidal interaction pattern (as depicted in the lower right panel of Fig. \ref{fig:contour_over}), and it is located in close in projection to a GC-overdensity in the 2D maps, as anticipated in Sect. \ref{sec:density2d},.

\section{Discussion}
\label{sec:discu}

The analysis of the GC system around the two bright galaxies, NGC\,3640 and its close-in-projection companion NGC\,3641, confirms known properties of the pair and expected characteristics of GC systems around bright galaxies. However, we also identified several new features.
\vspace{0.1cm}

From the analysis of the GCs' 2D density maps, we observe that the GC population appears to follow the diffuse light pattern originating from the past merging events involving NGC\,3640. Indeed, we notice a stretching in the distribution of GCs from southwest to northeast, aligning with the observed light pattern of merging features, including what appears being a dwarf galaxy with tidal features (Dwarf\,12, Fig. \ref{fig:contour_over}, lower right panel, and lower panel in Fig \ref{fig:lsb_ex}). 

The highest GC density peak is on NGC\,3641, despite it being less massive than NGC\,3640. Furthermore, there is no compelling evidence of a spatial displacement between the GCs density peak and the location of the galaxy photocenter. 

For both of the brightest galaxies in the field, the radial density profile of GCs candidates follows that of the galaxy light (see Fig. \ref{fig:radial_fit_both}). 
We find an over-density in radial density profile, that we associate to an intra-group GC component, possibly linked to the extended merging activity  of NGC\,3640. This GCs population could now be part of the dark matter halo of the galaxy, and gradually fades into the background  at galactocentric radius $R_{gal}\sim 30$ arcmin.

When studying the GCs color distributions, NGC\,3641 presents a bimodal distribution, quite typical in normal elliptical galaxies \citep{peng06}, while NGC\,3640 shows a broad unimodal color distribution. 

There are known cases that deviate from the typical bimodal distribution, and this is  generally attributed to specific evolutionary paths, like merging activity \citep[e.g. NGC\,1316,][]{goudfrooij01a,goudfrooij01b,richtler2012}, or low accretion fractions \citep[e.g. NGC\,1277,][]{Beasley18}. The unimodal $g-i$ distribution of GCs host in NGC\,3640 could possibly arise from an interaction that happened 3-5 Gyr ago, as pointed out by the shells around NGC\,3640, and which could have been accompanied by a star formation event. However, NGC\,3640 does not exhibit obvious evidence of recent intense star-formation activity, which could have led to the formation of a population of intermediate-color, relatively young (3-5 Gyr old) massive star clusters. Indeed, simple stellar population models predict $(g-i)>0.9$ for stellar populations with ages $t\geq3$ Gyr and $[Fe/H]\geq0.0$ \citep{raimondo05,cantiello24}. Therefore, even such a population of intermediate age GCs should not be the main driver of the unimodality. Only stellar populations with $t\leq2$ Gyr reach the blue colors required to fill the gap in the bimodal GC distribution. Additionally, unlike the field around NGC\,1316 \citep{iodice17}, we do not observe substantial diffuse dust patterns in our images \citep[see also][]{Brough2007}. 
One alternative explanation could be the dry merging of two GC systems with different color distributions, which, when combined, generate the broad color distribution we observe. 

Furthermore, one may notice that the $(u-r)$ color distribution reaches approximately the same red/metal-rich border for both galaxies. However, the $(g{-}i)$ color distribution of NGC\,3641 reveals the presence of a GC population $(g{-}i)\sim0.2$ mag redder than that in NGC\,3640. This is a partly unexpected result, given that more massive galaxies tend to host relatively redder GC systems compared to lower mass ones \citep{peng06}.  This is only specific for the $ugri$ matched catalog since, as mentioned above, using the $gri$ catalog we observe a broad unimodal $(g-i)$ color distribution which consistently reaches the same red limit for the two galaxies.  
An explanation for this behavior might arise from a combination of effects. On the observational side, we must note that the $(u-r)$ color is a more efficient tracer of metallicity than the $(g-i)$ because of the wider wavelength coverage, which spans a magnitude interval over a factor of two larger. 
Moreover, there is an additional observational bias due to the spatial segregation of red GCs and the different image depth at smaller galactocentric radii. Red GCs tend be concentrated around the core of the galaxy \citep{dirsch03,cantiello15,cantiello18fds}, hence in our dataset these have a lower detection efficiency compared to the blue GCs, because of the brighter detection limit around the center of NGC\,3640 compared to the smaller and fainter companion NGC\,3641 (see Fig. \ref{fig:comple_gclf_}).
Finally, the bluer bands in our observational dataset, namely the $u$ and $g$, are both shallower than the $r$-band relative to the TOM, hence the efficiency of detection for red GCs is relatively lower than for blue ones. 
The combination of such effects, plus the peculiar color distribution on NGC\,3640 as highlighted by its color unimodality, could conspire toward the observed bluer $(g-i)$ color in the more massive NGC3640 relative to NGC\,3641. 
Given the degeneracy between the age and metallicity for the optical colors, which is further complicated by the interaction history of the system \citep[i.e., different GC ages,][]{hempel07}, additional near-IR photometry or spectroscopic data may be necessary to fully explain the peculiar behavior of the GC system in NGC\,3640.

Analyzing the GCLF of both galaxies, we found that the turn-over magnitudes of the distribution on NGC\,3640 and NGC\,3641 are consistent with each other, suggesting that the galaxies lie roughly at the same distance ($\sim$ 27 Mpc). This result is consistent with existing distance estimates from  SBF measurements.

Furthermore, through the study of the GCLF, we estimated the total number of GC populations and the specific frequency of the galaxies. Despite the mentioned merging activity, NGC\,3640 exhibits a specific frequency aligned with values for galaxies of the same mass ($S_{\rm N}=2.0\pm0.6$), a result that likely support the dry-merging scenario. NGC\,3641 shows a specific frequency $S_{\rm N}=4.5\pm1.6$, about four times higher than expected for galaxies of its same magnitude \citep[$\langle S_N \rangle=1.1\pm0.9$ for galaxies with -18$\leq M_V \leq$-20]{harris13}, indicating a relatively rich GC population. Examining the behavior of the specific frequency as a function of absolute magnitude in \citet{harris13}, we found other galaxies, such as NGC\,1336, NGC\,4352, NGC\,1172, NGC\,1419, NGC\,1380A/B with -18$\leq M_V \leq$-20 and $S_N\geq3$. By inspecting the main properties (i.e. $R_e$, dynamical masses and GC mass)  of galaxies with $S_N\leq2$ and with $S_N\geq3$, we found no evidence of any differences between these objects, as all the inspected properties are consistent each other within the estimated uncertainties. We noted that all of the galaxies, but NGC\,1380A/B, live in isolated environments (i.e. distant from all the massive galaxies of the group/cluster). 
Hence, although exceptional, NGC\,3641 is not unique in terms of $S_{\rm N}$.
Summarizing, the high $S_{\rm N}$ of NGC\,3641 could be explained if this old, relatively compact elliptical (see Fig. \ref{fig:radial_fit_both}), experienced an intensive early starburst with rapid enrichment and high cluster formation efficiency (CFE), which also boosted its $S_{\rm N}$ and resulted in a more concentrated GC population.
The relatively low $S_{\rm N}$ of NGC\,3640 could be a result of the increased brightness of the galaxy due to the presence of an intermediate-age population, similar to the peculiar brightest galaxy in Fornax, NGC\,1316 \citep{harris13}. Hence, compared to its less massive companion, NGC\,3640 might have had a more gentle star formation history and efficiency, with a lower CFE, explaining its lower $S_{\rm N}$ value.


Finally, we also examined the content of GCs in dwarf galaxy candidates across the field. The results of this analysis indicate that there is no evidence for an over-density of GCs for these galaxies. Nevertheless, we find a couple of dwarf galaxies, particularly one that appears to be a tidally distorted dwarf (Dwarf\,12, Fig. \ref{fig:lsb_ex}), exhibiting a concentration of compact sources in their cores.. Further studies will be necessary to investigate in more detail the true origin of such sources.

Combined together, the observed 30\% larger recession velocity of NGC\,3641 compared to its more massive companion, the lack of any obvious evidence of merging features around NGC\,3641 and its high density of GCs seem to suggest that we are observing a not-too-close and fast encounter between the bright and loose galaxy NGC\,3640 and the denser and fainter companion, NGC\,3641, moving along our line of sight. The combination of the specific geometry of the encounter and dynamical friction effects might be the cause of the increased GC density (by a projection effect) and the lack of evidence for typical morphological irregularities in NGC\,3641. The high $S_{N}$ might further indicate that the GCs system has possibly experienced few external interaction events, similar to the case of NGC\,1336 in Fornax \citep{cantiello20}.

\section*{Summary}
\label{sec:summarize}
In this work, we presented a thorough analysis of the GC population around the interacting galaxy NGC\,3640 and its companion NGC\,3641. The analysis takes advantage of the multi-band ($ugri$) wide $\sim 1.5\times 1.5$ sq. degrees area coverage of the field from the VEGAS survey carried out with the VST telescope.
Our main results can be summarized as follows:

\begin{enumerate}

    \item The GC 2D density map seems to follow the galaxy's light patches resulting from past merging event; 
    \item The highest over-density peak of GCs is observed to be on NGC\,3641 rather than NGC\,3640, the more massive galaxy among the two;
    \item The surface density profiles of the GCs seem to show the presence of an intra-group GC component,  consistent with the scenario of ongoing merging; 
    \item The color distribution of NGC\,3641 shows a bimodal distribution  as expected for a normal elliptical galaxy of this magnitude, while a broad unimodal distribution is observed for the GCs in  NGC\,3640, possibly due to the past merging events;
    \item From the study of the GCLF, the two galaxies appear to be roughly located at the same distance ($\sim$ 27 Mpc); 
    \item The value of the specific frequency for NGC\,3640 aligns with expectations, while NGC\,3641 appears to host a very rich GC population.
    \item  The identification of 17 new dwarf galaxies around NGC\,3640.
    
\end{enumerate}

In the future, thanks to the upcoming facilities such as the Large Synoptic Survey Telescope (LSST), we anticipate significant advancements in our ability to further characterize GC systems like the one we have observed. The LSST, with its wide-field survey capabilities and unprecedented depth, will provide invaluable data that will enable us to delve deeper into the properties  of these systems. This enhanced observational capacity will offer new insights into the formation, evolution, and interactions of galaxies and their associated GC populations.
We are actively seeking new optical and IFU data for objects like NGC\,3641, where we observe extremely high concentrations of GCs. Such data are particularly interesting and ideal for the characterization of the peculiarly high $S_{\rm N}$ as observed in this galaxy.

\section*{Acknowledgements}

We thank the anonymous referee for her/his comments and
constructive suggestions. This work is based on visitor mode observations collected at the European Southern Observatory (ESO) La Silla Paranal Observatory within the VST Guaranteed Time Observations, Programme ID: 097.B-0806(A). 
MM acknowledges financial support from the INAF-OAAb, INFN institute of Naples, the University of Naples Federico II and the VST project (P.I. P. Schipani). 
EI, MS, MC acknowledges support by Italian Ministry for Education University and Research (MIUR) grant PRIN 2022 2022383WFT “SUNRISE”, CUP C53D23000850006, and by the VST funds. EI, MS, MC and RH acknowledge funding from the Italian National Institute of Astrophysics (INAF) through large grant PRIN 12-2022 "INAF-EDGE" (PI L. Hunt). 

This research has made use of the NASA/IPAC Extragalactic Database (NED), which is funded by the National Aeronautics and Space Administration and operated by the California Institute of Technology. We also acknowledge the usage of the Extragalactic Distance Database (EDD, \url{https://edd.ifa.hawaii.edu/}). We made extensive use of the softwares of SExtractor \citep{bertin96} and Topcat \citep[\url{https://www.star.bris.ac.uk/~mbt/topcat/};][]{taylor05}). This research has made use of the VizieR catalogue access tool, CDS, Strasbourg, France \citep{10.26093/cds/vizier}. The original description  of the VizieR service was published in \citet{vizier00}. This research has made use of the SDSS database \url{www.sdss.org}. Funding for the Sloan Digital Sky Survey V has been provided by the Alfred P. Sloan Foundation, the Heising Simons Foundation, the National Science Foundation, and the Participating Institutions. SDSS acknowledges support and resources from the Center for High-Performance Computing at the University of Utah. SDSS telescopes are located at Apache Point Observatory, funded by the Astrophysical Research Consortium and operated by New Mexico State University, and at Las Campanas Observatory, operated by the Carnegie Institution for Science. This work made use of Astropy (\url{http://www.astropy.org}) a community-developed core Python package and an ecosystem of tools and resources for astronomy \citep{astropy:2013, astropy:2018, astropy:2022}.

\bibliographystyle{aa.bst}

\bibliography{Paper_NGC3640_clean.bib}

\newpage
\appendix

\section{GC candidates}
We present an excerpt from the table of GC candidates selected as described in Sect. \ref{sec:3}.

\begin{table*}[htbp]
  \centering
  \makebox[\linewidth][c]{%
    \rotatebox{90}{%
      \begin{minipage}{\textheight}
        \centering 
        \caption{Extract of the GCs catalog.}
        \begin{tabular}{ccccccccccccccc}
         \hline
        \hline
         \\[-2ex]
            RA (J2000) & Dec (J2000) & $m_u$ & $m_g$ & $m_r$ & $m_i$ & CI & FWHM & F.R.. & Elong. & $E(B-V)$  \\
            (deg)&(deg)&(mag)&(mag)&(mag)&(mag)&(mag)&($arcsec$)&($arcsec$)&   &  \\
            (1)&(2)&(3)&(4)&(5)&(6)&(7)&(8)&(9)&(10)&(11)\\
            \\[-2ex]
        
         \hline 
         \\[-1.5ex]
         169.882473 &2.462693 & 22.64 $\pm$ 0.15 & 21.74 $\pm$ 0.04 & 21.16 $\pm$ 0.03 & 20.95 $\pm$  0.05 & -1.04 & 4.2  & 1.33 & 3.34 & 0.05 \\
         169.868631 &2.461746 & 22.19 $\pm$ 0.10  & 22.23 $\pm$ 0.05 & 21.90  $\pm$ 0.05 & 21.92 $\pm$  0.13 & -0.75 & 1.04 & 0.50  & 1.07 & 0.05 \\
         170.234739 &2.467318 & ... & 21.91 $\pm$ 0.04 & 21.27 $\pm$ 0.02 & 21.15 $\pm$  0.06 & -0.96 & 1.28 & 0.67 & 1.13 & 0.06 \\
         169.642611 &2.469429 & 22.68 $\pm$ 0.16 & 22.04 $\pm$ 0.03 & 21.60  $\pm$ 0.04 & 21.36 $\pm$  0.08 & -0.41 & 0.67 & 0.36 & 1.17 & 0.05 \\
         169.904968 &2.474853 & 22.51 $\pm$ 0.11 & 21.92 $\pm$ 0.03 & 21.96 $\pm$ 0.04 & 21.72 $\pm$  0.11 & -0.40 & 0.62 & 0.36 & 1.06 & 0.05 \\
         170.182124 &2.476701 & ... & 22.99 $\pm$ 0.09 & 22.66 $\pm$ 0.07 & 21.79 $\pm$  0.12 & -0.71 & 0.96 & 0.46 & 1.22 & 0.06 \\
         169.656861 &2.478907 & 22.97 $\pm$ 0.17 & 22.82 $\pm$ 0.06 & 22.03 $\pm$ 0.04 & 21.69 $\pm$  0.11 & -1.00  & 1.68 & 0.74 & 1.35 & 0.05 \\
         169.607314 &2.479872 & 22.67 $\pm$ 0.13 & 22.16  $\pm$ 0.04 & 21.24 $\pm$ 0.02 & 20.89 $\pm$  0.05 & -1.15 & 1.89 & 0.91 & 1.14 & 0.05 \\
         169.621129 &2.479677 & 22.50  $\pm$ 0.11 & 21.91 $\pm$ 0.03 & 21.10  $\pm$ 0.02 & 20.84 $\pm$  0.05 & -1.03 & 1.41 & 0.72 & 1.09 & 0.05 \\
         170.348844 &2.482354 & ... & 22.78  $\pm$ 0.06 & 22.23 $\pm$ 0.05 & 21.92 $\pm$  0.13 & -0.87 & 1.75 & 0.87 & 1.53 & 0.05 \\
         169.860197 &2.485210 & 22.68 $\pm$ 0.13 & 22.05  $\pm$ 0.03 & 21.33 $\pm$ 0.02 & 20.98 $\pm$  0.06 & -1.2  & 2.03 & 0.92 & 1.12 & 0.05 \\
    
         ... & ...& ...&... &... & ...& ...& ...& ...& ...&...\\
         170.0820998 & 2.487936	&22.61	$\pm$0.12	&21.66$\pm$	0.02&	21.23	$\pm$0.02&	21.08$\pm$	0.06	&-1.21	&3.43	&1.21	&2.02	&0.05\\
         170.324047 & 2.487348	&23.01	$\pm$0.18	&22.86$\pm$	0.07&	22.19	$\pm$0.05&	21.75$\pm$	0.11	&-0.56	&0.76	&0.43	&1.12	&0.05\\
         ... &...&...&...&...&...&...&...&...&...&...\\
         169.848508 & 2.487541	&22.88	$\pm$0.15	&21.78$\pm$	0.03&	21.13	$\pm$0.02&	21.05$\pm$	0.06	&-0.41	&0.57	&0.36	&1.07	&0.05\\
         169.877903 & 2.488781	&...	&22.75$\pm$	0.06&	22.36	$\pm$0.06&	22.16$\pm$	0.16	&-1.18	&1.97	&0.67	&1.37	&0.05\\
     \hline
     \\[-1.5ex]
     
        \end{tabular}
        \vspace{0.5cm} 
        \begin{minipage}{\textheight}
          \footnotesize 
          \textbf{Notes.} Columns list: (1) Right Ascension; (2) Declination; (3–6) $ugri$-band magnitudes with errors; (7)
concentration index; (8) FWHM in arcseconds; (9) Flux Radius in arcseconds; (10) Elongation, major-to-minor axis ratio; (11) Reddening from
Schlafly \& Finkbeiner (2011). All morphological quantities from Cols. (8-10) are derived from SExtractor.
        \end{minipage}
        
        \label{tab:gc_catalog}
    
      \end{minipage}%
    }%
  }%
\end{table*}

\newpage
\clearpage

\section{Completeness test}
We present the table with the magnitude limits obtained from the completeness function for 26 off-galaxy regions.

\begin{table}[h]

    \centering
    \begin{tabular}{cccccc}
    \hline
    \hline
Region&RA&DEC&$m_{80}$&$m_{50}$& $R_{gal}$\\
      & (deg)  &(deg)& (mag)&(mag)&($arcmin)$\\

      \hline
1 & 170.4062536 &3.2365302   &25.1  &26.09   & 7.66     \\
2 & 170.5679639 &3.2982038   &25.2  &25.96   &17.77    \\
3 & 170.7331735 &3.3511679   &25.0  &25.80   &28.15    \\
4 & 170.8722610 &3.4171864   &24.8  &25.73   &37.26    \\
5 & 170.9903627 &3.5198300   &24.5  &25.23   &46.00    \\
6 & 170.1462799 &3.3381091   &25.3  &26.07   &10.06    \\
7 & 170.0015789 &3.4293940   &25.2  &26.00   &20.30    \\
8 & 169.8074195 &3.4367707   &25.0  &25.79   &30.75    \\
9 & 169.6459389 &3.4404031   &24.8  &25.60   &39.90    \\
10 & 170.0763071 &3.1890070   &25.4  &26.07   &12.44    \\
11 & 169.8971542 &3.1162750   &25.1  &25.92   &23.96    \\
12 & 169.7134168 &2.9615195   &24.9  &25.73   &37.66    \\
13 & 170.5199555 &3.0954495   &25.2  &26.10   &16.72    \\
14 & 170.5915715 &2.9299464   &25.2  &25.94   &26.21   \\
15 & 170.5825757 &2.7968937   &24.9  &25.73   &31.98   \\
16 & 170.5928208 &2.6197908   &24.5  &25.19   &41.44  \\
17 & 170.1993570 &3.0438461   &25.1  &26.11   &12.40   \\
18 & 169.9063405 &2.7892282   &25.0  &25.72   &34.83   \\
19 & 169.7584146 &2.6294325   &24.5  &25.15   &47.88  \\
20 & 170.4395008 &3.4641150   &25.3  &26.06   &16.80  \\
21 & 170.5820283 &3.5928917   &25.2  &25.91   &28.16  \\
22 & 170.6654458 &3.7672842   &24.9  &25.67   &39.49  \\
23 & 170.3073802 &3.5839002   &25.2  &26.01   &21.01   \\
24 & 170.2276120 &3.7647546   &25.1  &25.79   &31.94   \\
25 & 170.0268196 &3.8727972   &24.7  &25.38   &41.14   \\
26 & 170.8122158 &3.8829242   &24.5  &25.04   &50.37  \\

\hline

    \end{tabular}
    \caption{Summary of 50\% and 80\% magnitude limits obtained for 26 off-galaxy regions around NGC\,3640, with corresponding positions of the center of the analyzed  regions and distances from NGC\,3640.  }
    \label{tab:off_gal_compl}

\end{table}

\newpage
\clearpage

\section{Dwarfs measured properties}

\newpage
\clearpage
\begin{table}[htbp]
  \centering
  
  \makebox[\textwidth][c]{%
    \rotatebox{90}{%
      \begin{minipage}{\textheight}
        \centering 
        \caption{Dwarfs galaxy properties}
        \begin{tabular}{cccccccccccccc}
         \toprule
        \toprule

              ID     & RA     & DEC  & $ \mu_{e,g}$  & $R_{e,g}$ & $n_g$& $\mu_{e,r}$ & $R_{e,r}$ & $n_r$     & $m_{g,T}$  & $m_{r,T}$&  $ (g-r)_{e,o}$ &Case & Note\\
                & (deg) & (deg) & ($mag/arcsec^2$)& (arcsec) &       &($mag/arcsec^2$)& (arcsec) &          & (mag) &(mag) & (mag) \\
            (1)&(2)&(3)&(4)&(5)&(6)&(7)&(8)&(9)&(10)&(11)&(12)&(13)&(14)\\
            \\
        
         \midrule 
     Dwarf 1  &170.828434 & 3.955088  & 22.6 $\pm$ 0.1  & 6.4  $\pm$ 0.1 & 0.8  $\pm$ 0.1 & 22.0 $\pm$ 0.1 & 6.0  $\pm$ 0.1 & 0.7  $\pm$ 0.1&     <16.6  & 14.9 $\pm$ 0.6   & 0.53 &  B  &    \\
Dwarf 2  &170.584866 & 3.900309  & 25.6 $\pm$ 0.1  & 12.7 $\pm$ 0.1 & 0.5  $\pm$ 0.1 & 25.5 $\pm$ 0.1 & 16.2 $\pm$ 0.1 & 0.6  $\pm$ 0.1&     <18.0   &           <17.5     &  0.23 &  B  &   \\
Dwarf 3  &169.711403 & 3.763879  & 23.8 $\pm$ 0.1  & 8.3  $\pm$ 0.1 & 0.5  $\pm$ 0.1 & 23.4 $\pm$ 0.1 & 8.3  $\pm$ 0.1 & 0.5  $\pm$ 0.1& 16.7 $\pm$ 0.3  & 16.4 $\pm$ 0.2  &  0.41 &  A  &    \\
Dwarf 4  &169.628457 & 3.646629  & 24.1 $\pm$ 0.1  & 4.8  $\pm$ 0.1 & 0.9  $\pm$ 0.1 & 23.6 $\pm$ 0.1 & 4.3  $\pm$ 0.1 & 1.0  $\pm$ 0.1& 18.5 $\pm$ 0.5  & 18.1 $\pm$ 0.5  &  0.37 &  C  &    \\
Dwarf 5  &171.035561 & 3.734798  & 25.3 $\pm$ 0.1  & 21.5 $\pm$ 0.1 & 2.4  $\pm$ 0.1 & 23.2 $\pm$ 0.1 & 7.8  $\pm$ 0.1 & 1.2  $\pm$ 0.1&       <16.7   &     <16.8         &  0.65 &  B  &  \\
Dwarf 6  &170.517235 & 3.614390  & 23.3 $\pm$ 0.1  & 3.9  $\pm$ 0.1 & 0.8  $\pm$ 0.1 & 22.5 $\pm$ 0.1 & 3.7  $\pm$ 0.1 & 0.8  $\pm$ 0.1& 17.9 $\pm$ 0.2  & 17.3 $\pm$ 0.2  &  0.61 &  A  &    \\
Dwarf 7  &170.278923 & 3.468104  & 24.4 $\pm$ 0.1  & 11.2 $\pm$ 0.1 & 0.7  $\pm$ 0.1 & 23.9 $\pm$ 0.1 & 11.6 $\pm$ 0.1 & 0.7  $\pm$ 0.1& 17.1 $\pm$ 0.1  & 16.6 $\pm$ 0.1  &  0.52 &  A  &  1 \\
Dwarf 8 &170.467360 & 3.405171  & 23.3 $\pm$ 0.1  & 15.6 $\pm$ 0.1 & 0.4  $\pm$ 0.1 & 23.0 $\pm$ 0.1 & 15.5 $\pm$ 0.1 & 0.5  $\pm$ 0.1& 16.5 $\pm$ 0.5  & 15.8 $\pm$ 0.5  &   0.24  & C  &  1  \\
Dwarf 9 &169.952101 & 3.456099  & 24.3 $\pm$ 0.1  & 2.9  $\pm$ 0.1 & 1.0  $\pm$ 0.1 & 23.7 $\pm$ 0.1 & 2.9  $\pm$ 0.1 & 1.1  $\pm$ 0.1& 19.4 $\pm$ 0.3  & 18.7 $\pm$ 0.3  &   0.59  &  A &     \\
Dwarf 10 &169.934790 & 3.437198  & 25.3 $\pm$ 0.1  & 2.9  $\pm$ 0.1 & 0.8  $\pm$ 0.1 & 25.0 $\pm$ 0.1 & 2.8  $\pm$ 0.1 & 0.7  $\pm$ 0.1& 20.5 $\pm$ 0.3  & 20.4 $\pm$ 0.2  &  0.22 &   A &  1   \\
Dwarf 11 &169.607596 & 3.345803  & 25.2 $\pm$ 0.1  & 3.2  $\pm$ 0.1 & 0.9  $\pm$ 0.1 & 24.4 $\pm$ 0.1 & 2.4  $\pm$ 0.1 & 0.8  $\pm$ 0.1& 20.5 $\pm$ 0.1  & 20.2 $\pm$ 0.2  &  0.40 &    A&    \\
Dwarf 12 &170.132592 & 3.232916  & 26.5 $\pm$ 0.1  & 35.6 $\pm$ 0.1 & 0.9  $\pm$ 0.1 & 26.0 $\pm$ 0.1 & 43.7 $\pm$ 0.1 & 0.9  $\pm$ 0.1& 17.2 $\pm$ 0.5  & 16.0 $\pm$ 0.5  &  0.77 &    C&  1-2 \\
Dwarf 13 &170.467546 & 3.240638   & 26.7 $\pm$ 0.1  & 12.7 $\pm$ 0.1 & 1.0  $\pm$ 0.1 & 25.8 $\pm$ 0.1 & 10.9 $\pm$ 0.1 & 0.8  $\pm$ 0.1& 19.5 $\pm$ 0.5  & 18.8 $\pm$ 0.3  & 0.63 &   A &  1-3 \\
Dwarf 14 &170.723288 & 3.247790  & 24.3 $\pm$ 0.1  & 6.0  $\pm$ 0.1 & 0.8  $\pm$ 0.1 & 23.6 $\pm$ 0.1 & 5.9  $\pm$ 0.1 & 0.7  $\pm$ 0.1& 17.9 $\pm$ 0.2  & 17.3 $\pm$ 0.2  &  0.56  &   A&  1   \\
Dwarf 15 &170.502653 & 3.124323  & 25.4 $\pm$ 0.1  & 2.3  $\pm$ 0.1 & 1.0  $\pm$ 0.1 & 24.6 $\pm$ 0.1 & 2.1  $\pm$ 0.1 & 0.9  $\pm$ 0.1& 21.1 $\pm$ 0.2  & 20.6 $\pm$ 0.2  &  0.63  &   A&     \\
Dwarf 16 &170.313761 & 3.094260  & 26.8 $\pm$ 0.1  & 5.8  $\pm$ 2.1 & 1.9  $\pm$ 0.1 & 25.0 $\pm$ 0.1 & 3.0  $\pm$ 0.1 & 1.0  $\pm$ 0.1&     <21.0   &           <20.6     &  0.64  &  B &      \\
Dwarf 17 &170.812218 & 3.071396  & 26.2 $\pm$ 0.1  & 9.08 $\pm$ 0.1 & 1.2  $\pm$ 0.1 & 25.3 $\pm$ 0.1 & 6.6  $\pm$ 0.1 & 1.0  $\pm$ 0.1& 19.4 $\pm$ 0.5  & 19.2 $\pm$ 0.5  &  0.47  &   C&      \\
Dwarf 18 &169.815282 & 3.006275  & 25.9 $\pm$ 0.1  & 3.6  $\pm$ 0.1 & 0.6  $\pm$ 0.1 & 25.5 $\pm$ 0.1 & 4.1  $\pm$ 0.1 & 0.6  $\pm$ 0.1& 21.6 $\pm$ 0.4  & 21.0 $\pm$ 0.3  &  0.46  &   A&  1  \\
Dwarf 19 &169.549790 & 3.114700  & 24.5 $\pm$ 0.1  & 2.5  $\pm$ 0.1 & 0.7  $\pm$ 0.1 & 23.9 $\pm$ 5.1 & 2.3  $\pm$ 0.1 & 0.6  $\pm$ 0.1& 20.6 $\pm$ 0.1  & 20.2 $\pm$ 0.1  &  0.47  &   A&      \\
Dwarf 20 &170.458639 & 2.945284  & 25.2 $\pm$ 0.1  & 7.3  $\pm$ 0.1 & 0.9  $\pm$ 0.1 & 24.3 $\pm$ 0.1 & 5.9  $\pm$ 0.1 & 0.9  $\pm$ 0.1& 18.9 $\pm$ 0.5  &         <18.5   &  0.54  &   B&  1-3 \\
Dwarf 21 &170.190186 & 2.988030  & 23.8 $\pm$ 0.1  & 3.8  $\pm$ 0.1 & 0.9  $\pm$ 0.1 & 23.6 $\pm$ 0.1 & 4.3  $\pm$ 0.1 & 0.9  $\pm$ 0.1& 18.9 $\pm$ 0.5  & 18.5 $\pm$ 0.5  &  0.29  &   C&  1  \\
Dwarf 22 &169.832724 & 2.791686  & 24.4 $\pm$ 0.1  & 7.3  $\pm$ 0.1 & 0.9  $\pm$ 0.1 & 23.6 $\pm$ 0.1 & 6.0  $\pm$ 0.1 & 0.7  $\pm$ 0.1& 17.5 $\pm$ 0.3  & 17.2 $\pm$ 0.2  &  0.55  &   A&  1-3 \\
Dwarf 23 &169.849007 & 2.528960  & 24.7 $\pm$ 0.1  & 8.0  $\pm$ 0.1 & 1.3  $\pm$ 0.1 & 24.3 $\pm$ 0.1 & 6.7  $\pm$ 0.1 & 1.2  $\pm$ 0.1& 18.2 $\pm$ 0.5  & 18.2 $\pm$ 0.5  &  0.27  &   C&    \\
Dwarf 24 &169.613070 & 2.509576  &&&&&&&&& \\ 
Dwarf 25 &169.646269 & 2.471812  &&&&&&&&& \\ 
Dwarf 26  &170.497251 & 3.495409  &&&&&&&&& \\   
Dwarf 27 &170.175635 & 2.933353  &&&&&&&&& \\

     \hline
        \end{tabular}
        \begin{minipage}{\textheight}
          \footnotesize 
          \textbf{Notes.} Columns list: (1) Right Ascension; (2) Declination; (4–9) Results of the fitting procedure of surfaces profile in the $g$- and $r$-band; (10) Total magnitude in the $g$-band; (11) Total magnitude in the $r$-band; (12) $(g{-}r)$ color computed within an aperture of one $R_e=(R_{e,r}+R_{e,g})/2$; (13) Adopted method to estimate the magnitude: Case A we assumed the  asymptotic value of the curve of growth, Case B we assumed an upper limit to the magnitude $m_{eq,T}$; Case C we integrate the light up to where the galaxy light profile approach to $\mu\sim27mag/arcsec^2$; (14) Notes: 1)dwarf also present in the MATLAS sample; 2) dwarf classified as UDG in \citet{Marleau21}; 3) spectroscopic information from \citet{Heesters23}.
        \end{minipage}
       
        \label{tab:lsb}
      \end{minipage}
    }
  }
\end{table}

\newpage
\clearpage

\section{Dwarf galaxy candidates in the field}
\label{sec:appendix_dw}

In this appendix we show the LSB dwarf candidates identified (see Sect. \ref{sec:lsb}). The error bars are estimated by propagating the error on the isophotal flux, background value, zeropoint and extinction. As uncertainty on the background we assumed the 30\% of the fluctuation estimated by Autoprof\footnote{ Autoprof estimates the sky noise as
the 68.3 percentile of flux values below the sky level. For more detail see \citet{stone21}.}. Judging from the images and radial profiles, Dwarf\,12 and 13 might be nucleated.

\begin{figure*}[h!!]
    \centering
       \caption[width=\textwidth]{Cutout around the the identified dwarf galaxies in the $r$-band (left panel), the surface brightness profile (middle panel) and the color profile or each galaxy (right panel). The green circle and blue dashed line represent the $R_e$ in the $r-band$, also reported in Table \ref{tab:lsb}.}
    \label{fig:lsb_plot}
    \includegraphics[width=\textwidth]{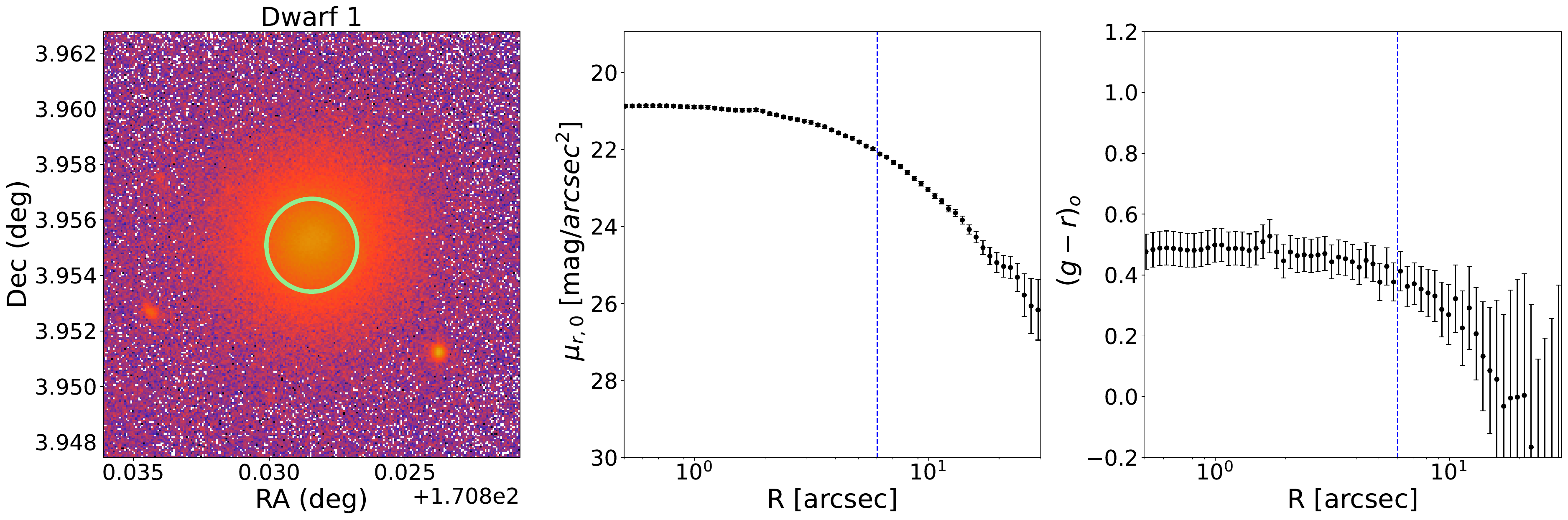}
    \includegraphics[width=\textwidth]{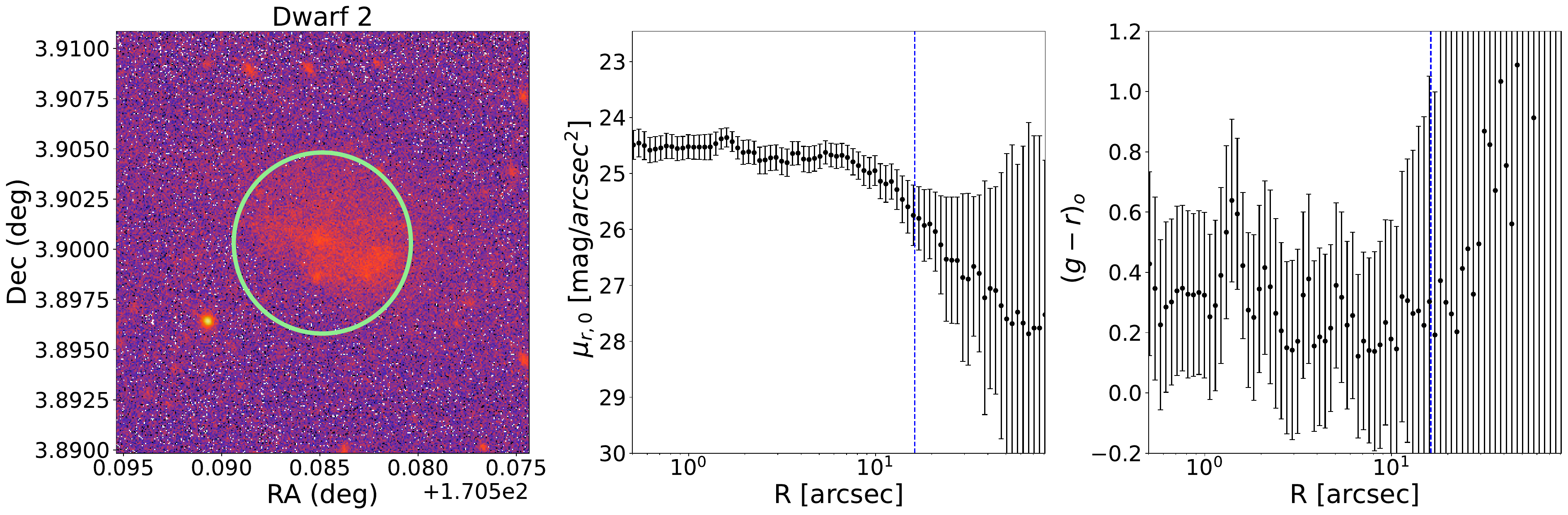}
    \includegraphics[width=\textwidth]{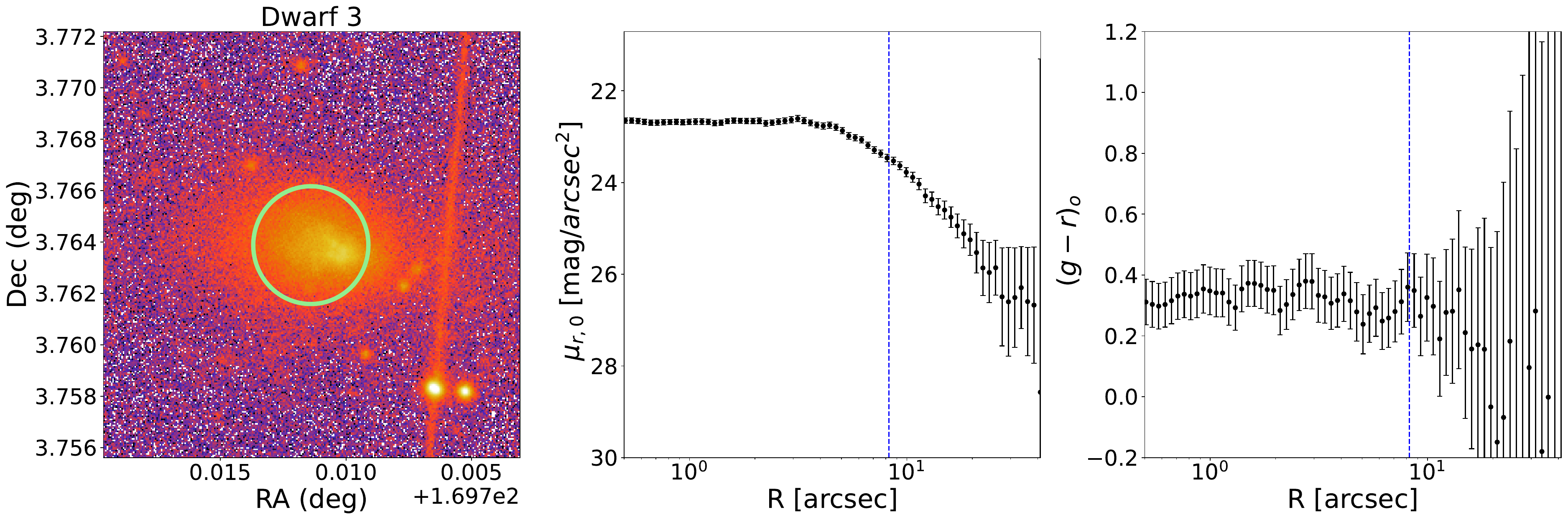}
     
\end{figure*}

\newpage
\clearpage

\begin{figure*}
    \centering
    \includegraphics[width=\textwidth]{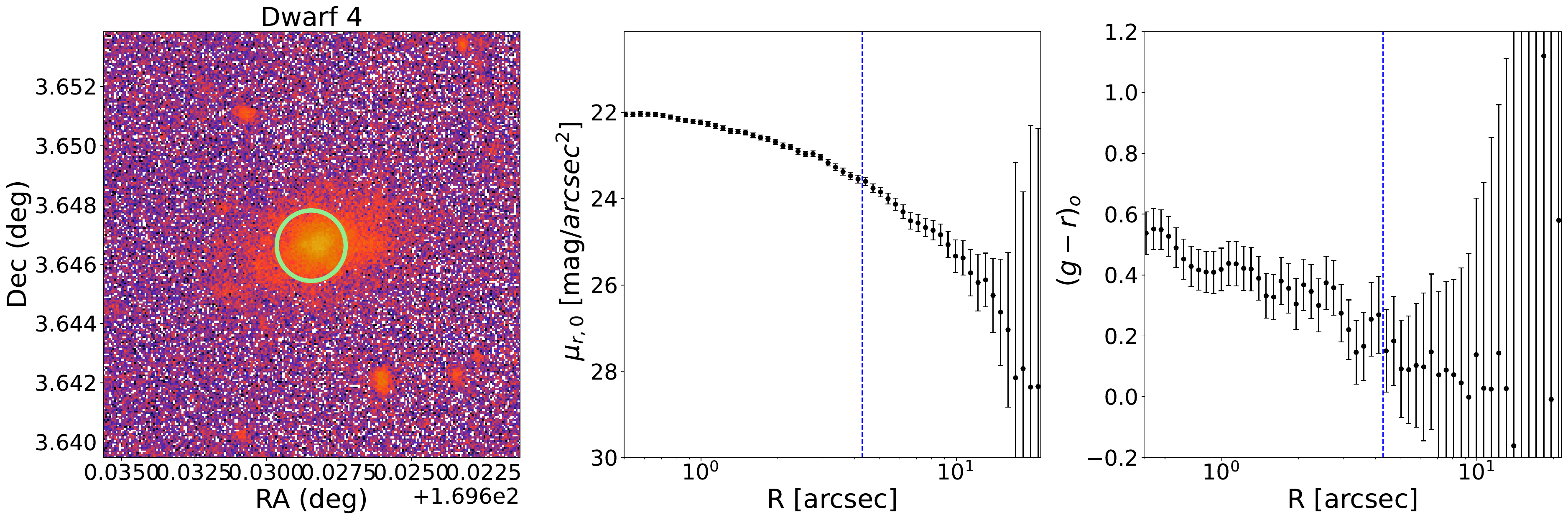}
    \includegraphics[width=\textwidth]{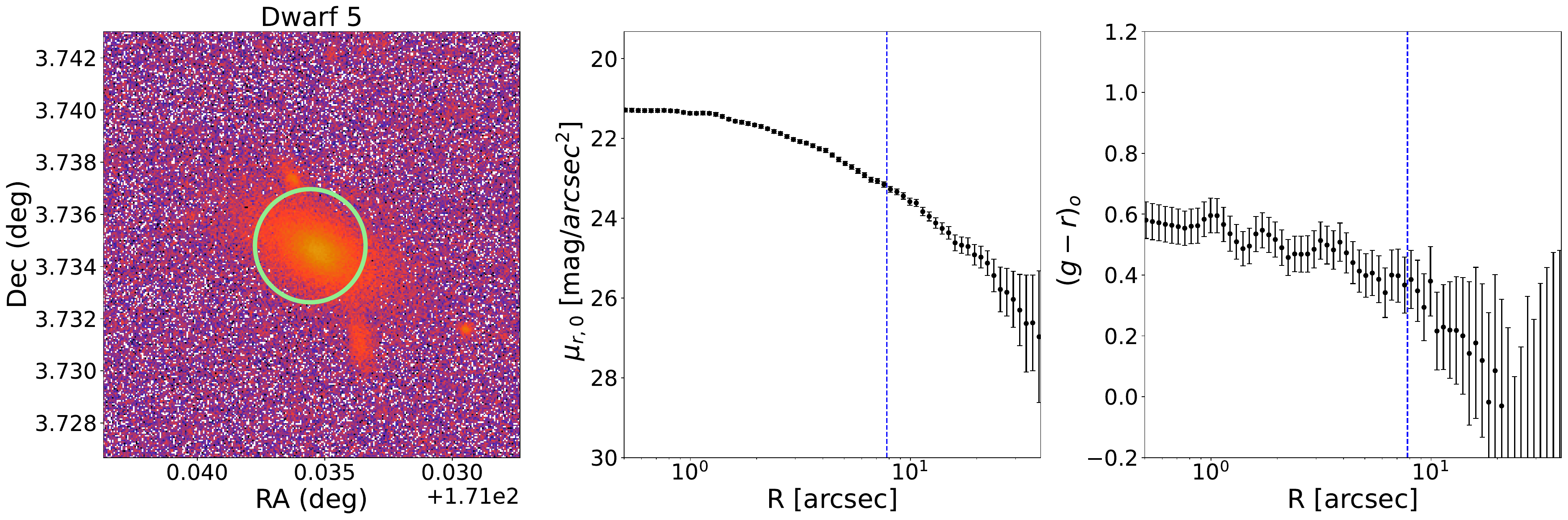}
    \includegraphics[width=\textwidth]{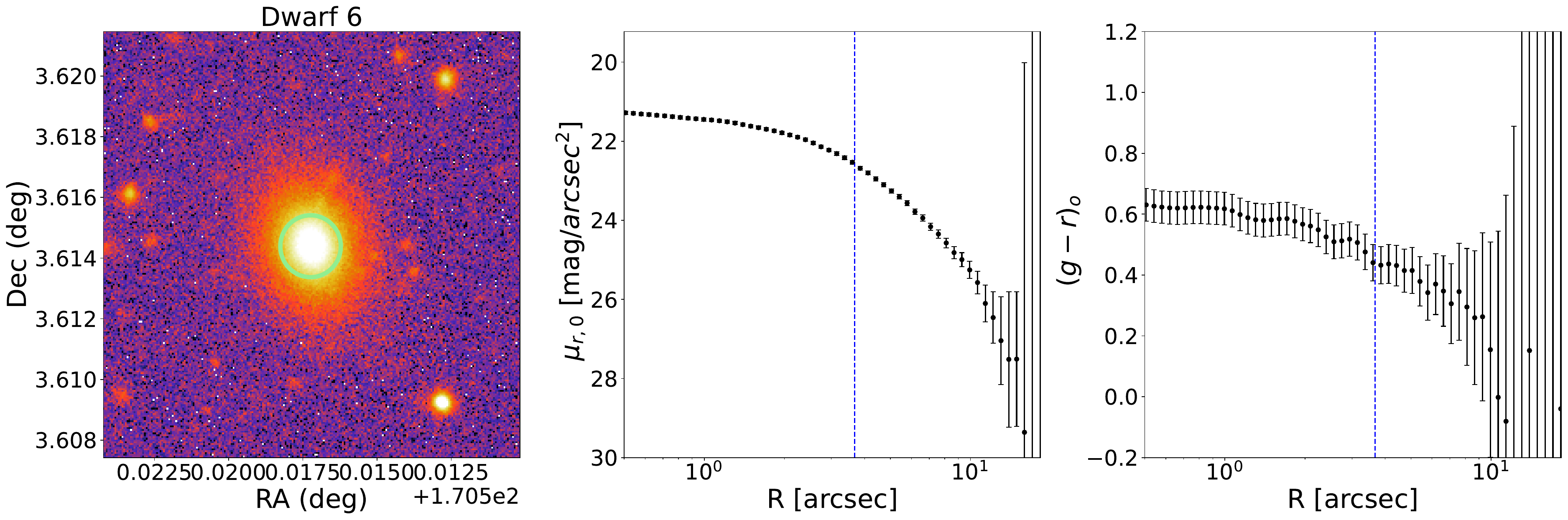}
    \includegraphics[width=\textwidth]{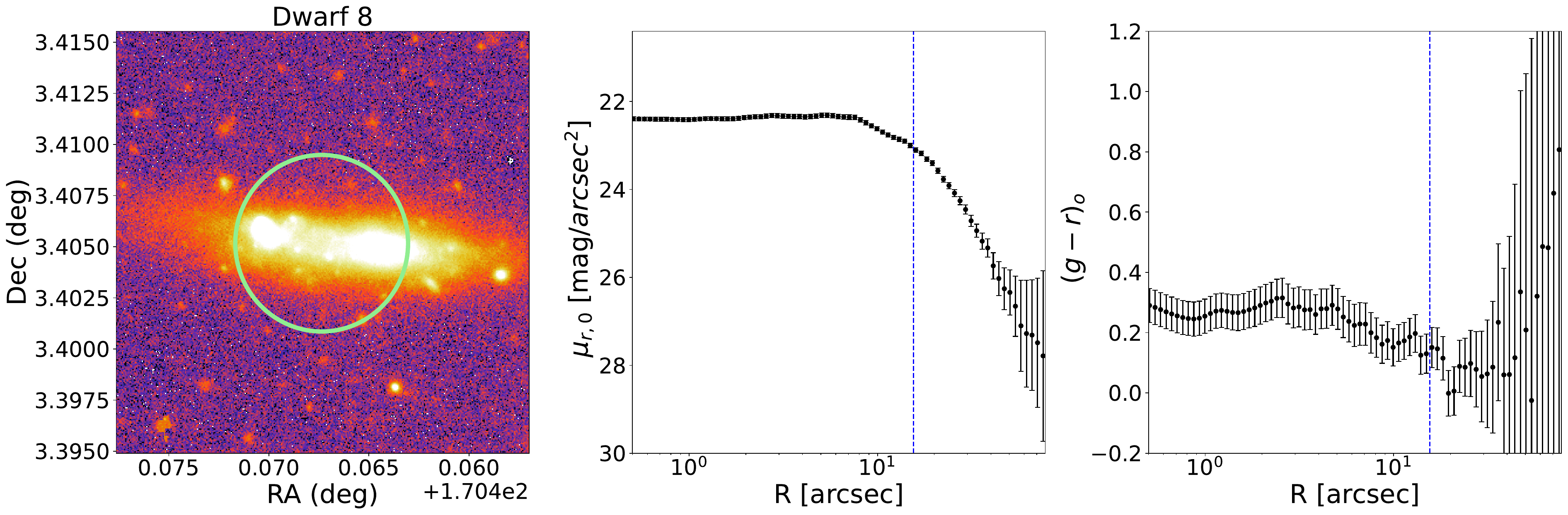}

\end{figure*}
\newpage
\clearpage

\begin{figure*}
    \centering
    \includegraphics[width=\textwidth]{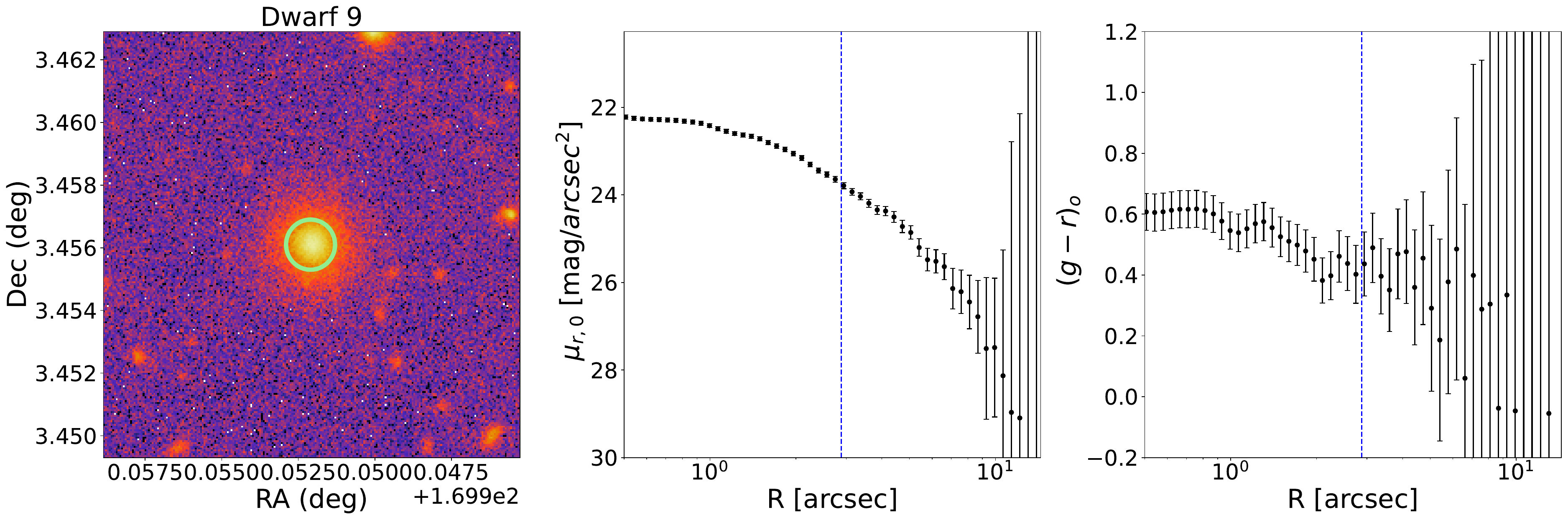}
    \includegraphics[width=\textwidth]{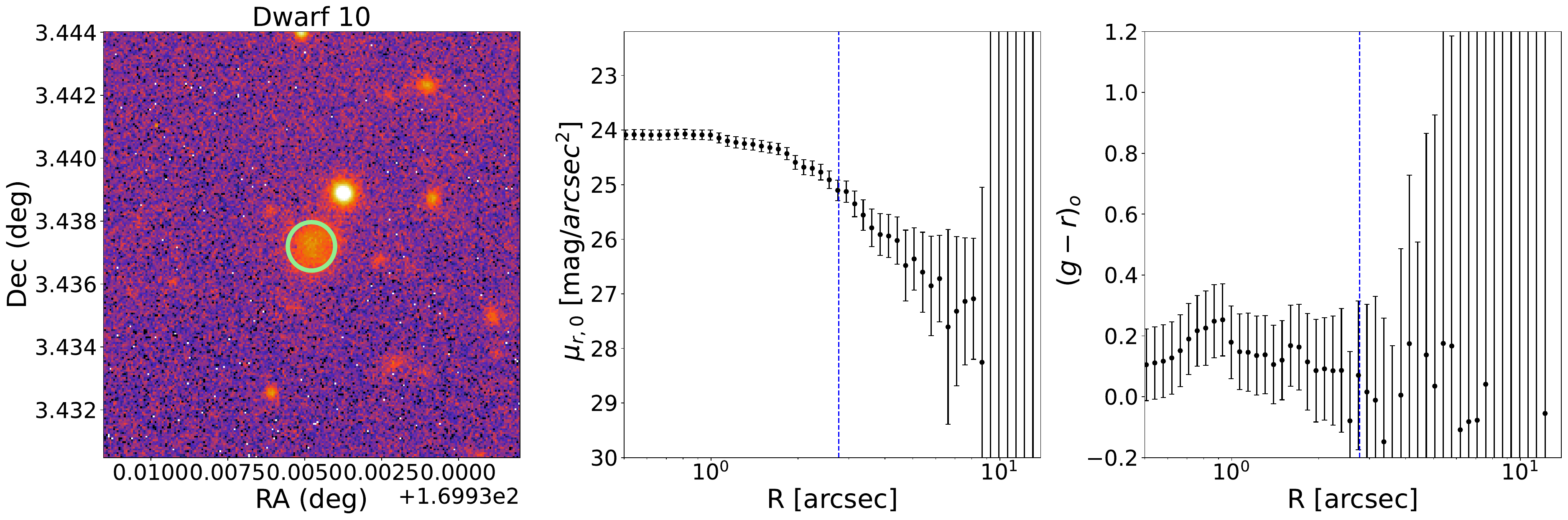}
    \includegraphics[width=\textwidth]{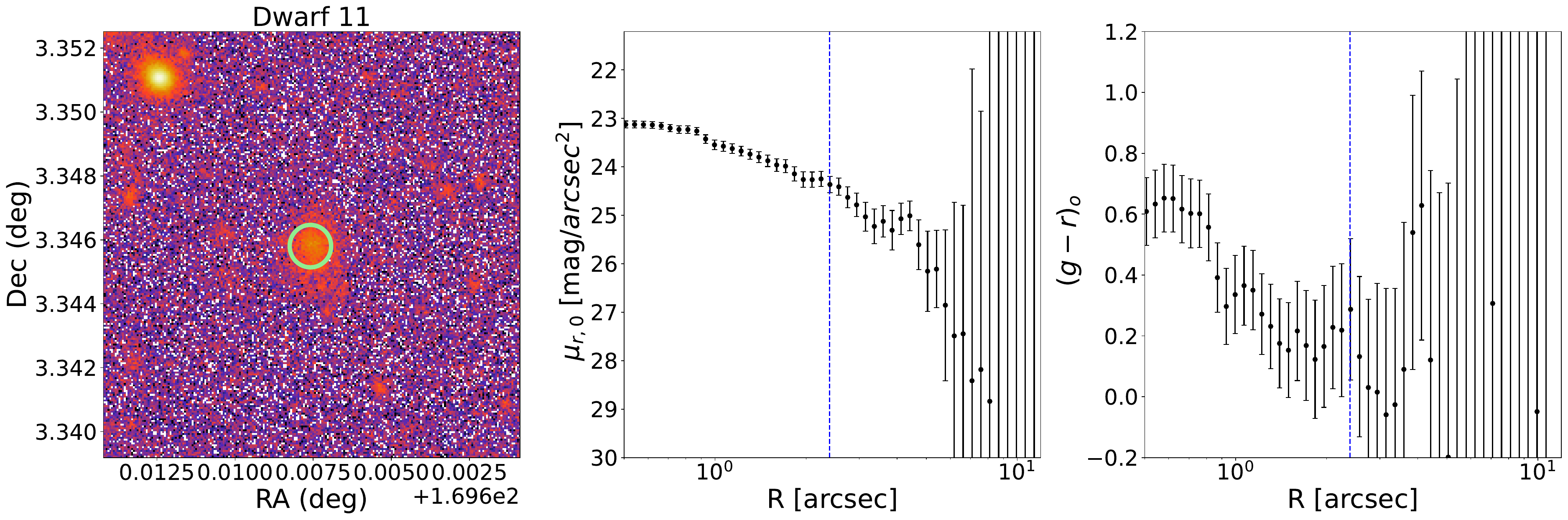}
    \includegraphics[width=\textwidth]{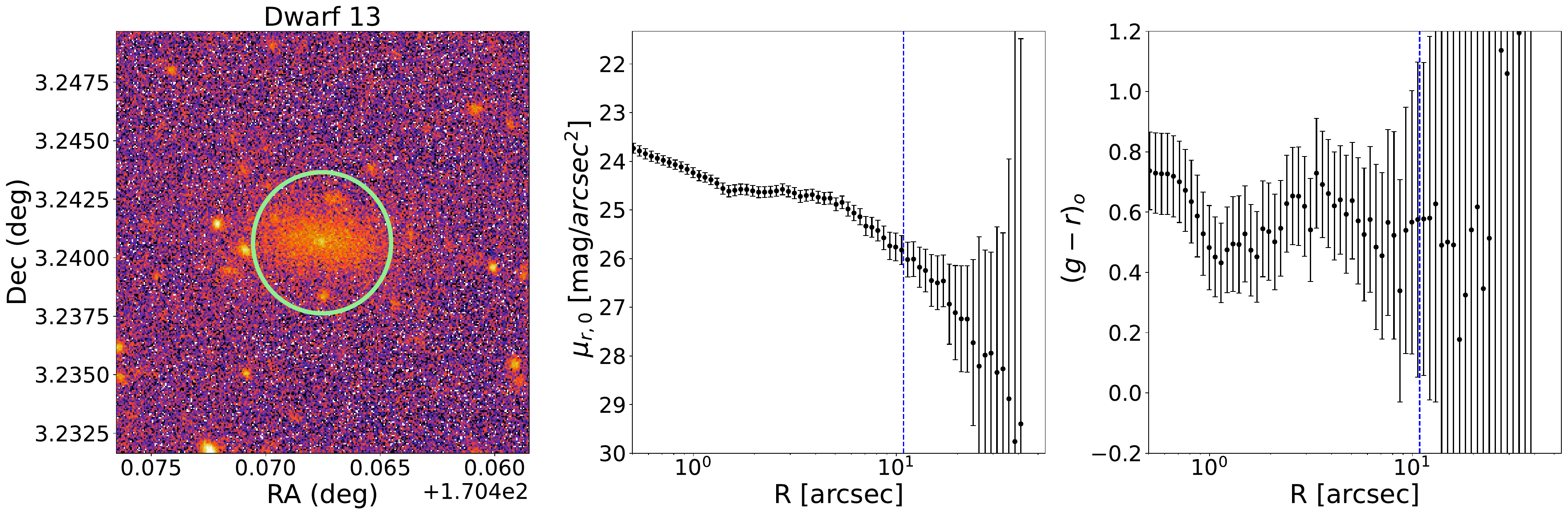}

\end{figure*}
\newpage
\clearpage

\begin{figure*}
    \centering
    \includegraphics[width=\textwidth]{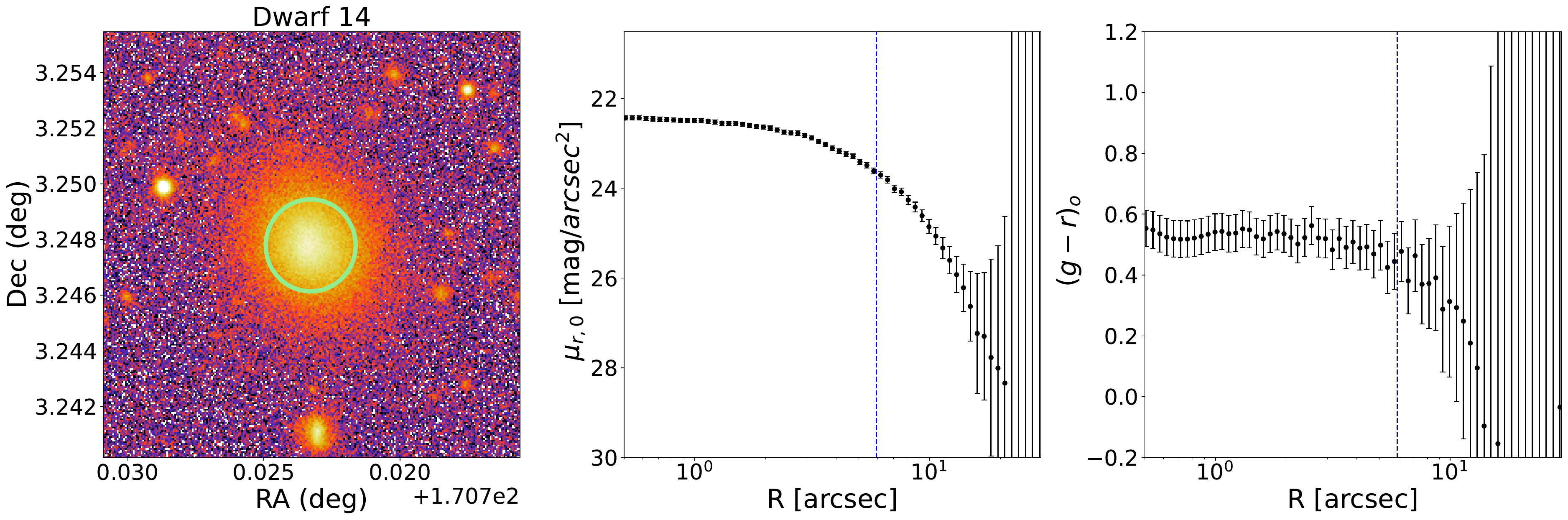}
    \includegraphics[width=\textwidth]{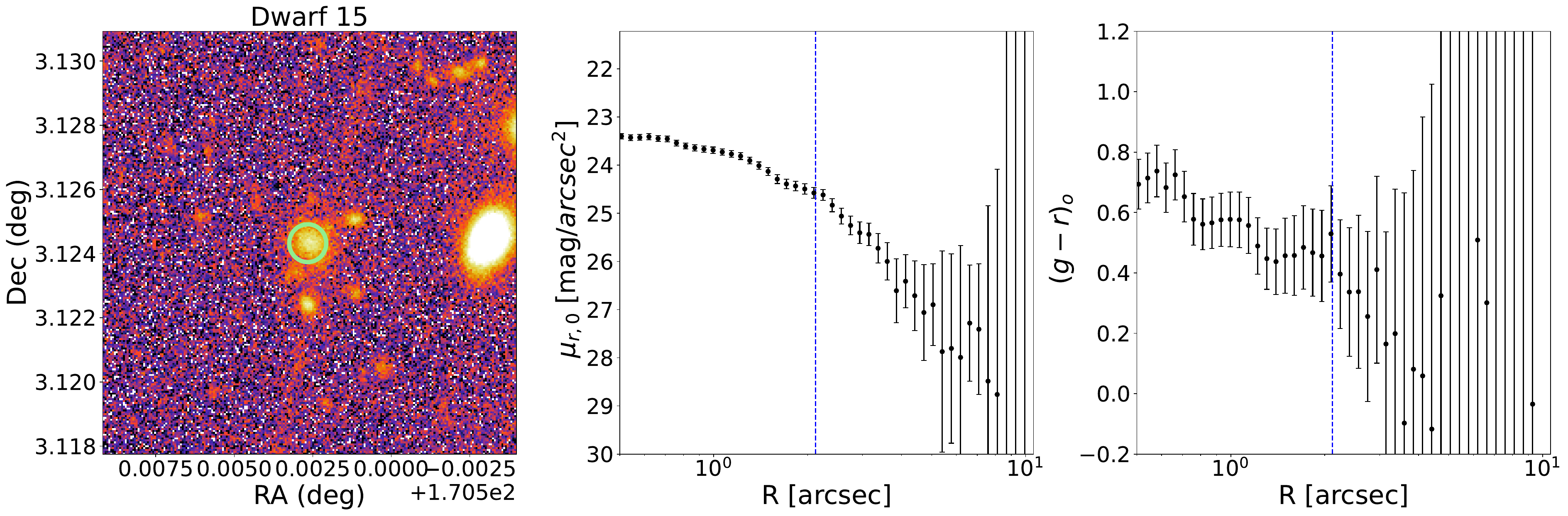}
    \includegraphics[width=\textwidth]{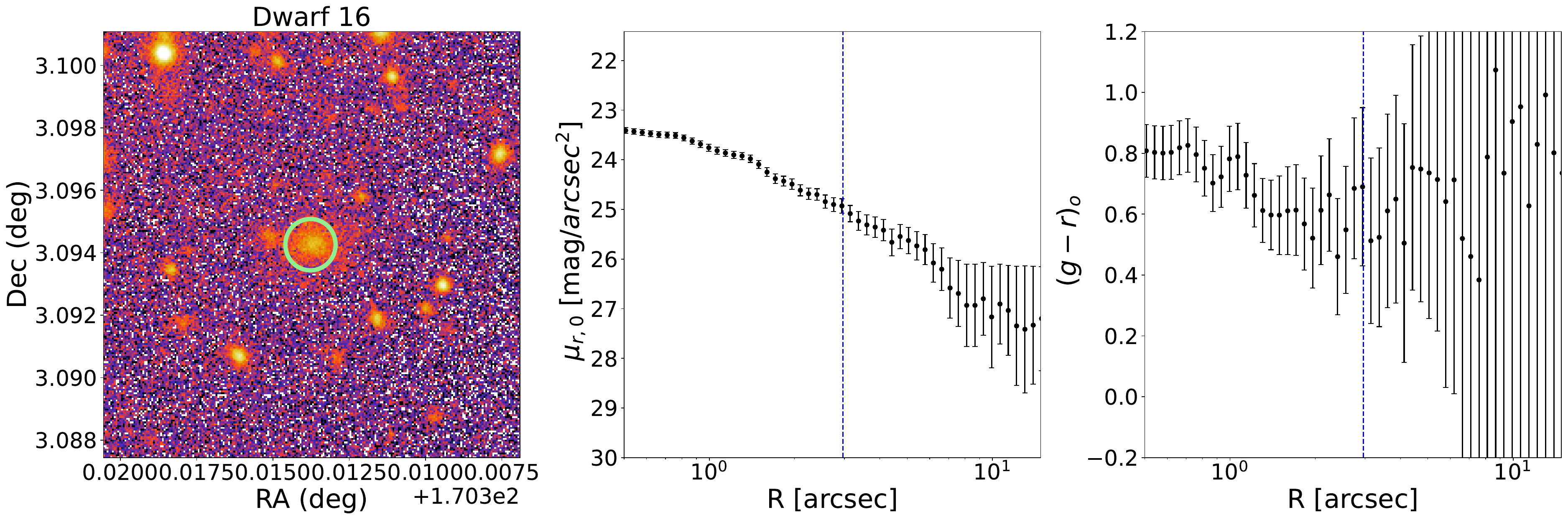}
    \includegraphics[width=\textwidth]{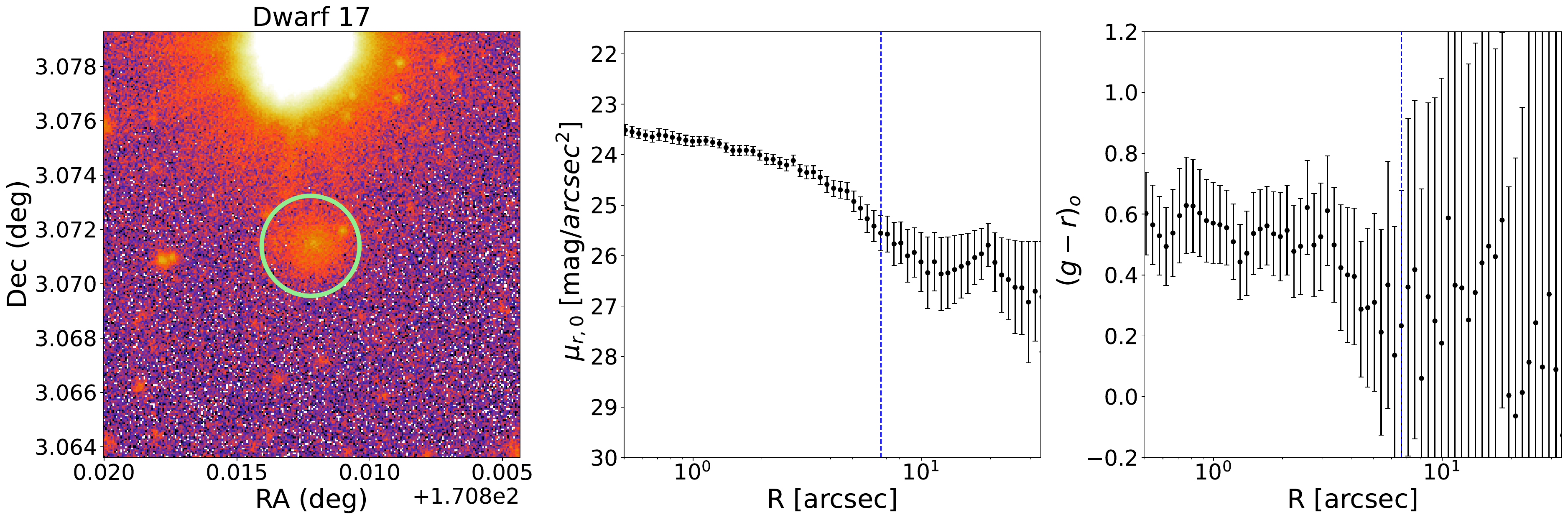}

\end{figure*}
\newpage
\clearpage

\begin{figure*}
    \centering
 
    \includegraphics[width=\textwidth]{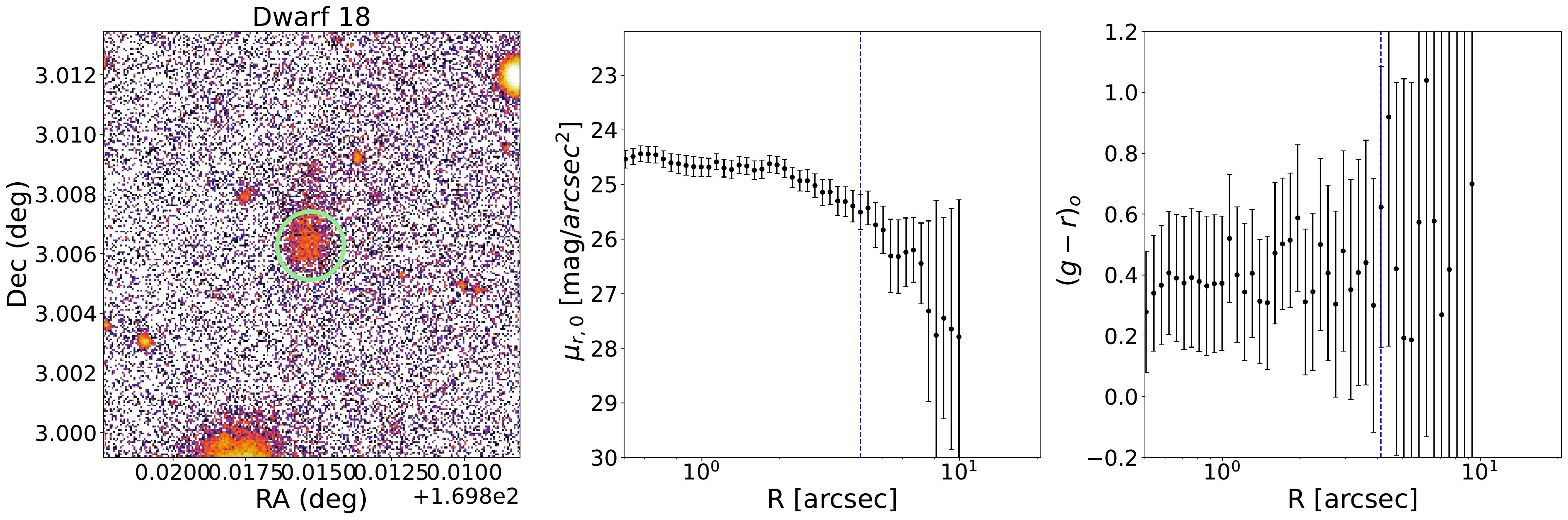}
    \includegraphics[width=\textwidth]{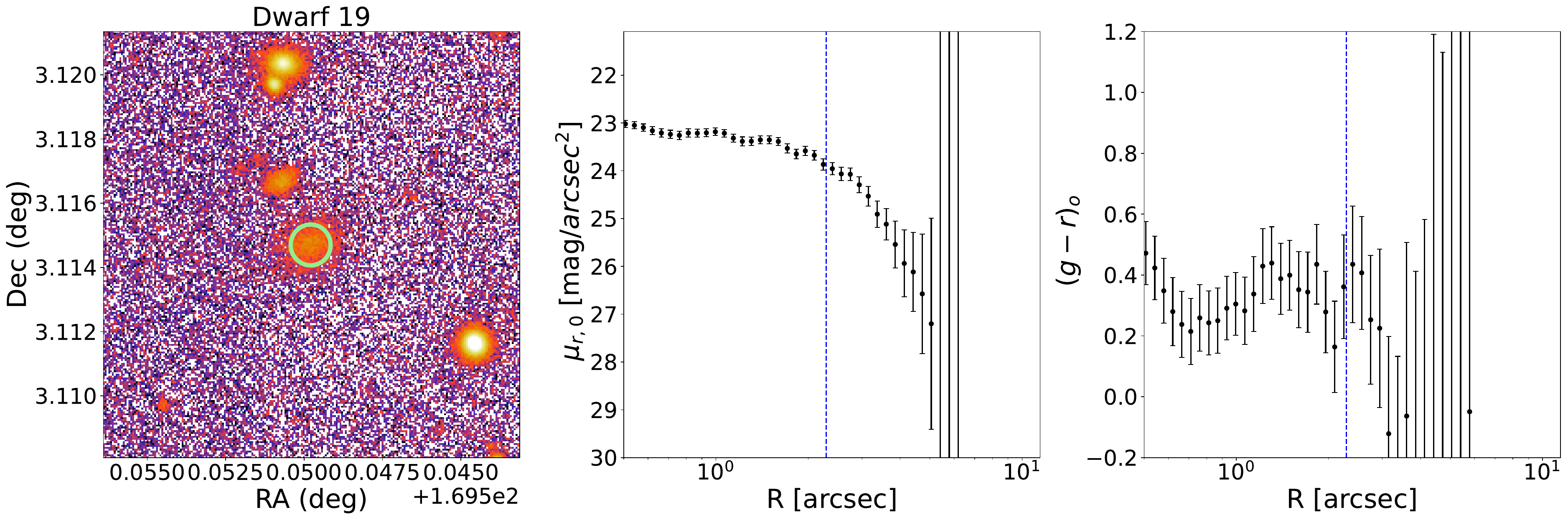}
    \includegraphics[width=\textwidth]{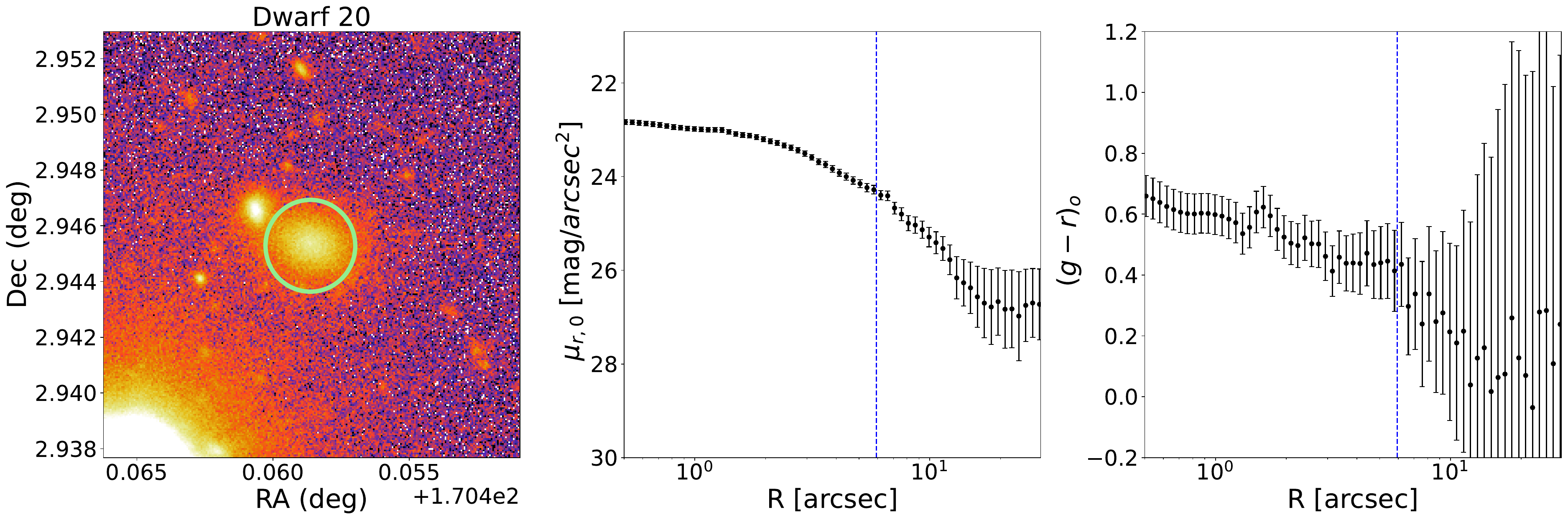}
    \includegraphics[width=\textwidth]{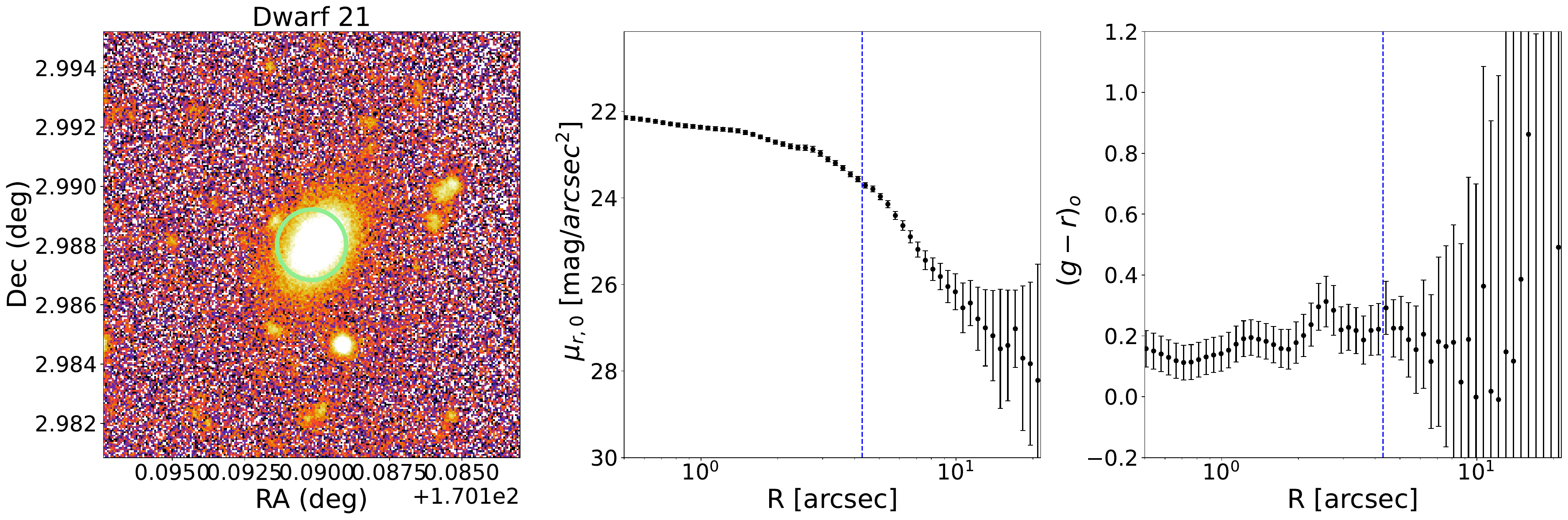}

\end{figure*}

\newpage
\clearpage

\begin{figure*}
    \centering
   
    \includegraphics[width=\textwidth]{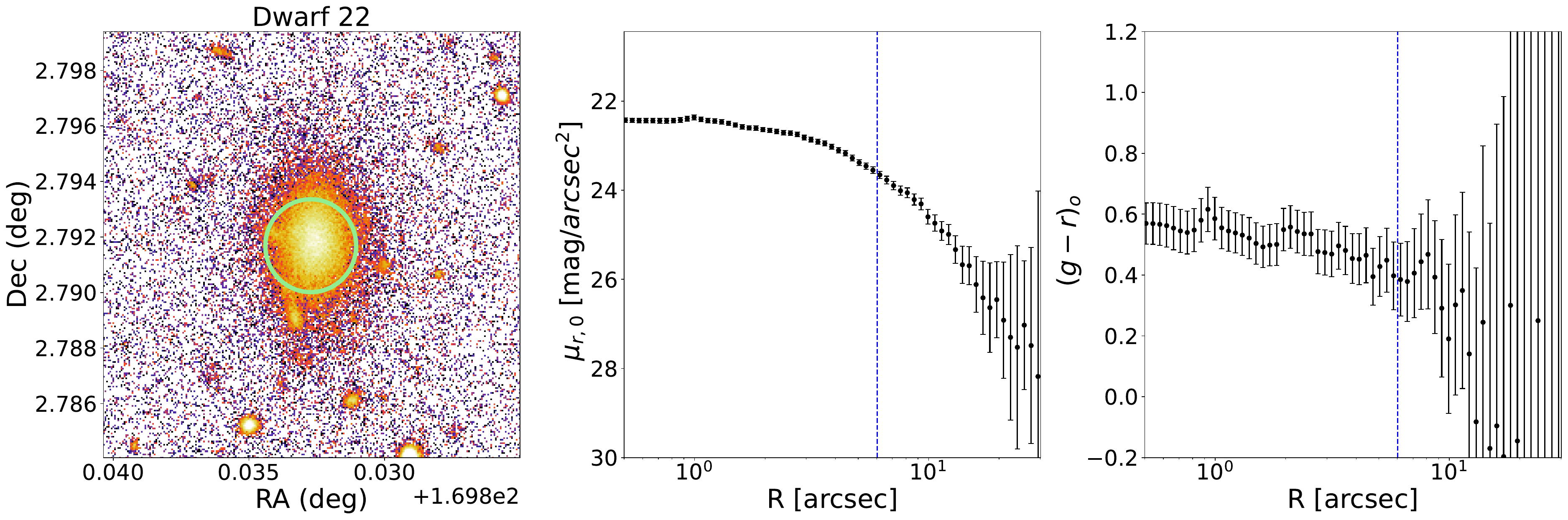}
    \includegraphics[width=\textwidth]{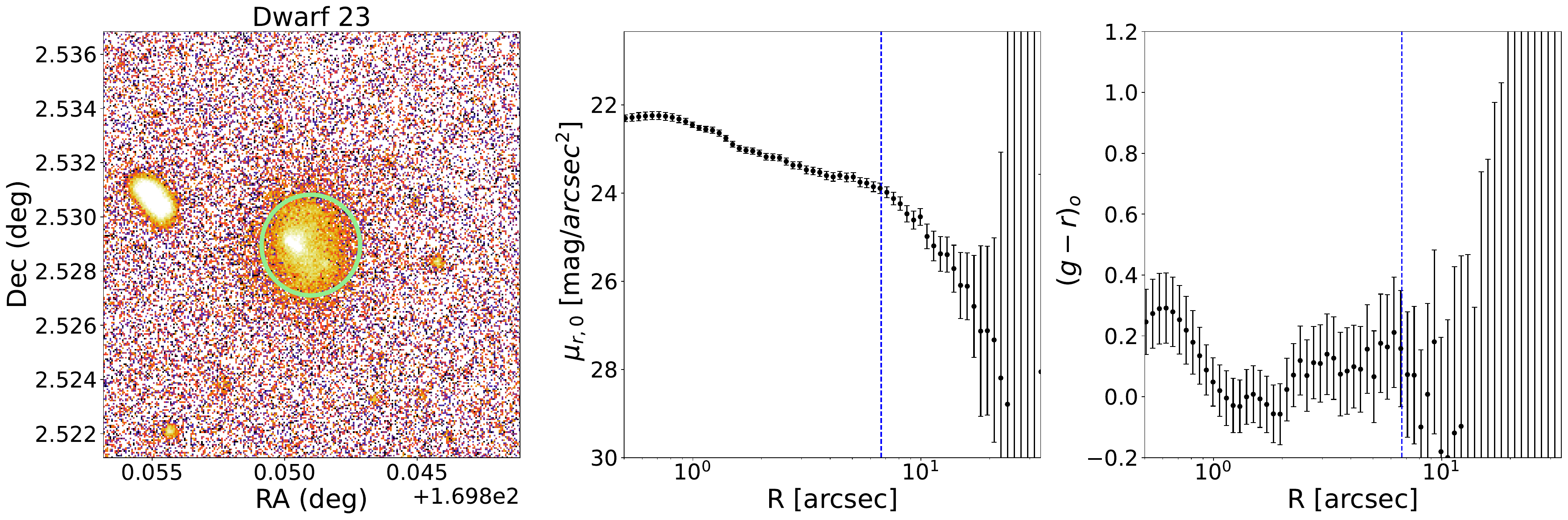}
   
\end{figure*}

\end{document}